\documentclass[preprint,superscriptaddress,amsmath,amssymb,aps,floatfix]{revtex4-1}

\usepackage{graphicx}
\usepackage{slashed}
\usepackage{bm}
\usepackage{hyperref}
\usepackage{graphicx,color}
\usepackage{multirow}
\usepackage{amsmath,amstext,array,amssymb}
\usepackage[normalem]{ulem}
\usepackage{graphicx}
\usepackage{xfrac}
\usepackage[T1]{fontenc}

\newcolumntype{C}{>{$}c<{$}}

\newcommand{\MP}[2]{ p_{#1}\cdot p_{#2} \, }
\newcommand{\trt}[2]{\mbox{Tr}(p_{#1} p_{#2})}
\newcommand{\trf}[4]{\mbox{Tr}(p_{#1} p_{#2} p_{#3} p_{#4})}
\newcommand{\trs}[6]{\mbox{Tr}(p_{#1} p_{#2} p_{#3} p_{#4} p_{#5} p_{#6})}

\begin{document}
\title{2-, 3- and 4-Body Decays in the Constructive Standard Model}
\author{Neil Christensen}
\email{nchris3@ilstu.edu}
\affiliation{Department of Physics, Illinois State University, Normal, IL 61790}
\author{Bryan Field}%
\email{bryan.field@farmingdale.edu}
\affiliation{Department of Physics, Farmingdale State College, Farmingdale, NY, 11735}
\author{Annie Moore}
\author{Santiago Pinto}
\affiliation{Department of Physics, Illinois State University, Normal, IL 61790}

\date{\today}

\begin{abstract}
We further develop the massive constructive theory of the Standard Model and use it to calculate the amplitude and squared amplitude for all $2$-body decays, a collection of weak $3$-body decays as well as Higgs decay to four neutrinos.  We compare our results with those from Feynman diagrams and find complete agreement.  We show that in all the cases considered here, the amplitudes of massive constructive theories are significantly simpler than those resulting from Feynman diagrams.  In fact, a naive counting of the number of calculations required for a matrix-element generator to compute a phase-space point is orders-of-magnitude smaller for the result coming from the constructive method suggesting that these generators might benefit from this method in the future even in the case of massive weak amplitudes. We also anticipate that our simpler expressions will produce numerically more stable expressions.
\end{abstract}

\maketitle

The Standard Model (SM) of particle physics is often written, and understood, as a quantum field theory for good reason.  When written as a Lagrangian, it is extraordinarily simple, elegant and even beautiful.  All the particles are incorporated into fields and, together with the interactions, are written in a way that manifestly satisfies all applied symmetries.  In fact, this formalism makes the symmetries so easy to satisfy, it is nearly trivial.  Indeed, the simplicity of the formalism allows the entire SM Lagrangian to be expressed in a compact, factorizable way, sometimes minimally written as $\mathcal{L} = -\frac{1}{4} F_{\mu\nu} F^{\mu\nu} + i\bar{\Psi} \slashed{D} \Psi + D_\mu\Phi^\dagger D^\mu\Phi - V(\Phi) + y_{ij} \bar{\Psi}_i \Phi \Psi_j +$ h.c. Furthermore, the field formalism is extraordinarily powerful, allowing for any scattering amplitude to be calculated perturbatively through the use of Feynman diagrams.  When the processes calculated include few external particles, the number of Feynman diagrams is small and each diagram has an implicit meaning.  The internal lines are taken to be intermediate virtual particles and the sum over diagrams is taken to be a sum over all possible intermediate (virtual) particle ``paths'' that connect the in and out states.  Although it is true that some processes have currently been found to be too difficult to calculate practically at some order in perturbation theory, this is not the fault of the field theory.  The rules apply to any process at any perturbative order even if practically too difficult.  Indeed, one may rightly wonder why any other formalism should be considered.  

As our particle colliders have become ever more powerful, the SM background has grown with it and the need to calculate scattering processes with greater numbers of final state particles has also increased to a point that Feynman diagrams have become impractical, even with modern computers.  To make the point a little more precise, calculating higher-order scattering amplitudes currently requires the use of perturbation theory which, although powerful, also mandates the introduction of considerable complications such as gauge-fixing terms, ghosts, loop integrals, renormalization coefficients, and the entire machinery of modern perturbation theory to obtain physical results. The beauty of the Lagrangian is of course still present in principle, but it is hidden behind a legion of practical computational tools. This has lead to a search for simpler formalisms to obtain scattering amplitudes.  In possibly the most profound pure gluodynamics calculation to date, Park and Taylor\cite{Parke:1986gb} showed that the maximally helicity-violating (MHV) gluon amplitude at tree level could be written (in modern notation) as $\lbrack 12 \rbrack^4/(\lbrack 12 \rbrack\lbrack 23 \rbrack \cdots \lbrack n1 \rbrack)$. In this example, gluon~1 and 2 have positive helicity while all the other $n-2$ gluons have negative helicity, and this expression is valid no matter how many final-state gluons are present.  Compare this with the same calculation utilizing Feynman diagrams. If the number of external gluons $n$ is $4$, $5$, $6$, $7$, $8$, $9$ and $10$, then the number of squared Feynman diagrams required to calculate the scattering amplitude\cite{Feng:2011np} is $4$, $25$, $220$, $2~485$, $34~300$, $559~405$ and $10~525~900$, respectively.  But, it is not only the number of Feynman diagrams that increases, but the complexity of the expressions for each individual diagram. For example, CalcHEP\cite{Belyaev:2012qa} calculates just one of the squared diagrams for $2$~gluons scattering to $5$~gluons ($n=7$) and writes computer code for that expression at approximately $3~500$ lines of code.  We emphasize that this is just $1$ out of the $2~485$ diagrams.  Furthermore, since any individual gluon diagram is not gauge invariant, it cannot represent a physical particle ``path'', so its interpretation is not clear. It is not an exaggeration to state that this one result from Parke and Taylor stunned and astounded the physics community at the time. Following this result, it took time to absorb and generalize these techniques to develop an approach that reached beyond the novelty of this initial expression.  A completely recursive formalism for calculating any pure-gluon amplitude at tree level was developed by Britto, Cachazo, Feng and Witten\cite{Britto:2005fq}.  This new method was also essentially diagrammatic, but it had many orders of magnitude fewer diagrams (in fact only one diagram for maximally helicity-violating processes) and the resulting expressions were orders-of-magnitude simpler, with every diagram being trivially gauge invariant and meaningful.  Along these lines, it only allowed diagrams with on-shell internal particles, albeit in complex momentum space.  This began a renewed drive to search for a ``constructive'' theory of particle physics, based on the old S-Matrix theory\cite{Eden:1966dnq} that derived the scattering amplitudes from the properties of the amplitude itself.  For a pedagogical introduction, see \cite{Elvang:2015rqa}.  This process has been extraordinarily successful for massless theories\cite{Dixon:1996wi, DelDuca:1999rs, Dixon:2013uaa} but, until recently, extensions to massive theories were unwieldy and lacked the simplicity of the massless theories.

Arkani-Hamed, Huang and Huang\cite{Arkani-Hamed:2017jhn} (AHH) extended the massless formalism in a natural way when they introduced the spin-spinor.  As Weinberg\cite{Weinberg:1995mt} shows, the scattering amplitude transforms under boosts and rotations under the little-group subgroup of the Lorentz group. He further shows that the little group for massless particles is the helicity group and amounts to a phase change under rotations while the little group for massive particles is the SU$(2)$-spin group.  The amplitude, then, transforms as a direct product of the little group for each external particle in the amplitude.  The reason the helicity spinors $|i\rangle$ and $|i\rbrack$ are so useful for massless amplitudes is that they each transform under a direct product of the helicity group and the SL$(2,\mathbb{C})$ Lorentz group.  When an inner product is taken, $\langle ij \rangle$ or $\lbrack ij \rbrack$, the SL$(2,\mathbb{C})$ transformations exactly cancel resulting in a Lorentz invariant product, whereas the helicity transformations add.  That is to say, this product transforms under a direct product of the helicity transformation for particle $i$ and particle $j$.  This is exactly what is needed to write an amplitude involving these particles in a minimal way.  The insight afforded by AHH was to generalize the helicity-spinor to a tensor where each column separately transforms under an SL$(2,\mathbb{C})$ Lorentz transformation, but each row transforms under spin-SU$(2)$.  That is to say, they introduced the spin spinors, $|\mathbf{i}^{\mathrm{I}} \rangle$ and $|\mathbf{i}^{\mathrm{I}} \rbrack$, which transform under a direct product of spin and SL$(2,\mathbb{C})$-Lorentz transformations. Just as in the case of the helicity spinors, products of spin spinors such as $\langle\mathbf{i}^{\mathrm{I}} \mathbf{j}^{\mathrm{J}}\rangle$ or $\lbrack\mathbf{i}^{\mathrm{I}} \mathbf{j}^{\mathrm{J}}\rbrack$ are Lorentz invariant but still transform under spin SU$(2)$.  This is perfect because these spinor products can be used as building blocks for massive scattering amplitudes as they transform under the same little group.  Furthermore, mixtures of helicity spinors and spin spinors, such as $\langle i\mathbf{j}^{\mathrm{J}} \rangle$, transform under a product of helicity for particle $i$ and spin for particle $j$.  So, including both helicity spinors for the massless particles and spin spinors for the massive particles allows us to write the scattering amplitude for any process, at least in principle.  

AHH\cite{Arkani-Hamed:2017jhn} also derived the most general $3$-point vertices and set up the rules to create a constructive theory using these vertices.  Following this, \cite{Christensen:2018zcq} found the complete set of constructive SM $3$-point vertices, found their high-energy limits and compared them with the massless version.  Other authors have looked at how spin-spinors can be used outside of path integral constructions\cite{Boels:2017gyc}, how to use this formalism to study the spin structure of QCD\cite{Ochirov:2018uyq}, as well as studies involving gravity\cite{Afkhami-Jeddi:2018apj}. There has also been a great deal of work on building effective theories utilizing on-shell amplitudes\cite{Shadmi:2018xan, Ma:2019gtx, Aoude:2019tzn, Durieux:2019eor}. There still remains much to do to fully establish this massive constructive theory to determine what kind of simplifications are manifest in a massive theory when compared to Feynman diagrams.  Let us name a few examples.  Although the $3$-point vertices of the SM have been enumerated\cite{Christensen:2018zcq}, the $4$-point vertices have not.  There are certain terms in the $3$-point vertices of \cite{Christensen:2018zcq} that have not yet been eliminated based purely on constructive techniques.  A complete set of rules, including all the technical details, for constructing higher-point amplitudes using the constructive vertices is still not fully understood.  In fact, a complete squaring of a scattering amplitude with spin-spinors will be presented here. There is still much to be done within a constructive research program.

In the present paper, we aim to fill one of these missing pieces.  We will show in detail how to construct $4$-point and $5$-point amplitudes using the $3$-point vertices given in \cite{Christensen:2018zcq} and using the rules suggested by \cite{Arkani-Hamed:2017jhn}. Our explicit $5$-point amplitude was calculated using the massive constructive method outlined in \cite{Arkani-Hamed:2017jhn} and was completely reduced to a minimal form that only included spinors and no momenta in the numerator. There are, of course, many amplitudes to choose from within the SM.  We will use the SM decays as our organizing principle for our choice of amplitudes.  We calculate all five $2$-body decays and the three weak $3$-body decays in the SM using the constructive method. We also set out to calculate a $4$-body decay in the SM.  In order to keep this calculation in line with the others, we choose the simplest, which is Higgs decay to $4$ neutrinos. The use of on-shell methods to study the Higgs couplings to gluons can be found in Ref.~\cite{Shadmi:2018xan}.

We will also work out the full technical details for squaring the amplitude and will apply it to all the amplitudes in this paper.  These details can be found in our appendix for future reference. In Appendix~\ref{sec:spinor conjugation}, we determine how to conjugate spin-spinors and spinor-chains.  We also find outer products of spinors (for example, $|\mathbf{i}\rangle \langle\mathbf{i}|$) and show how traces of momenta appear in squared amplitudes as well as calculate those traces.  In Appendix~\ref{sec:generalized Schouten}, we generalize the Schouten identities~\cite{Dixon:1996wi} to spinor-chains with multiple momenta sandwiched between the spinors at each end. 

With these results in hand, we will have an opportunity to compare the results of massive constructive calculations with those of Feynman diagrams.  We do this at low energies as well as at high energies and find agreement in the cases we consider.  It also allows us to determine whether a simplification of the final result is possible in the massive case like it is in the massless case.  We show that for the $4$-point and $5$-point amplitudes presented here, the constructive result is indeed much simpler. In fact, we will do a naive estimate of the number of calculations required to compute the amplitude for a phase-space point using the constructive method versus the Feynman diagram method and propose that there is an orders-of-magnitude savings possible. Based on these initial results, we speculate that these simpler expressions will lead to numerically more stable expressions for use in phase-space generators, but will leave a detailed study of these effects for a future effort. We emphasize that, although this was already known for purely massless theories, our examples are massive and in the weak sector and require the full spin-spinor structure introduced in \cite{Arkani-Hamed:2017jhn, Christensen:2018zcq}.  In order to accomplish all these objectives, we review, develop, and generalize important identities which are used to reduce and simplify the amplitude and its square.  

In Sec.~\ref{sec:general techniques}, we outline the techniques and identities that we will use throughout this paper.  In Sec.~\ref{sec:2-body decays}, we calculate the squared amplitude for neutrino decays of the Z~boson (Subsec.~\ref{sec:Z->vv}), massive fermion decays of the Z~boson (Subsec.~\ref{sec:Z->ff}), leptonic decays of the W~boson (Subsec.~\ref{sec:W->lv}), quark decays of the W~boson (Subsec.~\ref{sec:W->qq}) and fermionic decay of the Higgs boson (Subsec.~\ref{sec:h->ff}).  In Sec.~\ref{sec:3-body decays}, we work out the $4$-point amplitudes in the $3$-body decays of the SM and compare the square with Feynman diagrams.  We do this for leptonic muon decay (Subsec.~\ref{sec:m->vev}), quark decay of the tau (Subsec.~\ref{sec:m->vqq}) and quark decay of the top quark (Subsec.~\ref{sec:t->bqq}).  In Sec.~\ref{sec:4-body decays}, we calculate the $5$-point amplitude for Higgs decay to $4$ neutrinos.  We also show how simple the final result is, although the intermediate states are massive Z~bosons and require the full massive spin-spinor formalism.  In this subsection, we also show the Feynman diagram results for comparison.  In Sec.~\ref{sec:conclusions}, we summarize our results and conclude. 

\section{\label{sec:general techniques}Some Techniques for Massive Constructive Amplitudes and their Squares}
Before we begin the calculations, we will describe some of the steps used to square the amplitude,  and reduce the amplitude to a suitable form but reserve a full treatment for our Appendices.  This will involve several identities, some of which are known and some of which are novel.  

In order to square the amplitude, we will need a complex conjugate of the amplitude.  This will involve the conjugation of spinor-chains which begin and end with a spinor but also may have one or more momenta sandwiched between the spinors. In App.~\ref{sec:spinor conjugation}, we will derive the conjugate of a spinor-chain.  As we do this, we will find that it is convenient to also find the transpose.  Furthermore, we will find the conjugate of each individual spinor.  Although the square spinor was defined to be the conjugate of the angle spinor, that only applies directly to the helicity-spinors and the angle (square) spin-spinor with upper (lower) spin-indices.  We will see that there is a sign involved when the angle (square) spin-spinor has a lower (upper) spin-index.  

Once the amplitude has been complex conjugated and the amplitude has been multiplied by its conjugate, the spins, and therefore the spin-indices, must be summed over as experiments do not typically measure the spins of the final state particles.  In order to replace the resulting spinor-chains with an expression which only involves masses and traditional kinematic variables such as $p_i \cdot p_j$, we derive a set of identities involving the product of two spinors with their spin-indices ``contracted''.  This will be when two indices are the same and summed over with one up and one down in analogy with Lorentz indices of four-vectors.  One of these identities is already well-known as it gives the momentum when one spinor is an angle spinor while the other is a square spinor.  However, we will also derive in App.~\ref{sec:spinor conjugation} the case where the summation is over two angle spinors or two square spinors.  As we will see, in this case we will obtain the mass of the particle rather than their momentum and this mass will be multiplied by a Kronecker delta function on the spinors' Lorentz indices.  We will also enumerate all the cases with mixed angle and square spinors with their indices in different positions, which give the momentum of the particle with different signs. 

As we replace spinor products with momenta or mass (times a Kronecker delta function), we will obtain traces of momenta.  We must remember that these momenta are Hermitian two-by-two matrices.  Therefore, we will need identities relating these traces over momenta to the more traditional four-vector products $p_i \cdot p_j$.  We will derive these in App.~\ref{sec:spinor conjugation} and see that they are very similar mathematically to the traces of momenta times gamma matrices.  With this, we will have all the tools we need to square any spinor amplitude and compare with the square of Feynman diagrams using traditional methods.

Beyond the methods required to square an amplitude, we will also need to know the technical details of how to form an amplitude with four or more external particles using the $3$-point vertices of \cite{Christensen:2018zcq}.  In principle, the rules were outlined in \cite{Arkani-Hamed:2017jhn}.  However, we find that this process is not completely trivial. In particular, in the limited cases we have analyzed so far, we have only found agreement with the expressions of Feynman diagrams when we were able to reduce the amplitude all the way to a point that does not contain any momenta in the numerator.  On the other hand, it is acceptable and expected that momenta will be present in a propagator denominator.  That is to say, we have only found agreement with Feynman diagrams when the numerator contained no more than masses and spinor products with no momenta sandwiched between them, for example $\langle \mathbf{i}j \rangle$ or $\lbrack k\mathbf{l} \rbrack$ and other similar spinor products.  But, we have not yet found agreement when we have had $\langle i|p_j|\mathbf{k}\rbrack$ and other spinor-chains with explicit momenta.  We have also not yet found agreement in cases where an explicit $p_i\cdot p_j$ was present in the numerator. 

We believe the reason for this is that the constructive amplitude rules require the internal lines of a diagram to be on-shell during all intermediate steps with the sole exception of a propagator denominator.  Once the amplitude has been calculated, the internal lines are then allowed to go off the mass-shell.  The reason this is important is that, in most cases, the internal lines of a diagram cannot be on-shell if the external momenta are real, on-shell, and satisfy momentum conservation.  For a very simple example related to the calculations done here, the decay of a muon is via an intermediate far off-shell W~boson.  For another example, an electron-positron collision that results in an intermediate off-shell photon.  And, of course, there are a plethora of other examples.  Nevertheless, even in these cases, the final amplitude using constructive techniques requires these intermediate particles to be put on the mass-shell during intermediate steps of the calculation and they are only allowed to go off-shell at the end of the calculation.  To do this, two or more of the external momenta\cite{Elvang:2015rqa} are extended as complex momenta.  This allows both the external and internal lines to be on-shell, while also satisfying momentum conservation.  This complexification is only done during intermediate steps and the momenta are constrained to be real again at the end of the calculation.  

It is well-known how to do this complexification in the massless case \cite{Elvang:2015rqa}, however, it has not yet been determined how to complexify the momenta in the massive case.  In particular, it is not known how to modify the massive spin-spinor with a spin-index for a complex momentum.  The authors have not yet solved this issue satisfactorily and we will leave the complexification of the momenta for a future publication.  Without an understanding of this complexification, we do not expect to find the desired agreement with Feynman diagrams when momenta still persist in the amplitude numerator. It is important to note that if we are able to completely remove the momenta from the numerator using on-shell identities for the spinors and momenta, then, in every case we have studied, we find exact agreement with Feynman diagrams.  In some cases, we have found much simpler, but equivalent, expressions using constructive techniques, as we will see in Secs.~\ref{sec:3-body decays} and \ref{sec:4-body decays}.  

There are several identities available to us for this reduction.  Some are well-known such as momentum conservation.  We will take all momenta to be incoming, therefore, momentum conservation will take the form of replacing one momentum with minus the sum of the others.  Another important set of identities are mass-shell identities including the well-known $p_i^2 = m_i^2$ but also including all the spinor mass-shell identities\cite{Arkani-Hamed:2017jhn, Christensen:2018zcq} such as $p_i|\mathbf{i}\rangle = - m_i|\mathbf{i}\rangle$.  We remind the reader of these in App.~\ref{sec:spinor conjugation}.  We will also need to transpose momenta in a spinor-chain in order to get the momenta into different positions and we also remind the reader of these rules in the same subsection.  Finally, we will find it essential to use the Schouten identity\cite{Dixon:1996wi} in order to rearrange a product of two spinor-chains.  Although the Schouten identity is well-known, we derive generalizations of it that are useful for the reduction of these spinor-chains in App.~\ref{sec:generalized Schouten} and describe a mnemonic for remembering the generalized Schouten identity that we find very useful in our calculations.

\section{\label{sec:2-body decays}2-Body Decays}
In this section, we consider the $2$-body decays of the SM.  The amplitudes are already given by the $3$-point vertices in \cite{Christensen:2018zcq}.  What we do in this section is two things.  The first is that we work out the amplitude for each combination of physical spins of the external particles.  In some cases, where more than one of the particles is massive, we may have a matrix or even a third-rank tensor of amplitudes for each spin combination.  We also note that the symmetry factor of $1/2$ is replaced by $1/\sqrt{2}$ when dealing with the explicit spin-$0$ component of the amplitude.

The second thing we do is describe the squaring of a massive constructive amplitude.  We do this in two ways.  The first is that we take the explicit amplitude for each spin combination, as described in the previous paragraph, and simply square the absolute value  and sum over all spin combinations.  This is  what we would expect following the usual rules of Quantum Mechanics.  However, although this brute-force method is enlightening when we desire to know the amplitude and squared amplitude for each spin combination and may even be put to powerful use by some matrix-element calculators such as MadGraph, Herwig, Sherpa and Whizard\cite{Alwall:2011uj, Maltoni:2002qb, Bellm:2015jjp, Gleisberg:2008ta, Kilian:2007gr}, when calculating analytic expressions for the squared amplitude which are summed over spins, there is a better way.  This more efficient method is similar to what we do with Feynman diagrams.  We conjugate the amplitude, multiply the amplitude with the conjugate amplitude, contract the spin-indices corresponding to the same particle and sum over them, and finally use the spinor identities in Sec.~\ref{sec:general techniques} to replace spinor contractions with momenta or masses.  This leads directly to an expression involving masses and traces of momenta.  The traces are evaluated in a way analogous to traces of gamma matrices.  After evaluating these, we have the final expression in terms of masses and standard $4$-vector momentum products such as $p_1 \cdot p_2$.  

Once we have calculated the squared amplitude using the massive constructive methods, we compare them with each other and with Feynman diagrams.  We obtain our Feynman diagram result from CalcHEP\cite{Belyaev:2012qa}, which has a built-in analytic squared-Feynman-diagram calculator and is able to export its results directly to Mathematica\cite{Mathematica}.

The $2$-body decays of the SM include $Z\to \nu\bar{\nu}$, $Z\to f\bar{f}$, $W\to l\bar{\nu}$, $W\to q\bar{q}$ and $h\to f\bar{f}$, where $l$ is a charged lepton, $\nu$ is a neutrino, $f$ is a massive fermion and $q$ is a quark.  We must calculate each of these separately since their amplitudes are fundamentally different.  Since the neutrino is massless, its spinor is a helicity-spinor with two components whereas the spinor of the massive fermions is a spin-spinor with four components (two for each spin).  Therefore, we consider them each in turn in the rest of this section. For simplicity, we will ignore the color structures in the strong sector as they are simple to restore as clearly shown for the massless case in Ref~\cite{Dixon:1996wi}.  

\subsection{\label{sec:Z->vv}$Z \rightarrow \nu\bar{\nu}$}
We begin with the decay of the Z~boson to a neutrino~anti-neutrino pair. We remember that the $3$-point amplitude is given by \cite{Christensen:2018zcq},
\begin{equation}
    \mathcal{M} =g_{Z\nu\bar{\nu}} \frac{\langle\mathbf{3}1\rangle\lbrack2\mathbf{3}\rbrack}{M_Z}
    \label{eq:M(Z->vv)}
\end{equation}
where the neutrino is particle $1$, the anti-neutrino is particle $2$ and the (massive) Z~boson is particle $3$ and its spin-indices are implicit.  We have used the notation outlined in \cite{Arkani-Hamed:2017jhn} where a massive spin-spinor is distinguished from a massless helicity-spinor by bold facing the particle number.  Thus, in this amplitude, the $3$ is bold while the $1$ and $2$ of the neutrinos are not.  

\subsubsection{Explicit}
We begin by calculating each spin component of the Z~boson decay amplitude.  Afterwards, we will explicitly square and add them together to obtain the squared amplitude.  We start by writing the explicit form of the spinor products required in this $3$-point amplitude.  We do this in the rest frame of the Z~boson.  We could choose our coordinates such that the neutrino momenta are in the $\pm$~$z$-direction, however, we think it will be more instructive to allow the neutrinos to propagate in any direction.  Therefore, using Eqs.~(\ref{eq:[ij^J] explicit}) and (\ref{eq:<i^Ij> explicit}) we obtain,
\begin{equation}
\lbrack 2\mathbf{3}^{\mathrm{I}} \rbrack 
  = M_Z \left(
        \begin{array}{c}
          -c_2 \\ -s^*_2 
        \end{array}
        \right)
    \quad \mbox{and} \quad
\langle \mathbf{3}^{\mathrm{I}} 1 \rangle
  = M_Z \left(
        \begin{array}{cc}
          -s_1 \\ \phantom{+}c_1
        \end{array}
        \right),
\end{equation}
where, spin $-1/2$ is at the top and spin $+1/2$ is at the bottom and, $c_i = \cos(\theta_i/2)$ and $s_i=\sin( \theta_i/2 )e^{i\phi_i}$.  Putting these together, we find the amplitude is given by,
\begin{equation}
\mathcal{M}^{s_z} = 
    \frac{g_{Z\nu\bar{\nu}}}{M_Z}
    \left(\begin{array}{c}
       \langle \mathbf{3}^{1} 1 \rangle
       \lbrack 2\mathbf{3}^{1} \rbrack \\
    (
       \langle \mathbf{3}^{1} 1 \rangle
       \lbrack 2\mathbf{3}^{2} \rbrack +
       \langle \mathbf{3}^{2}1 \rangle
       \lbrack 2\mathbf{3}^{1} \rbrack
    ) / \sqrt{2} \\
    \langle \mathbf{3}^{2}1 \rangle
           \lbrack 2\mathbf{3}^{2} \rbrack
    \end{array}\right)
  = g_{Z\nu\bar{\nu}} M_Z 
    \left(
    \begin{array}{c}
       \sin^2 (\theta/2) \exp (i\phi) \\
      - \sin\theta/\sqrt{2} \\
       \cos^2 (\theta/2) \exp (-i\phi)
    \end{array}
    \right),
    \label{eq:M(Z->vv) explicit}
\end{equation}
where a superscript of $1$ represents the $-1/2$-spin component and a superscript of $2$ represents the $+1/2$-spin component and, from the \textsc{cm} frame, we have taken the neutrino~anti-neutrino pair to decay back-to-back which simplifies our angles as $\theta_2=\pi-\theta$ and $\phi_2=\pi+\phi$.  Each row gives the $3$-point amplitude for the $-1$-, $0$- and $+1$-spin Z~boson, respectively.  We note that when we symmetrize the explicit indices, we must use a factor of $1/\sqrt{2}$ for the spin-$0$ term instead of the $1/2$ that we will use when the indices are left implicit.  This is to obtain the right normalization for the spin-$0$ state. Next we take the square of the absolute value of each term and add them to obtain,
\begin{equation}
 \sum|\mathcal{M}|^2 = |g_{Z\nu\bar{\nu}}|^2 M_Z^2. 
 \label{eq:M^2(Z->vv) explicit}
\end{equation}

\subsubsection{Implicit}
We now go back to the original amplitude and square it with the spin-indices left unspecified.  We do this by multiplying by the complex conjugate of the amplitude and then summing over indices.  We begin by symmetrizing the indices of the Z~boson,
\begin{equation}
  \mathcal{M}^{\mathrm{IJ}} =
    g_{Z\nu\bar{\nu}}
    \left( 
        \frac{ \langle \mathbf{3}^{\mathrm{I}}1 \rangle
               \lbrack 2\mathbf{3}^{\mathrm{J}} \rbrack}
             {2M_Z} +
        \frac{ \langle \mathbf{3}^{\mathrm{J}}1 \rangle
               \lbrack 2\mathbf{3}^{\mathrm{I}} \rbrack}
             {2M_Z}
    \right),
    \label{eq:M(Z->vv) 1/2}
\end{equation}
where we note that we have used a factor of $1/2$ in our symmetrization for all spin-index values, rather than the $1/\sqrt{2}$ that we used when explicitly calculating the spin-$0$ component.  This is required to achieve the correct squared amplitude.  Now, in order to square, we multiply by the complex conjugate using the rules of Eqs.~(\ref{eq:app:(<>)*}) to (\ref{eq:app:([])*}).  We then sum over the spin of the indices giving,
\begin{eqnarray}
    \sum | \mathcal{M} |^2 &=& 
    -|g_{Z\nu\bar{\nu}}|^2
    \left(
       \frac{ \langle \mathbf{3}^{\mathrm{I}}1 \rangle
              \lbrack 2\mathbf{3}^{\mathrm{J}} \rbrack}
            {2M_Z} +
        \frac{ \langle \mathbf{3}^{\mathrm{J}}1 \rangle
               \lbrack 2\mathbf{3}^{\mathrm{I}} \rbrack}
            {2M_Z}
    \right)
        \frac{ \lbrack 1\mathbf{3}_{\mathrm{I}} \rbrack
               \langle \mathbf{3}_{\mathrm{J}}2 \rangle}
             {M_Z} \nonumber \\
    &=&
    -|g_{Z\nu\bar{\nu}}|^2
    \left(
        \frac{ \lbrack 1\mathbf{3}_{\mathrm{I}} \rbrack
               \langle \mathbf{3}^{\mathrm{I}}1 \rangle
               \lbrack 2\mathbf{3}^{\mathrm{J}} \rbrack
               \langle \mathbf{3}_{\mathrm{J}}2 \rangle}
             {2M_Z^2} +
        \frac{ \langle 2\mathbf{3}_{\mathrm{J}} \rangle
               \langle \mathbf{3}^{\mathrm{J}}1 \rangle
               \lbrack 2\mathbf{3}^{\mathrm{I}} \rbrack
               \lbrack \mathbf{3}_{\mathrm{I}}1 \rbrack}
             {2M_Z^2}
    \right),
\end{eqnarray}
where we have expanded in the second term.  Next we use the rules for contracted spinors given in Eqs.~(\ref{eq:app:p=|>[|}) through (\ref{eq:App:|][|=-m}), (\ref{eq:app:<ji>[ij]=2pipj massless}) and (\ref{eq:app:<ji>[ij]=2pipj}) to obtain,
\begin{eqnarray}
  \sum |\mathcal{M}|^2 &=&
  |g_{Z\nu\bar{\nu}}|^2
    \left(
       \frac{\mbox{Tr}(p_1p_3) \, \mbox{Tr}(p_2p_3)}
            {2M_Z^2}
      + \frac{\mbox{Tr}(p_1 p_2)}{2}
    \right)
  \nonumber\\
  &=&|g_{Z\nu\bar{\nu}}|^2
    \left(
       \frac{2p_1\cdot p_3 \, p_2\cdot p_3}
            {M_Z^2}
      + p_1 \cdot p_2
    \right).
\end{eqnarray}
In the \textsc{cm} frame, where the Z~boson is at rest, we can take $p_3 = (M_Z, 0, 0, 0)$, $p_1 = (M_Z/2)(1, 0, 0, 1)$ and $p_2 = (M_Z/2)(1, 0, 0, -1)$ to obtain
\begin{equation}
    \sum |\mathcal{M}|^2 = | g_{Z\nu\bar{\nu}} |^2 M_Z^2.
    \label{eq:M^2(Z->vv)}
\end{equation}
As we can see, this agrees with the result we found from explicitly calculating the amplitude for each spin of the Z~boson.  It also agrees with the result of Feynman diagrams as output by CalcHEP.

\subsection{\label{sec:Z->ff}$Z \rightarrow f\bar{f}$}
We next consider the decay of the Z~boson to a massive fermion and its antiparticle.  The $3$-point amplitude for this is given by \cite{Christensen:2018zcq},
\begin{equation}
  \mathcal{M} = 
    \frac{g_L \langle\mathbf{31}\rangle
              \lbrack\mathbf{23}\rbrack
        + g_R \lbrack\mathbf{31}\rbrack
              \langle\mathbf{23}\rangle}
         {M_Z},
    \label{eq:M(Z->ff)}
\end{equation}
where the Z~boson is again particle $3$, the fermion is particle $1$ and the antifermion is paricle $2$.  We can see that, since the fermions are massive and require spin-spinors rather than the helicity-spinors of the previous subsection, all three particle numbers are bold faced. This means they have implicit spin-indices.  We will make these indices explicit when necessary in the rest of this subsection.

\subsubsection{Explicit}
We begin by explicitly calculating the amplitude for each spin combination.  There are three massive particles this time, so we will use two matrices to represent all the spin combinations.  We use the explicit formulas for spinor products given in Eqs.~(\ref{eq:<i^Ij^J> explicit}) and (\ref{eq:[i^Ij^J] explicit}).  With these, we obtain,
\begin{eqnarray}
\mathcal{M}^{-\sfrac{1}{2} \, s_3 \, s_2}
&=&
   \left(\begin{array}{cc}
      g_L \langle \mathbf{3}^1\mathbf{1}^1 \rangle \lbrack\mathbf{2}^1\mathbf{3}^1 \rbrack
    + g_R \lbrack \mathbf{3}^1\mathbf{1}^1 \rbrack \langle\mathbf{2}^1\mathbf{3}^1 \rangle
    & g_L \langle \mathbf{3}^1\mathbf{1}^1 \rangle \lbrack\mathbf{2}^2\mathbf{3}^1 \rbrack
    + g_R \lbrack \mathbf{3}^1\mathbf{1}^1 \rbrack \langle\mathbf{2}^2\mathbf{3}^1 \rangle \\
      g_L
   ( \langle \mathbf{3}^1\mathbf{1}^1 \rangle \lbrack \mathbf{2}^1\mathbf{3}^2 \rbrack
         +\langle \mathbf{3}^2\mathbf{1}^1 \rangle \lbrack \mathbf{2}^1\mathbf{3}^1 \rbrack
   )/\sqrt{2}
    & g_L ( \langle \mathbf{3}^1\mathbf{1}^1 \rangle \lbrack \mathbf{2}^2\mathbf{3}^2 \rbrack
                +\langle \mathbf{3}^2\mathbf{1}^1 \rangle \lbrack \mathbf{2}^2\mathbf{3}^1 \rbrack
          )/\sqrt{2} \\
      +g_R ( \lbrack \mathbf{3}^1\mathbf{1}^1 \rbrack \langle \mathbf{2}^1\mathbf{3}^2 \rangle
                +\lbrack \mathbf{3}^2\mathbf{1}^1 \rbrack \langle \mathbf{2}^1\mathbf{3}^1 \rangle
          )/\sqrt{2}
    & +g_R ( \lbrack \mathbf{3}^1\mathbf{1}^1 \rbrack \langle \mathbf{2}^2\mathbf{3}^2 \rangle
                +\lbrack \mathbf{3}^2\mathbf{1}^1 \rbrack \langle \mathbf{2}^2\mathbf{3}^1 \rangle
          )/\sqrt{2} \\
      g_L \langle \mathbf{3}^2\mathbf{1}^1 \rangle \lbrack \mathbf{2}^1\mathbf{3}^2 \rbrack
    + g_R \lbrack \mathbf{3}^2\mathbf{1}^1 \rbrack \langle \mathbf{2}^1\mathbf{3}^2 \rangle
    & g_L \langle \mathbf{3}^2\mathbf{1}^1 \rangle \lbrack \mathbf{2}^2\mathbf{3}^2 \rbrack
    + g_R \lbrack \mathbf{3}^2\mathbf{1}^1 \rbrack \langle \mathbf{2}^2\mathbf{3}^2 \rangle
\end{array}\right)/M_Z \nonumber\\
&=& \frac{1}{2}
\left(
  \begin{array}{cc}
     - (g_L + g_R) \, m_f \, \sin\theta \, \exp(2i\phi)
     &
     B \sin^2(\theta/2) \exp(i\phi) \\
      \sqrt{2} (g_L+g_R) \, m_f \, \cos\theta \, \exp(i\phi)
     & 
     - B \sin\theta/\sqrt{2} \\
     (g_L + g_R) \, m_f \, \sin\theta
     &
     B \cos^2(\theta/2) \, \exp(-i\phi)
  \end{array}
\right)
    \label{eq:M(Z->ff) explicit 1}\\
\mathcal{M}^{+\sfrac{1}{2} \, s_3 \, s_2} 
&=& \frac{1}{2}
\left(
  \begin{array}{cc}
    A \cos^2(\theta/2) \, \exp(i\phi)
   & 
   (g_L + g_R) \, m_f \, \sin\theta \\
   A \sin\theta/\sqrt{2}
   & 
   -\sqrt{2} \, (g_L+g_R) \,
   m_f \, \cos\theta \, \exp(-i\phi) \\
   A \sin^2(\theta/2) \exp(-i\phi)
   &  
   -(g_L+g_R) \, m_f \, \sin\theta \,
   \exp(-2 i\phi)
   \end{array}
\right) 
    \label{eq:M(Z->ff) explicit 2}
\end{eqnarray}
where $A = g_L ( 2p_f - M_Z ) - g_R ( M_Z + 2p_f )$, $B = g_R ( M_Z - 2p_f ) + g_L ( M_Z + 2p_f )$ and $p_f = \sqrt{ M_Z^2 - 4 m_f^2}/2$.  The top matrix is for a $-1/2$-spin fermion while the bottom matrix is for a $+1/2$-spin fermion.  The rows, once again, represent the spins of the Z~boson, beginning with $-1$-spin at the top and increasing by $1$ for lower rows.  The left column is for a $-1/2$-spin antifermion while the right column contains the results for a $+1/2$-spin antifermion.  We have once again assumed the rest frame of the Z~boson but allowed the fermion-anti-fermion pair to propagate in any direction.  To obtain the much simpler result when the $z$-direction is taken to lie along the motion of the fermion, simply set $\theta = 0$ and $\phi = 0$.  If we multiply each element by its complex conjugate and add them all together, we obtain,
\begin{equation}
\sum |\mathcal{M}|^2 =
  ( g_L^2 + g_R^2 ) ( M_Z^2 - m_f^2 ) + 6 \, g_L \, g_R \, m_f^2.
  \label{eq:M^2_Zff}
\end{equation}

\subsubsection{Implicit}
We next square the amplitude with general spin-indices and compare with the previous result.  After symmetrizing in the spin-indices of the Z~boson and multiplying by the complex conjugate, we obtain,
\begin{eqnarray} \nonumber
  \sum |\mathcal{M}|^2  
&=& 
  \frac{g_L^2}{2M_Z^2}
  (
    \langle \mathbf{31} \rangle^{HI}
    \lbrack \mathbf{23} \rbrack^{JK}
   +\langle \mathbf{31} \rangle^{KI}
    \lbrack \mathbf{23} \rbrack^{JH}
  )
  \lbrack \mathbf{13} \rbrack_{IH}
  \langle \mathbf{32} \rangle_{KJ} \\ \nonumber
&+&
  \frac{g_R^2}{2M_Z^2}
  (
    \lbrack \mathbf{31} \rbrack^{HI}
    \langle \mathbf{23} \rangle^{JK}
   +\lbrack \mathbf{31} \rbrack^{KI}
    \langle \mathbf{23} \rangle^{JH}
  )
  \langle \mathbf{13} \rangle_{IH}
  \lbrack \mathbf{32} \rbrack_{KJ} \\ \nonumber
&+&
  \frac{g_L g_R}{2M_Z^2}
  (
    \langle \mathbf{31} \rangle^{HI}
    \lbrack \mathbf{23} \rbrack^{JK}
   +\langle \mathbf{31} \rangle^{KI}
    \lbrack \mathbf{23} \rbrack^{JH}
  )
  \langle \mathbf{13} \rangle_{IH}
  \lbrack \mathbf{32} \rbrack_{KJ} \\
 &+&  
  \frac{g_R g_L}{2M_Z^2} 
  (
    \lbrack \mathbf{31} \rbrack^{HI}
    \langle \mathbf{23} \rangle^{JK}
   +\lbrack \mathbf{31} \rbrack^{KI}
    \langle \mathbf{23} \rangle^{JH}
  )
  \lbrack \mathbf{13} \rbrack_{IH}
  \langle \mathbf{32} \rangle_{KJ},
\end{eqnarray}
where the the factor of $2$ is the symmetrization factor.  We next expand and apply the identities found in Eqs.~(\ref{eq:app:p=|>[|}) through (\ref{eq:App:|][|=-m}) and (\ref{eq:app:<ji>[ij]=2pipj}) to obtain
\begin{eqnarray}
\sum |\mathcal{M}|^2  &=&
(
    \frac{g_L^2+g_R^2}{2M_Z^2}
  )
  \left[ \mbox{Tr}(p_1 p_3) \, \mbox{Tr}(p_2 p_3) + M_Z^2\mbox{Tr}( p_1 p_2)
  \right]
 + 6 \, g_L \, g_R \, m_f^2
 \nonumber\\
  &=& (
    \frac{g_L^2+g_R^2}{M_Z^2}
  )
  ( 2 p_1 \cdot p_3 \, p_2 \cdot p_3 + M_Z^2 p_1 \cdot p_2
  ) 
 + 6 \, g_L \, g_R \, m_f^2.
\end{eqnarray}
If we take the momenta to be $p_3=M_Z(1,0,0,0), p_1 = ( M_Z/2, 0, 0, p_f )$ and $p_ 2 =( M_Z/2, 0, 0, -p_f )$ with $2p_f=\sqrt{M_Z^2-4m_f^2}$, we obtain
\begin{equation}
\label{eq:M^2(Z->ff) implicit}
\sum |\mathcal{M}|^2 = 
  (
    g_L^2 + g_R^2
  )
  (
    M_Z^2 - m_f^2
  )
  + 6 \, g_L \, g_R \, m_f^2,
\end{equation}
which, as the reader can see, is exactly the same as Eq.~(\ref{eq:M^2_Zff}).  We also find agreement with the Feynman diagram result coming from CalcHEP.

\subsection{\label{sec:W->lv}$W \rightarrow l \bar{\nu}$}
We next turn to the decay of the W~boson and first consider the leptonic decay.  The $3$-point amplitude for this process is,
\begin{equation}
\mathcal{M} = 
 \frac{g_{Wl\nu}}{M_W} 
   \langle \mathbf{31} \rangle
   \lbrack 2\mathbf{3} \rbrack,
   \label{eq:M(W->lv)}
\end{equation}
where particle 1 is the charged lepton and is represented by a bold-faced spin-spinor, particle $2$ is the antineutrino and is represented by a helicity-spinor and particle $3$ is the W~boson and is represented by two bold-faced spin-spinors whose spin-indices are (implicitly) symmetrized.  

\subsubsection{Explicit}
As we have before, we begin by calculating the amplitude for each explicit spin combination.  Since there are two massive particles, we can display the amplitudes as a matrix.  We take the rows to give the spins of the W~boson and the columns to give the spins of the charged lepton.  We again use the rest frame of the W~boson but allow the leptons to propagate in any direction for illustration.  The amplitudes are,
\begin{equation}
\mathcal{M}^{s_3 \, s_1} 
= g_{Wl\nu} \sqrt{M_W^2-M_l^2}
  \left(
    \begin{array}{cc}
     \sin^2(\theta/2) \exp(i\phi)
      &
      m_l \sin\theta / (2M_W)  \\
     -\sin\theta / \sqrt{2}
      & 
     -m_l \cos\theta \, \exp(-i\phi)/(2M_W) \\
      \cos^2(\theta/2) \, \exp(-i\phi)  
      &
     -m_l \sin\theta \, \exp(-2i\phi)/(2M_W) 
\end{array}
\right),
   \label{eq:M(W->lv) explicit}
\end{equation}
where, as before, we are required to use a factor of $1/\sqrt{2}$ rather than $1/2$ when symmetrizing for the spin-$0$ W~boson. Summing over the square of the absolute value of each of these gives, 
\begin{equation}
\sum |\mathcal{M}|^2 =
  \frac{g_{Wl\nu}^2}{2M_W^2}
  (
    2 M_W^2 + m_l^2
  )
  (
    M_W^2 - m_l^2
  ).
   \label{eq:M^2(W->lv) explicit}
\end{equation}

\subsubsection{Implicit}
We next square the amplitude and sum over spins without calculating the spin combinations explicitly.  After making the spin-indices visible, we have
\begin{equation}
\sum |\mathcal{M}|^2  
 = -\frac{g_{Wl\nu}^2}{2M_W^2}
    (
      \langle \mathbf{3}^L \mathbf{1}^I \rangle
      \lbrack 2\mathbf{3}^K \rbrack
     +\langle \mathbf{3}^K \mathbf{1}^I \rangle
      \lbrack 2\mathbf{3}^L \rbrack
    )
    \lbrack \mathbf{1}_I \mathbf{3}_L \rbrack
    \langle \mathbf{3}_K 2 \rangle.
\end{equation}
Expanding and using the identities in Eqs.~(\ref{eq:app:p=|>[|}) through (\ref{eq:App:|][|=-m}), (\ref{eq:app:<ji>[ij]=2pipj massless}) and (\ref{eq:app:<ji>[ij]=2pipj}), we obtain,
\begin{equation}
\sum |\mathcal{M}|^2 =
  \frac{g_{Wl\nu}^2}{M_W^2}
  (
    2 p_1 \cdot p_3 \, p_2 \cdot p_3
   + M_W^2 p_1 \cdot p_2 
  ).
\end{equation}
Taking the decay to occur in the \textsc{cm} frame, we take the momenta to be $p_3 = (M_W, 0, 0, 0)$,  $p_1 = (M_W^2+m_l^2, 0, 0, M_W^2-m_l^2)/(2M_W)$, $p_2 = (M_W^2 - m_l^2, 0, 0, m_l^2 - M_W^2)/(2M_W)$.  With this, we obtain,
\begin{equation}
\label{eq:M^2(W->lv)}
\sum|\mathcal{M}|^2 =
  \frac{g_{Wl\nu}^2}{2M_W^2}
  (
    2 M_W^2 + m_l^2
  )
  (
    M_W^2 - m_l^2
  ).
\end{equation}
We find perfect agreement with the explicit spin-combination method and, indeed, with Feynman diagrams as output by CalcHEP.

\subsection{\label{sec:W->qq}$W \rightarrow q\bar{q}$}
We now need to do the quark decay of the W~boson.  Since the quarks are all massive, the amplitude includes all spin-spinors.  It is given by,
\begin{equation}
\mathcal{M} = 
  \frac{g_{Wqq}}{M_W} 
    \langle \mathbf{31} \rangle
    \lbrack \mathbf{23} \rbrack,
    \label{eq:M(W->qq)}
\end{equation}
where the particles are the W~boson (3), up-type quark (2) and down-type quark (1).  Of course, we could include a non-trivial CKM matrix element if we would like.  But, that does not affect the spinor algebra performed here and we will leave it out to obtain simpler expressions.

\subsubsection{Explicit}
As before, we begin by finding the explicit amplitude for each spin combination.  Since there are three massive particles, a single matrix will not suffice.  We will split the amplitudes into two matrices.  The first will be for a $-1/2$-spin up-type quark while the second matrix will be for a $+1/2$-spin up-type quark.  As usual, we will take the rows to give the spin of the W~boson and we will take the column to give the spin of the down-type quark.  We find,
\begin{eqnarray}
\mathcal{M}^{-\sfrac{1}{2} \, s_3 \, s_2} &=&
   \frac{1}{2}g_{Wqq}
   \left(
   \begin{array}{cc}
    2 \, C \sin^2(\theta/2) \exp(i\phi)
    &
    - D \sin\theta \exp(2i\phi)  \\
    -\sqrt{2} \, C \sin\theta
    & 
    \sqrt{2} D \cos\theta \exp(i\phi) \\
    2 \, C \cos^2(\theta/2) \exp(-i\phi)
    &
    D \sin\theta
   \end{array}
   \right) 
    \label{eq:M(W->qq) explicit 1}\\
   \mathcal{M}^{+\sfrac{1}{2} \, s_3 \, s_2} &=&
  -\frac{1}{2}g_{Wqq}
   \left(
   \begin{array}{cc}
     - A \sin\theta 
     &  
     2 B \cos^2(\theta/2) \exp(i\phi)  \\
     \sqrt{2} A \cos\theta \exp(-i\phi)
     & 
     \sqrt{2} B \sin\theta \\
    A \sin\theta \exp(-2i\phi)
     & 
     2 B \sin^2(\theta/2) \exp(-i\phi) 
   \end{array}
   \right),
    \label{eq:M(W->qq) explicit 2}
\end{eqnarray}
where $A = \sqrt{(E_{u} - p_q) (E_{d} + p_q)}$, $B = \sqrt{(E_{u} - p_q) (E_{d} - p_q)}$, $C = \sqrt{(E_{u} + p_q) (E_{d} + p_q)}$ and $D = \sqrt{(E_{u} + p_q) (E_{d} - p_q)}$, where we take $u$ to represent the up-type quark of any generation and $d$ to represent the down-type quark of any generation.  
If we square the absolute value of each component and add them all together, we obtain,
\begin{equation}
\sum |\mathcal{M}|^2 = 
   g_{Wqq}^2
   (E_{u} E_{d} + p_q^2
   ).
\end{equation}
In the \textsc{cm} frame, we can take the energies and momentum of the quarks to be,
\begin{align}
E_{u} &= \frac{1}{2M_W}
          ( M_W^2 + M_{u}^2 - M_{d}^2
          ) \label{eq:E_q1} \\
E_{d} &= \frac{1}{2M_W}
          ( M_W^2 + M_{d}^2 - M_{u}^2
          ) \label{eq:E_q2} \\
p_q &= \frac{1}{2M_W}
       \sqrt{ M_W^4 - 2 M_W^2
       ( M_{u}^2 + M_{d}^2
       )
      +( M_{u}^2 - M_{d}^2
       )}.
\label{eq:p_q}
\end{align} 
This gives us,
\begin{equation}
\sum |\mathcal{M}|^2 =
   g_{Wqq}^2
   \left[
      M_W^2 
      - \frac{1}{2}
      ( M_{u}^2 + M_{d}^2
      )
      -\frac{
       ( M_{u}^2 - M_{d}^2
       )^2}{2 M_W^2}
   \right].
   \label{eq:M^2(W->qq)}
\end{equation}

\subsubsection{Implicit}
If we explicitly symmetrize the spin-indices and multiply by its complex conjugate, we obtain,
\begin{equation}
\sum |\mathcal{M}|^2 =
    \frac{g_{Wqq}^2}{2M_W^2}
    (
      \langle \mathbf{3}^L \mathbf{1}^I \rangle
      \lbrack \mathbf{2}^J \mathbf{3}^K \rbrack
     +\langle \mathbf{3}^K \mathbf{1}^I \rangle
      \lbrack \mathbf{2}^J \mathbf{3}^L \rbrack
    )
    \lbrack \mathbf{1}_I \mathbf{3}_L \rbrack
    \langle \mathbf{3}_K \mathbf{2}_J \rangle.
\end{equation}
Expanding and using the identities in Eqs.~(\ref{eq:app:p=|>[|}) through (\ref{eq:App:|][|=-m}) and (\ref{eq:app:<ji>[ij]=2pipj}), we find,
\begin{equation}
  \sum |\mathcal{M}|^2 =
  \frac{g_{Wqq}^2}{M_W^2}
  ( 2 p_1 \cdot p_3 \,
           p_2\cdot p_3 + M_W^2 \, p_1 \cdot p_2
  ).
\end{equation}
After inserting the energy and momentum as in Eqs.~(\ref{eq:E_q1}) through (\ref{eq:p_q}), we obtain, 
\begin{equation}
\sum |\mathcal{M}|^2 =
   g_{Wqq}^2
   \left(
      M_W^2 - \frac{1}{2}
      ( M_{u}^2 + M_{d}^2
      )
      -\frac{
       ( M_{u}^2 - M_{d}^2
       )^2}{2M_W^2}
   \right),
   \label{eq:M^2(W->qq) implicit}
\end{equation}
in perfect agreement with Eq.~(\ref{eq:M^2(W->qq)}) and with Feynman diagrams as given by CalcHEP. We further notice that in the limit that $M_{d} \to 0$ and $M_{u} \to M_l$, we obtain the same result as in Eq.~(\ref{eq:M^2(W->lv) explicit}).  On the other hand, if we take $M_{u} \to m_f$ and $M_{d} \to m_f$, so they are the same, we obtain the $g_L$ part of Eq.~(\ref{eq:M^2(Z->ff) implicit}).

\subsection{\label{sec:h->ff}$h \rightarrow f\bar{f}$}
There is only one $2$-body decay left in the SM and that is the Higgs decay to two massive fermions.  The $3$-point amplitude is given by,
\begin{equation}
\mathcal{M} = 
  -\frac{m_f}{v}
  (
    \langle \mathbf{12} \rangle 
   +\lbrack \mathbf{12} \rbrack
  ),
  \label{eq:M(h->ff)}
\end{equation}
where $v = 246$~GeV is the vacuum expectation value (\textsc{vev}) of the Higgs boson.  Although we do not have fields with vacuum expectation values, it is still a parameter in our constructive model that determines the strength of the Yukawa coupling.

\subsubsection{Explicit}
Since the Higgs is spinless, we can represent the amplitudes as a matrix in the spins of the fermions.  The explicit $3$-point amplitude in the \textsc{cm} frame is given by,
\begin{equation}
\mathcal{M}^{s_1 \, s_2} = 
   \frac{m_f}{v}
   \sqrt{M_h^2 - 4 m_f^2}
   \left(
   \begin{array}{cc}
    -\exp(i\phi) & 0 \\
   0             & \exp(-i\phi) \\
  \end{array}
  \right).
  \label{eq:M(h->ff) explicit}
\end{equation}
As usual, the Higgs only couples to same-spin fermions, so the off-diagonal terms are zero.  If we sum the absolute value of each element squared, we obtain,
\begin{equation}
\sum |\mathcal{M}|^2 = 
   \frac{2 m_f^2}{v^2}
   (M_h^2 - 4 m_f^2 ).
   \label{eq:M^2(h->ff)}
\end{equation}

\subsubsection{Implicit}
After multiplying by the complex conjugate of the amplitude and summing over spins, we obtain,
\begin{equation}
\sum |\mathcal{M}|^2 = 
   \frac{m_f^2}{v^2}
   (
      \langle \mathbf{2}_J \mathbf{1}_I \rangle
      \lbrack \mathbf{1}^I \mathbf{2}^J \rbrack
    + \lbrack \mathbf{2}_J \mathbf{1}_I \rbrack
      \lbrack \mathbf{1}^I \mathbf{2}^J \rbrack
    + \langle \mathbf{2}_J \mathbf{1}_I \rangle
      \langle \mathbf{1}^I \mathbf{2}^J \rangle
    + \lbrack \mathbf{2}_J \mathbf{1}_I \rbrack
      \langle \mathbf{1}^I \mathbf{2}^J \rangle
   ).
\end{equation}
Using the identities in Eqs.~(\ref{eq:app:p=|>[|}) through (\ref{eq:App:|][|=-m}) and (\ref{eq:app:<ji>[ij]=2pipj}), we obtain,
\begin{equation}
\sum |\mathcal{M}|^2 = 
  \frac{4 m_f^2}{v^2}
  ( p_1 \cdot p_2 - m_f^2
  ).
  \label{eq:M^2(h->ff) implicit}
\end{equation}
In the \textsc{cm} frame, $E_f = M_h/2$ and $p_f = \sqrt{E_f^2 - m_f^2}$ giving us,
\begin{equation}
\sum |\mathcal{M}|^2 = 
   \frac{2 m_f^2}{v^2}
   (M_h^2 - 4 m_f^2
   ),
\end{equation}
in agreement with Eq.~(\ref{eq:M^2(h->ff)}) and with Feynman diagrams.

\section{\label{sec:3-body decays}3-Body Decays}
We have now exhausted the $2$-body decays of the SM and we move onto $3$-body decays. We will do all the 3-body decays that do not involve the radiation of a photon or gluon from a $2$-body decay.  From now on, we will not describe the process of squaring the amplitude since it is exactly the same as in the previous section.  The only new element is having another external particle to consider.    

As we move onto $3$-body decays, our focus shifts to constructing higher-point amplitudes using 3-point vertices and propagators.  As described in Sec.~\ref{sec:general techniques}, we have only found agreement with Feynman diagrams when we have been able to reduce higher-point amplitudes all the way to the point that there are no momenta left in the numerator.  Our main tools for doing this are momentum conservation, the Schouten identity (see App.~\ref{sec:generalized Schouten}) and the spinor identities (see App.~\ref{sec:spinor conjugation}).  We believe this requirement is related to the requirement that the internal momenta are on-shell.  Any momenta that are left over at the end, not being complex, will not satisfy the on-shell condition for internal particles.  In this section, we will demonstrate in detail the way this is done with the examples of the SM $3$-body decays.  

Furthermore, since all momenta are completely removed from the numerators, the amplitudes can only be functions of spinor products, masses and propagator denominators.  It turns out this allows us to guess the structure of the final amplitude in some cases where there are few building blocks available for the amplitude.  For each example, we will begin by determining the most general structure allowed for the amplitude.  We will then calculate it using the constructive techniques and compare with our expectation.

\subsection{\label{sec:m->vev}$\mathbf{\mu \rightarrow \nu_{\mu} \, e \, \bar{\nu}_e}$}
We begin with leptonic muon decay, which requires the connection of two $3$-point vertices.  Since there is only one diagram and each vertex only has one term, it is quite simple.  We take the ordering of the momenta to be the same as the order of the particles in the process.  The muon will be $p_1$, $p_2$ for the muon neutrino, $p_3$ for the electron, and $p_4$ for the electron anti-neutrino.  We take all particles to be incoming ($p_1+p_2+p_3+p_4=0$) during the calculation and only flip the momenta of the final states at the point of comparison with Feynman diagrams.  Since the (incoming) muon anti-neutrino must have $+1/2$-helicity, the final amplitude must have a left- or right-facing $|2\rbrack$ and since the (incoming) electron neutrino must have $-1/2$-helicity, the final amplitude must have a left- or right-facing $|4\rangle$.  Since the muon and electron are $1/2$-spin, there must be one spinor of either type for each of them.  There are only two nonzero combinations of these four spinors.  Since the helicity-spinors cannot be contracted with each other, they must each be contracted with the spinors of the electron or muon.  Therefore, we expect the amplitude to be of the form,
\begin{equation}
    \mathcal{M} = g_{Wff}^2\frac{\mathcal{A} \lbrack 2\mathbf{3} \rbrack
            \langle \mathbf{1}4 \rangle
     +\mathcal{B} \langle\mathbf{3}4 \rangle\lbrack\mathbf{1}2\rbrack 
   }{(p_1+p_2)^2 - M_W^2}.
   \label{eq:M(m->vlv) expected}
\end{equation}
Since a $4$-particle amplitude is dimensionless, we see that $\mathcal{A}$ and $\mathcal{B}$ must be dimensionless products of the masses $M_W, m_\mu$ and $m_e$.  In order to determine what these coefficients are, we must now calculate the amplitude using the constructive vertices.

The two vertices, taken from \cite{Christensen:2018zcq} are,
\begin{equation}
\frac{ig_{Wl\nu}}{M_W}
  \lbrack 2\mathbf{P}_{12}^I \rbrack
  \langle \mathbf{P}_{12}^J \mathbf{1} \rangle 
\quad \mbox{and} \quad
-\frac{ig_{Wl\nu}}{M_W}
  \langle 4\mathbf{P}_{12 \, K} \rangle
  \lbrack \mathbf{P}_{12 \, L} \mathbf{3} \rbrack,
  \label{eq:vertices m->vlv}
\end{equation}
where the spinors with a bold $\mathbf{P}_{12}$ are spin-spinors with momentum $p_{12}=p_1+p_2$.  These are the spinors of the W~boson and are taken to be on-shell, which brings us to a very important detail that we discussed in Sec.~\ref{sec:general techniques}.  If the W~boson is off-shell, we do not have a spinor formalism for it.  On the other hand, if all the external momenta are real, on-shell and satisfy momentum conservation, the internal line is usually not on-shell and is certainly not on-shell in the case of muon decay.  Nevertheless, the rules of the constructive method \cite{Arkani-Hamed:2017jhn, Elvang:2015rqa} require that the internal line is taken to be on-shell during the intermediate steps of the calculation (with the exception of the propagator denominator).  This normally requires at least two of the momenta to be momentarily taken complex.  Although this complexification is well understood for purely massless theories\cite{Britto:2005fq}, it is not currently present in the literature for massive constructive theories.  In particular, the modification of the spin-spinors for the complex momentum case is not yet known.  We have worked on this massive complexificaiton, but have not satisfactorily resolved it yet.  Without a detailed understanding of the complexification, it may appear that we are not able to calculate these amplitudes at all.  However, we will see that, at least in the cases presented here, the internal momenta and internal spinors can be completely removed from the amplitude numerator using a combination of momentum conservation, the on-shell conditions and the Schouten identities.  Since we do not know which of the external momenta might be complexified, we see that it is imperative that we remove all traces of the momenta from the numerator.  This is precisely what we will do in all the amplitudes presented in this work and we will save the details of complexification for a later work.

Our next step is to symmetrize over the spin-indices of the W~boson, multiply these vertices and sum over the internal spins.  When we do this, there will be two spin contractions coming from the two orders of the symmetrized spin-indices.  We also divide by the propagator denominator to obtain, 
\begin{equation}
\mathcal{M} =
  g_{Wl\nu}^2
  \frac{
  (
    \lbrack 2 \mathbf{P}_{12}^I \rbrack
    \lbrack \mathbf{P}_{12 \, I} \mathbf{3} \rbrack
    \langle \mathbf{1} \mathbf{P}_{12}^J \rangle
    \langle \mathbf{P}_{12 \, J} 4 \rangle
   +\lbrack 2\mathbf{P}_{12}^I \rbrack
    \langle \mathbf{P}_{12 \, I} 4 \rangle
    \langle \mathbf{1} \mathbf{P}_{12}^J \rangle
    \lbrack \mathbf{P}_{12 \, J} \mathbf{3} \rbrack
    )
    }{2M_W^2 ((p_1+p_2)^2 - M_W^2)},
   \label{eq:M(m->vlv) unsimplified}
\end{equation}
where we have flipped the order of the two angle bracket products and the signs cancel for these flips (i.e. $\langle ij\rangle=-\langle ji\rangle$ and similarly for square brackets).  We now use the identities in Eqs.~(\ref{eq:app:p=|>[|}) through (\ref{eq:App:|][|=-m}) to obtain,
\begin{equation}
\mathcal{M} = 
   g_{Wl\nu}^2
   \frac{
   (-
      M_W^2 \lbrack 2\mathbf{3} \rbrack
            \langle 4\mathbf{1} \rangle
     +m_\mu \lbrack 2|p_1|4 \rangle 
            \lbrack \mathbf{13} \rbrack
     + \lbrack 2|p_1|4 \rangle
       \langle \mathbf{1}|p_2| \mathbf{3} \rbrack
   )}{2M_W^2 ((p_1+p_2)^2 - M_W^2)}.
\end{equation}
As we can see, at this point, we still have momenta left in the numerator.  Therefore, we cannot consider our job done. We next simplify this amplitude by using the Schouten identities from App.~\ref{sec:generalized Schouten}.  We begin with the third term,
\begin{eqnarray}
    \lbrack 2|p_1|4 \rangle
       \langle \mathbf{1}|p_2| \mathbf{3} \rbrack &=&
       \lbrack\mathbf{3}|p_1|4\rangle\langle \mathbf{1}|p_2|2\rbrack -
       \langle \mathbf{1}|p_2p_1|4 \rangle\lbrack\mathbf{3}2\rbrack.
\end{eqnarray}
Since the neutrino is taken to be massless, $p_2|2\rbrack=0$ and we are left with only the last term.  We further use $p_2p_1 = (p_1+p_2)^2-m_{\mu}^2-p_1p_2 = M_W^2-m_{\mu}^2-p_1p_2$, where we have used the on-shell property for the internal line, to obtain,
\begin{eqnarray}
    \lbrack 2|p_1|4 \rangle
       \langle \mathbf{1}|p_2| \mathbf{3} \rbrack &=&
        -(M_W^2-m_{\mu}^2)\langle \mathbf{1}4 \rangle\lbrack\mathbf{3}2\rbrack
       +m_{\mu}\lbrack \mathbf{1}|p_2|4 \rangle\lbrack\mathbf{3}2\rbrack,
\end{eqnarray}
where we have also used $\langle\mathbf{1}|p_1 = m_{\mu}\lbrack\mathbf{1}|$.
Combining this with our full amplitude gives,
\begin{equation}
\mathcal{M} = 
   g_{Wl\nu}^2
   \frac{
   \left[-
      2M_W^2 \lbrack 2\mathbf{3} \rbrack
            \langle 4\mathbf{1} \rangle
     +m_{\mu}^2\lbrack 2\mathbf{3} \rbrack
            \langle 4\mathbf{1} \rangle
     +m_\mu (
     \lbrack 2|p_1|4 \rangle 
            \lbrack \mathbf{13} \rbrack
     + \lbrack \mathbf{1}|p_2|4 \rangle\lbrack\mathbf{3}2\rbrack
       )
   \right]}{2M_W^2 ((p_1+p_2)^2 - M_W^2)}.
\end{equation}
We next use conservation of momentum $p_2=-(p_1+p_3+p_4)$ on the last term,
\begin{eqnarray}
    \lbrack \mathbf{1}|p_2|4 \rangle\lbrack\mathbf{3}2\rbrack &=&
    -m_\mu \langle\mathbf{1}4\rangle\lbrack\mathbf{3}2\rbrack
    -\lbrack \mathbf{1}|p_3|4 \rangle\lbrack\mathbf{3}2\rbrack
\end{eqnarray}
where we have also used the masslessness condition of the neutrino $p_4|4\rangle=0$.
This leaves us with,
\begin{equation}
\mathcal{M} = 
   g_{Wl\nu}^2
   \frac{
   \left[-
      2M_W^2 \lbrack 2\mathbf{3} \rbrack
            \langle 4\mathbf{1} \rangle
     -m_\mu (
     \lbrack 2|p_3|4 \rangle 
            \lbrack \mathbf{13} \rbrack
     +\lbrack \mathbf{1}|p_3|4 \rangle\lbrack\mathbf{3}2\rbrack
       )
   \right]}{2M_W^2 ((p_1+p_2)^2 - M_W^2)}.
\end{equation}
We now apply the Schouten identity to the last term,
\begin{eqnarray}
    \lbrack \mathbf{1}|p_3|4 \rangle\lbrack\mathbf{3}2\rbrack &=&
    \lbrack2|p_3|4 \rangle\lbrack\mathbf{31}\rbrack -
    m_e\langle\mathbf{3}4 \rangle\lbrack2\mathbf{1}\rbrack,
\end{eqnarray}
by using $\lbrack\mathbf{3}|p_3=m_e\langle\mathbf{3}|$.
This cancels the middle term and leaves us with,
\begin{equation}
\mathcal{M} = 
   g_{Wff}^2
   \frac{
      2M_W^2 \lbrack 2\mathbf{3} \rbrack
            \langle \mathbf{1}4 \rangle
     -m_\mu m_e \lbrack\mathbf{1}2\rbrack \langle\mathbf{3}4 \rangle
   }{2M_W^2 ((p_1+p_2)^2 - M_W^2)}.
   \label{eq:M(m->vlv)}
\end{equation}
As promised, we find that each term is composed of different combinations of a square helicity-spinor for the muon neutrino which is helicity $+1/2$, an angle helicity-spinor for the anti-electron neutrino which is helicity $-1/2$ and a spin-spinor of either square or angle type for the muon and the electron.   In fact, we see that one term has an angle spinor for the muon and a square spinor for the electron while the other term has the opposite.  These are all the combinations that are possible while giving the correct helicities for the neutrinos.  The muon and electron spinors could be contracted with each other but the neutrino spinors cannot.  Moreover, we notice that the spinor products that connect opposite sides of the propagator have a $M_W^2$ coefficient while the products that only connect the same side of the propagator have external-line masses.  We will see more of this in the next two sections.  We will comment further on this structure later.

Before squaring this amplitude, we point out that it is significantly simpler than the Feynman diagram amplitude.  The propagator denominator and overall factor will, of course, be the same, but the number of computations required to calculate the amplitude for a given spin combination of the muon and electron is considerably larger.  For comparison, we write the numerator of the Feynman-diagram result, in unitary gauge, up to an overall constant,
\begin{equation}
    \mathcal{M}_{\text{FD}} \propto
    \bar{\psi}_{\nu_{\mu}}\gamma^\alpha P_L\psi_{\mu}
    \left(-g_{\alpha\beta}+\frac{p_{W\alpha}p_{W\beta}}{M_W^2}\right)
    \bar{\psi}_{e}\gamma^{\beta}P_L\psi_{\nu_{e}}.
    \label{eq:M(m->evv) FD}
\end{equation}
For any spin combination, first the 4-component spinors $\bar{\psi}_{\nu_{\mu}}, \psi_{\mu}, \bar{\psi}_{e}$ and $\psi_{\nu_{e}}$ must be constructed for each spin.  Then, these 4-component spinors must be multiplied by the $4\times4$ matrices $\gamma^\alpha P_L$ in $\bar{\psi}_{\nu_{\mu}}\gamma^\alpha P_L\psi_{\mu}$ for each $\alpha$ and $\bar{\psi}_{e}\gamma^{\beta}P_L\psi_{\nu_{e}}$ for each $\beta$.  Finally, the result must be multiplied by the $4\times4$ matrix $\left(-g_{\alpha\beta}+p_{W\alpha}p_{W\beta}/M_W^2\right)$.  There are, no doubt, some improvements over this simple-minded picture, but the number of calculations cannot be reduced by much using this Feynman-diagram method and this is exactly what is done by most of the matrix element generators on the market.  We hasten to note that, our goal is not to criticize the matrix element generators, which are spectacular in our opinion.  Rather, it is to point out the potential improvement in efficiency in the constructive technique relative to Feynman diagrams.  Now, compare this with the calculation of the amplitude given in Eq.~(\ref{eq:M(m->vlv)}).  We need only calculate the inner products $\lbrack 2\mathbf{3} \rbrack$, $\langle \mathbf{1}4 \rangle$, $\lbrack\mathbf{1}2\rbrack$, and $ \langle\mathbf{3}4 \rangle$.  Lest the reader thinks there is a large amount of unseen complexity in these calculations, we remind the reader that we have written down all the relevant cases in App.~\ref{app:explicit spinor products}.  We also note that the matrices in this appendix contain the results for all the spin combinations of each spinor.  Therefore, to compare with our earlier discussion of Feynman diagrams, in order to calculate one spin combination, we need only calculate one element of each of these arrays for each spinor product.  We then combine the results with $2M_W^2$ and $m_{\mu}m_e$ and add to obtain the numerator.  This is trivial in comparison.   

Moreover, the improvement in the simplicity of the formulas also extends to the squared amplitude. Following the rules described in App.~\ref{sec:spinor conjugation} and demonstrated in Sec.~\ref{sec:2-body decays}, we square the amplitude, summing over spins, to obtain,
\begin{equation}
    \sum |\mathcal{M}|^2 =
     g_{Wff} ^{4} \frac{4M_W^{4} p_1 \cdot p_4 \, p_2 \cdot p_3 
                      + 2m_e^2 m_\mu^2 M_W^2 \, p_2 \cdot p_4 
                      + m_e^2 m_\mu^2 \, p_1 \cdot p_2 \, p_3 \cdot p_4 }
                      {M_W^4 \left[ (p_{1} +p_{2} )^2-M_W^2 \right]^2},
     \label{eq:M^2(m->evv)}
\end{equation}
where there are three terms coming from the squaring of the first term of Eq.~(\ref{eq:M(m->vlv)}), the second term and the cross term.  
On the other hand, after squaring the Feynman diagram and summing over spins, we obtain something like,
\begin{align} \nonumber
    \mbox{Tr} & \left[\slashed{p}_{\nu_{\mu}}\gamma^\alpha P_L\left(\slashed{p}_{\mu}+m_{\mu}\right)\gamma^\epsilon\right]
    \left(-g_{\alpha\beta}+\frac{p_{W\alpha}p_{W\beta}}{M_W^2}\right) \\ 
    & \qquad
    \times \left(-g_{\epsilon\eta}+\frac{p_{W\epsilon}p_{W\eta}}{M_W^2}\right)
    \mbox{Tr}\left[\left(\slashed{p}_{e}+m_e\right)
             \gamma^\beta P_L\slashed{p}_{\nu_{e}}\gamma^\eta\right].
\end{align}
Of course, after tracing the gamma matrices, some simplification will be possible, but it is not likely an automated program will find a path to Eq.~(\ref{eq:M^2(m->evv)}).  In fact, the result coming from Feynman diagrams, as given by CalcHEP, is much more complicated.  After reversing the momentum of the final three particles, it can be written,
\begin{align} \nonumber
    \sum |\mathcal{M}|^2 &=
    \frac{e^4}
         {4\sin^4\theta_W
          M_W^4 \left[ \left( p_1 + p_2 \right)^2 - M_W^2 \right]^2} \\ \nonumber
    & \quad \times
      \Big[ m_e^4 m_\mu^2 \left( m_\mu^2 +
                          \left( 2 M_W^2 + 
                          \left( p_1 \cdot p_3 + p_1 \cdot p_4 \right)\right)\right) 
           +4 M_W^4 p_1 \cdot p_4 \left( m_\mu^2 + 2 p_1 \cdot p_4 \right) \\ \nonumber
    & \qquad + m_e^2 m_\mu^6 -4M_W^4 m_e^2 p_1 \cdot p_4 +m_\mu^4 m_e^2 \left(2 M_W^2 + 3\left(p_1 \cdot p_3 +p_1 \cdot p_4 \right)\right) \\
    & \qquad + 2m_\mu^2 m_e^2 \left(2M_W^2 p_1 \cdot p_3 +\left(p_1 \cdot p_3 + p_1 \cdot p_4 \right)^{2}   \right)
    \Big].
    \label{eq:M^2(m->evv) Feynman}
\end{align}
We find agreement between our result and that coming from CalcHEP after using standard momentum conservation and on-shell mass conditions on both results.  Nevertheless, we can see that the result naturally coming from the constructive method is, once again, much simpler.  In both these cases, this result and the ones that follow are strongly suggestive that a matrix-element generator or a squared amplitude calculator based on this constructive approach would lead to much more efficient calculations of the scattering cross sections needed for current and future scattering experiments.  Of course, in order to achieve this, the constructive method will need to be further developed.  We will discuss this in more detail in the conclusions.  

\subsection{\label{sec:m->vqq}$\mathbf{\tau\to \nu_{\tau}q\bar{q}}$}

In this section, we consider tau decay to two quarks and a neutrino which again requires one diagram with the connection of two three-point vertices.  Since each vertex still only has one term, it is still quite simple.  We take the ordering of the momenta to be the same as the order of the particles in the process.  $p_1$ will be for the tau, $p_2$ for the tau neutrino, $p_3$ for the down-type quark and $p_4$ for the up-type quark.  We continue to take all particles incoming during the calculation and only flip the momenta of the external particles when we compare with Feynman diagrams at the end.

As in the previous subsection, we begin by determining the allowed form of the final amplitude based on the spin structure of the external particles.  Since the (incoming) tau anti-neutrino is helicity $+1/2$, there must be a $|2\rbrack$ facing in either direction.  However, since all three other particles are massive and spin $1/2$, they can be either angle or square brackets in either direction.  There are many more possibilities this time.  There are three combinations of the $|2\rbrack$ with each of the other particle's spinors.  The other two particle's spinors are then contracted with each other as either square or angle brackets.  Therefore, the amplitude must be a linear combination of the terms
\begin{eqnarray}
    \lbrack\mathbf{1}2\rbrack\langle\mathbf{34}\rangle, \quad \quad
    \lbrack2\mathbf{3}\rbrack\langle\mathbf{14}\rangle, \quad \quad
    \lbrack2\mathbf{4}\rbrack\langle\mathbf{13}\rangle, 
    \label{eq:mu->nu,q,q:prods1}\\
    \lbrack\mathbf{1}2\rbrack\lbrack\mathbf{34}\rbrack, \quad \quad
    \lbrack2\mathbf{3}\rbrack\lbrack\mathbf{14}\rbrack, \quad \quad
    \lbrack2\mathbf{4}\rbrack\lbrack\mathbf{13}\rbrack.
    \label{eq:mu->nu,q,q:prods2}
\end{eqnarray}
However, not all three of the products in the bottom row are unique.  One of them can be rewritten as a linear combination of the other two using a Schouten identity.  Below, we will replace the last with the first two.  So, we should be able to simplify this amplitude to no more than 5 of these and possibly less depending on the vertices.    The coefficients of these terms will be a dimensionless product of the masses $M_W, m_\tau, m_u, m_d$ divided by the W-boson propagator denominator.

The two vertices, taken from \cite{Christensen:2018zcq} are,
\begin{equation}
\frac{ig_{Wl\nu}}{M_W}
  \lbrack 2\mathbf{P}_{12}^I \rbrack
  \langle \mathbf{P}_{12}^J \mathbf{1} \rangle 
\quad \mbox{and} \quad
-\frac{ig_{Wqq}}{M_W}
  \langle \mathbf{4}\mathbf{P}_{12 \, K} \rangle
  \lbrack \mathbf{P}_{12 \, L} \mathbf{3} \rbrack,
  \label{eq:vertices m->vqq}
\end{equation}
where $|\mathbf{P}_{12}\rangle$ and $|\mathbf{P}_{12}\rbrack$ are the spin-spinors for a massive particle of momentum $p_{12} = p_1 + p_2$.  As we discussed in the previous subsection, since we do not yet have a detailed understanding of the complexification of these spin-spinors, we must completely remove all traces of the internal spin-spinors and all traces of momenta in the numerator.  We must reduce the amplitude numerator to a form that only has products of external spinors and masses.

We next symmetrize over the spin indices, multiply these vertices and sum over the internal spins.  There are two spin contractions due to the symmetrization.  We also divide by the propagator denominator to obtain,
\begin{equation}
\mathcal{M} =
  g_{Wl\nu}g_{Wqq}
  \frac{
  (
    \lbrack 2 \mathbf{P}_{1,2}^I \rbrack
    \lbrack \mathbf{P}_{1,2 \, I} \mathbf{3} \rbrack
    \langle \mathbf{1} \mathbf{P}_{1,2}^J \rangle
    \langle \mathbf{P}_{1,2 \, J} \mathbf{4} \rangle
   +\lbrack 2\mathbf{P}_{1,2}^I \rbrack
    \langle \mathbf{P}_{1,2 \, I} \mathbf{4} \rangle
    \langle \mathbf{1} \mathbf{P}_{1,2}^J \rangle
    \lbrack \mathbf{P}_{1,2 \, J} \mathbf{3} \rbrack
    )
    }{2M_W^2 ((p_1+p_2)^2 - M_W^2)},
  \label{eq:M(m->vqq) unsimplified}
\end{equation}
where we have flipped the order of the two angle bracket products and the signs cancel for these flips.  We now use the identities in Eqs.~(\ref{eq:app:p=|>[|}) through (\ref{eq:App:|][|=-m}) to obtain,
\begin{equation}
\mathcal{M} = 
   g_{Wl\nu}g_{Wqq}
   \frac{
   (-
      M_W^2 \lbrack 2\mathbf{3} \rbrack
            \langle \mathbf{4}\mathbf{1} \rangle
     +m_\tau \lbrack 2|p_1|\mathbf{4} \rangle 
            \lbrack \mathbf{13} \rbrack
     + \lbrack 2|p_1|\mathbf{4} \rangle
       \langle \mathbf{1}|p_2| \mathbf{3} \rbrack
   )}{2M_W^2 ((p_1+p_2)^2 - M_W^2)}.
\end{equation}
We next simplify this amplitude by using the Schouten identities in App.~\ref{sec:generalized Schouten}.  We begin with the third term,
\begin{eqnarray}
    \lbrack 2|p_1|\mathbf{4} \rangle
       \langle \mathbf{1}|p_2| \mathbf{3} \rbrack &=&
       \lbrack\mathbf{3}|p_1|\mathbf{4}\rangle\langle \mathbf{1}|p_2|2\rbrack -
       \langle \mathbf{1}|p_2p_1|\mathbf{4} \rangle\lbrack\mathbf{3}2\rbrack.
\end{eqnarray}
Since the neutrino is taken to be massless, $p_2|2\rbrack = 0$ and we are left with only the last term.  We further use $p_2 p_1 = (p_1 + p_2)^2 - m_{\tau}^2 - p_1 p_2 = M_W^2 - m_{\tau}^2 - p_1 p_2$, where we have used the on-shell property for the internal line, to obtain,
\begin{eqnarray}
    \lbrack 2|p_1|\mathbf{4} \rangle
       \langle \mathbf{1}|p_2| \mathbf{3} \rbrack &=&
        -(M_W^2-m_{\tau}^2)\langle \mathbf{14} \rangle\lbrack\mathbf{3}2\rbrack
       +m_{\tau}\lbrack \mathbf{1}|p_2|\mathbf{4} \rangle\lbrack\mathbf{3}2\rbrack,
\end{eqnarray}
where we have also used $\langle\mathbf{1}|p_1 = m_{\tau}\lbrack\mathbf{1}|$.
Combining this with our full amplitude gives,
\begin{equation}
\mathcal{M} = 
   g_{Wl\nu}g_{Wqq}
   \frac{
   \left[-
      2M_W^2 \lbrack 2\mathbf{3} \rbrack
            \langle \mathbf{4}\mathbf{1} \rangle
     +m_{\tau}^2\lbrack 2\mathbf{3} \rbrack
            \langle \mathbf{4}\mathbf{1} \rangle
     +m_\tau (
     \lbrack 2|p_1|\mathbf{4} \rangle 
            \lbrack \mathbf{13} \rbrack
     + \lbrack \mathbf{1}|p_2|\mathbf{4} \rangle\lbrack\mathbf{3}2\rbrack
       )
   \right]}{2M_W^2 ((p_1+p_2)^2 - M_W^2)}.
\end{equation}
We next use conservation of momentum $p_2=-(p_1+p_3+p_4)$ on the last term,
\begin{eqnarray}
    \lbrack \mathbf{1}|p_2|\mathbf{4} \rangle\lbrack\mathbf{3}2\rbrack &=&
    -m_\tau \langle\mathbf{1}\mathbf{4}\rangle\lbrack\mathbf{3}2\rbrack
    -\lbrack \mathbf{1}|p_3|\mathbf{4} \rangle\lbrack\mathbf{3}2\rbrack
    +m_u\lbrack \mathbf{1}\mathbf{4} \rbrack\lbrack\mathbf{3}2\rbrack.
\end{eqnarray}
We also apply momentum conservation to the third term.
This leaves us with,
\begin{align} \nonumber
\mathcal{M} &= 
   \frac{g_{Wl\nu}g_{Wqq}}
        {2M_W^2 ((p_1+p_2)^2 - M_W^2)} \\ \nonumber
   &\times
   [-
      2M_W^2 \lbrack 2\mathbf{3} \rbrack
            \langle \mathbf{4}\mathbf{1} \rangle
     +m_\tau m_u (
     \lbrack2\mathbf{3}\rbrack\lbrack \mathbf{4}\mathbf{1} \rbrack
     +\lbrack 2\mathbf{4} \rbrack 
            \lbrack \mathbf{13} \rbrack ) \\ 
    & \qquad
     -m_\tau (
     \lbrack 2|p_3|\mathbf{4} \rangle 
            \lbrack \mathbf{13} \rbrack
     +\lbrack \mathbf{1}|p_3|\mathbf{4} \rangle\lbrack\mathbf{3}2\rbrack
       )
   ].
\end{align}
We now apply the Schouten identity to the last term,
\begin{eqnarray}
    \lbrack \mathbf{1}|p_3|\mathbf{4} \rangle\lbrack\mathbf{3}2\rbrack &=&
    \lbrack2|p_3|\mathbf{4} \rangle\lbrack\mathbf{31}\rbrack -
    m_d\langle\mathbf{34} \rangle\lbrack2\mathbf{1}\rbrack.
\end{eqnarray}
This cancels the second-to-last term and leaves us with,
\begin{equation}
\mathcal{M} = 
   g_{Wl\nu}g_{Wqq}
   \frac{
   \left[-
      2M_W^2 \lbrack 2\mathbf{3} \rbrack
            \langle \mathbf{4}\mathbf{1} \rangle
     +m_\tau m_u(
     \lbrack2\mathbf{3}\rbrack\lbrack \mathbf{4}\mathbf{1} \rbrack
     +\lbrack 2\mathbf{4} \rbrack 
            \lbrack \mathbf{13} \rbrack
    )
     +m_\tau m_d \langle\mathbf{3}\mathbf{4} \rangle\lbrack2\mathbf{1}\rbrack 
   \right]}{2M_W^2 ((p_1+p_2)^2 - M_W^2)}.
\end{equation}
Finally, we apply the Schouten identity to the third term,
\begin{equation}
\lbrack 2\mathbf{4}\rbrack \lbrack\mathbf{13}\rbrack =
    \lbrack\mathbf{34}\rbrack\lbrack\mathbf{1}2\rbrack -
    \lbrack\mathbf{14}\rbrack\lbrack\mathbf{3}2\rbrack.
\end{equation}
Plugging in, we finally have,
\begin{equation}
\mathcal{M} = 
   g_{Wff}^2
   \frac{
      2M_W^2 \lbrack 2\mathbf{3} \rbrack
            \langle \mathbf{1}\mathbf{4} \rangle
     +m_\tau m_u
     \lbrack\mathbf{1}2\rbrack\lbrack\mathbf{34}\rbrack
     -m_\tau m_d \lbrack\mathbf{1}2\rbrack\langle\mathbf{3}\mathbf{4} \rangle 
   }{2M_W^2 ((p_1+p_2)^2 - M_W^2)}.
  \label{eq:M(m->vqq)}
\end{equation}
In the end, we only needed two spinor products from Eq.~(\ref{eq:mu->nu,q,q:prods1}) and one from Eq.~(\ref{eq:mu->nu,q,q:prods2}).  Interestingly, the term that mixes spinors from opposite sides of the propagator have a purely numerical coefficient (the $M_W^2$ in the numerator cancels the $M_W^2$ in the denominator.)  On the other hand, the spinor products that only contain spinors from the same side of the propagator contain an $m_u$ if it has a $\lbrack\mathbf{34}\rbrack$ while it has an $m_d$ if it has a $\langle\mathbf{34}\rangle$.  On the other hand, it only has an $m_\tau$ and a $\lbrack\mathbf{1}2\rbrack$.  We suspect that if the neutrino were massive, we would also have a $m_\nu\langle\mathbf{12}\rangle$.  Indeed, we will find that this is the case in the next subsection.

Once again, we see that our constructive result for the amplitude given in Eq.~(\ref{eq:M(m->vqq)}) is significantly simpler than the form given by Feynman diagrams.  The Feynman diagram case is nearly identical to Eq.~(\ref{eq:M(m->evv) FD}) but with the electron and electron-neutrino replaced with the up- and down-quarks.  The Feynman diagram calculation is just as complicated as before but the constructive amplitude calculation is nearly as simple as before.  Now, instead of the $2$ terms we obtained in Eq.~(\ref{eq:M(m->vlv)}), we get $3$ terms.  But, each term is a relatively efficient calculation as we show in App.~\ref{app:explicit spinor products} for a variety of cases. 

We have also obtained the squared expression given by,
\begin{align} \nonumber
    \sum |\mathcal{M}|^2 =&
     g_W^4 \Big[\frac{4 M_W^4 p_1 \cdot p_4 \,
                              p_2 \cdot p_3 
                    + m_\tau^2 m_u^2 \left( 2M_W^2 p_2 \cdot p_3 + p_1 \cdot p_2 \, p_3 \cdot p_4 \right)}
                    {M_W^4 \left[\left( p_1 + p_2 \right)^2 - M_W^2 \right]^2}
     \\
     & \qquad +\frac{m_d^2 m_\tau^2 \left(2m_u^2 p_1 \cdot p_2 + \left(2M_W^2 p_2 \cdot p_4 + p_1 \cdot p_2 \,  p_3 \cdot p_4 \right)\right)}{M_W^4\left[\left( p_1 + p_2 \right)^2-M_W^2 \right]^2}\Big],
     \label{eq:M^2(m->vud)}
\end{align}
and comared with Feynman diagrams coming from CalcHEP and find agreement.  The expression, coming from Feynman diagrams as given by CalcHEP has 19 terms by comparison and is not very illuminating so we do not include it here.  However, once again, we find the formula obtained by the constructive technique gives a much simpler result with only $6$ individual terms.  Again, a calculation of scattering amplitudes using the constructive technique leads to a more efficient computation when compared with Feynman diagrams, whether before squaring or after.

\subsection{\label{sec:t->bqq}$\mathbf{t\to b q\bar{q}}$}
There is one further $3$-body decay in the SM.  It is a second- or third-generation quark decaying to lower generation quarks.  We will consider top-quark decay to a bottom quark and two other quarks which requires the connection of two three-point vertices.  Since each vertex still only has one term, it is still quite simple.  We take the ordering of the momenta to be the same as the order of the particles in the process.  The top quark will be $p_1$, $p_2$ for the anti-bottom quark, $p_3$ for the anti-up quark and $p_4$ for the down quark. 

As in the previous section, we begin by determining the form of the final amplitude.  Since all the particles are massive and spin $+1/2$, they can be either angle or square brackets in either direction, so there are more possibilities than last time.  There are three combinations of the momenta and three different combinations of angle and square spinors.  Therefore, the amplitude must be a linear combination of the terms
\begin{align}
    \langle\mathbf{12}\rangle\langle\mathbf{34}\rangle, \quad \quad
   &\langle\mathbf{13}\rangle\langle\mathbf{24}\rangle, \quad \quad
    \langle\mathbf{14}\rangle\langle\mathbf{23}\rangle, 
    \label{eq:t->b,q,q:prods1}\\
    \lbrack\mathbf{12}\rbrack\langle\mathbf{34}\rangle, \quad \quad
   &\lbrack\mathbf{13}\rbrack\langle\mathbf{24}\rangle, \quad \quad
    \lbrack\mathbf{14}\rbrack\langle\mathbf{23}\rangle, 
    \label{eq:t->b,q,q:prods2}\\
    \langle\mathbf{12}\rangle\lbrack\mathbf{34}\rbrack, \quad \quad
   &\langle\mathbf{13}\rangle\lbrack\mathbf{24}\rbrack, \quad \quad
    \langle\mathbf{14}\rangle\lbrack\mathbf{23}\rbrack, 
    \label{eq:t->b,q,q:prods3}\\
    \lbrack\mathbf{12}\rbrack\lbrack\mathbf{34}\rbrack, \quad \quad
   &\lbrack\mathbf{13}\rbrack\lbrack\mathbf{24}\rbrack, \quad \quad
    \lbrack\mathbf{14}\rbrack\lbrack\mathbf{23}\rbrack. 
    \label{eq:t->b,q,q:prods4}
\end{align}
However, not all three of the products in the top and bottom row are unique.  One of them can be rewritten as a linear combination of the other two using a Schouten identity.  So, we should be able to simplify this amplitude to no more than $10$ of these and possibly less depending on the vertices.  The coefficients, other than the propagator and the coupling constant, will be a dimensionless product of the masses $M_W, m_t, m_b, m_u, m_d$.

The two vertices, taken from \cite{Christensen:2018zcq} are,
\begin{equation}
\frac{ig_{Wff}}{M_W}
  \lbrack \mathbf{2}\mathbf{P}_{1,2}^I \rbrack
  \langle \mathbf{P}_{1,2}^J \mathbf{1} \rangle 
\quad \mbox{and} \quad
-\frac{ig_{Wff}}{M_W}
  \langle \mathbf{4}\mathbf{P}_{1,2 \, K} \rangle
  \lbrack \mathbf{P}_{1,2 \, L} \mathbf{3} \rbrack,
\end{equation}
where all momenta are incoming and we will flip the sign of the outgoing momenta at the end.   

We next symmetrize over the spin-indices, multiply these vertices and sum over the internal spins.  There are two spin contractions due to the symmetrization.  We also divide by the propagator denominator to obtain,
\begin{equation}
\mathcal{M} =
  g_{Wff}^2
  \frac{
  (
    \lbrack \mathbf{2} \mathbf{P}_{1,2}^I \rbrack
    \lbrack \mathbf{P}_{1,2 \, I} \mathbf{3} \rbrack
    \langle \mathbf{1} \mathbf{P}_{1,2}^J \rangle
    \langle \mathbf{P}_{1,2 \, J} \mathbf{4} \rangle
   +\lbrack \mathbf{2}\mathbf{P}_{1,2}^I \rbrack
    \langle \mathbf{P}_{1,2 \, I} \mathbf{4} \rangle
    \langle \mathbf{1} \mathbf{P}_{1,2}^J \rangle
    \lbrack \mathbf{P}_{1,2 \, J} \mathbf{3} \rbrack
    )
    }{2M_W^2 ((p_1+p_2)^2 - M_W^2)},
\end{equation}
where we have flipped the order of the two angle bracket products and the signs cancel for these flips.  We now use the identities in Eqs.~(\ref{eq:app:p=|>[|}) through (\ref{eq:App:|][|=-m}) to obtain,
\begin{align} \nonumber
\mathcal{M} &=
   \frac{g_{Wff}^2}
        {2M_W^2 ((p_1+p_2)^2 - M_W^2)} \\ \nonumber
    & \times
   (
      M_W^2 \lbrack \mathbf{2}\mathbf{3} \rbrack
            \langle \mathbf{14} \rangle
     +m_t m_b \langle \mathbf{2}\mathbf{4} \rangle 
            \lbrack \mathbf{13} \rbrack
     +m_t \lbrack \mathbf{2}|p_1|\mathbf{4} \rangle 
            \lbrack \mathbf{13} \rbrack \\
    & \qquad
     + \lbrack \mathbf{2}|p_1|\mathbf{4} \rangle
       \langle \mathbf{1}|p_2| \mathbf{3} \rbrack
     + m_b \langle \mathbf{2}\mathbf{4} \rangle
       \langle \mathbf{1}|p_2| \mathbf{3} \rbrack
   ).
\end{align}
The first two terms are done, but we need to simplify the other terms using the Schouten identities in App.~\ref{sec:generalized Schouten}. We begin with the fourth term,
\begin{eqnarray}
    \lbrack \mathbf{2}|p_1|\mathbf{4} \rangle
       \langle \mathbf{1}|p_2| \mathbf{3} \rbrack &=&
       \lbrack\mathbf{3}|p_1|\mathbf{4}\rangle\langle \mathbf{1}|p_2|\mathbf{2}\rbrack -
       \langle \mathbf{1}|p_2p_1|\mathbf{4} \rangle\lbrack\mathbf{32}\rbrack.
\end{eqnarray}
Since the bottom quark is massive, we use $p_2|\mathbf{2}\rbrack=-m_b|\mathbf{2}\rangle$.  We further use the on-shell condition for the internal line to switch the order of $p_2$ and $p_1$ as in $p_2 p_1 = (p_1 + p_2)^2 - m_t^2 - m_b^2 - p_1 p_2 = M_W^2 - m_t^2 - m_b^2 - p_1 p_2$ to obtain,
\begin{align}
    \lbrack \mathbf{2}|p_1|\mathbf{4} \rangle
       \langle \mathbf{1}|p_2| \mathbf{3} \rbrack =
        -(M_W^2-m_t^2-m_b^2)\langle \mathbf{14} \rangle\lbrack\mathbf{32}\rbrack
        -m_b\lbrack\mathbf{3}|p_1|\mathbf{4}\rangle\langle \mathbf{1}\mathbf{2}\rangle
       +m_t\lbrack \mathbf{1}|p_2|\mathbf{4} \rangle\lbrack\mathbf{32}\rbrack,
\end{align}
where we have also used $\langle\mathbf{1}|p_1 = m_t\lbrack\mathbf{1}|$.
Combining this with our full amplitude gives,
\begin{align}
\mathcal{M} = 
   g_{Wff}^2\Bigg[&
   \frac{
      (2M_W^2-m_t^2-m_b^2) \lbrack \mathbf{23} \rbrack
            \langle \mathbf{14} \rangle
     +m_t m_b \langle \mathbf{2}\mathbf{4} \rangle 
            \lbrack \mathbf{13} \rbrack
   }{2M_W^2 ((p_1+p_2)^2 - M_W^2)}\nonumber\\
   &+\frac{
     m_t (
     \lbrack \mathbf{2}|p_1|\mathbf{4} \rangle 
            \lbrack \mathbf{13} \rbrack
     + \lbrack \mathbf{1}|p_2|\mathbf{4} \rangle\lbrack\mathbf{32}\rbrack
       )
     + m_b( \langle \mathbf{2}\mathbf{4} \rangle
       \langle \mathbf{1}|p_2| \mathbf{3} \rbrack
        -\lbrack\mathbf{3}|p_1|\mathbf{4}\rangle\langle \mathbf{1}\mathbf{2}\rangle
        )
   }{2M_W^2 ((p_1+p_2)^2 - M_W^2)}
   \Bigg].
\end{align}

We now use conservation of momentum $p_1+p_2+p_3+p_4=0$ on the third, fourth and sixth terms to obtain,
\begin{align} \nonumber
    \lbrack \mathbf{2}|p_1|\mathbf{4} \rangle 
            \lbrack \mathbf{13} \rbrack
     + \lbrack \mathbf{1}|p_2|\mathbf{4} \rangle\lbrack\mathbf{32}\rbrack 
     & =
    -m_b\langle \mathbf{2}\mathbf{4} \rangle 
            \lbrack \mathbf{13} \rbrack
     - m_t\langle \mathbf{1}\mathbf{4} \rangle\lbrack\mathbf{32}\rbrack 
     +m_d\lbrack \mathbf{2}\mathbf{4} \rbrack
            \lbrack \mathbf{13} \rbrack \\ 
    & \qquad +m_d\lbrack \mathbf{1}\mathbf{4} \rbrack\lbrack\mathbf{32}\rbrack
     -\lbrack \mathbf{2}|p_3|\mathbf{4} \rangle 
            \lbrack \mathbf{13} \rbrack
     - \lbrack \mathbf{1}|p_3|\mathbf{4} \rangle\lbrack\mathbf{32}\rbrack, \\ \nonumber
      \langle \mathbf{2}\mathbf{4} \rangle
       \langle \mathbf{1}|p_2| \mathbf{3} \rbrack
        -\lbrack\mathbf{3}|p_1|\mathbf{4}\rangle\langle \mathbf{1}\mathbf{2}\rangle 
    & =
        m_u\langle\mathbf{3}\mathbf{4}\rangle\langle \mathbf{1}\mathbf{2}\rangle
        -m_d\lbrack\mathbf{3}\mathbf{4}\rbrack\langle \mathbf{1}\mathbf{2}\rangle \\ 
    & \qquad +\langle \mathbf{2}\mathbf{4} \rangle
       \langle \mathbf{1}|p_2| \mathbf{3} \rbrack
        +\lbrack\mathbf{3}|p_2|\mathbf{4}\rangle\langle \mathbf{1}\mathbf{2}\rangle,
\end{align}
Plugging these in gives us,
\begin{align} \nonumber
\mathcal{M} &= 
   \frac{ g_{Wff}^2 }{2M_W^2 ((p_1+p_2)^2 - M_W^2)}  \\ \nonumber
  & \qquad \times \Bigg[ (2M_W^2-m_b^2) \lbrack \mathbf{23} \rbrack
                             \langle \mathbf{14} \rangle
      + m_b m_u \langle\mathbf{1}\mathbf{2}\rangle \langle\mathbf{3}\mathbf{4}\rangle
      - m_b m_d \langle\mathbf{1}\mathbf{2}\rangle \lbrack\mathbf{3}\mathbf{4}\rbrack
   \nonumber \\
   & \qquad \qquad
    +m_tm_d(\lbrack \mathbf{2}\mathbf{4} \rbrack
            \lbrack \mathbf{13} \rbrack
     +\lbrack \mathbf{1}\mathbf{4} \rbrack\lbrack\mathbf{32}\rbrack)
    -m_t (
     \lbrack \mathbf{2}|p_3|\mathbf{4} \rangle 
            \lbrack \mathbf{13} \rbrack
     + \lbrack \mathbf{1}|p_3|\mathbf{4} \rangle\lbrack\mathbf{32}\rbrack
       )
   \nonumber\\
   & \qquad \qquad
     + m_b( \langle \mathbf{2}\mathbf{4} \rangle
       \langle \mathbf{1}|p_2| \mathbf{3} \rbrack
        +\lbrack\mathbf{3}|p_2|\mathbf{4}\rangle\langle \mathbf{1}\mathbf{2}\rangle
        )
   \Bigg].
\end{align}
We now apply the Schouten identity to the last two lines,
\begin{align}
    \lbrack \mathbf{2}\mathbf{4} \rbrack
            \lbrack \mathbf{13} \rbrack
     +\lbrack \mathbf{1}\mathbf{4} \rbrack\lbrack\mathbf{32}\rbrack &=
     \lbrack \mathbf{2}\mathbf{4} \rbrack
            \lbrack \mathbf{13} \rbrack
    +\lbrack \mathbf{2}\mathbf{4} \rbrack\lbrack\mathbf{31}\rbrack
    -\lbrack \mathbf{3}\mathbf{4} \rbrack\lbrack\mathbf{21}\rbrack =
    \lbrack\mathbf{12}\rbrack\lbrack \mathbf{3}\mathbf{4} \rbrack, \\
    \lbrack \mathbf{2}|p_3|\mathbf{4} \rangle 
            \lbrack \mathbf{13} \rbrack &=
            m_u\langle\mathbf{3}\mathbf{4}\rangle\lbrack\mathbf{12}\rbrack -
            \lbrack\mathbf{32}\rbrack\lbrack\mathbf{1}|p_3|\mathbf{4}\rangle, \\
    \langle \mathbf{2}\mathbf{4} \rangle
       \langle \mathbf{1}|p_2| \mathbf{3} \rbrack &=
    -\lbrack\mathbf{3}|p_2|\mathbf{4}\rangle\langle\mathbf{12}\rangle
    -m_b\lbrack\mathbf{3}\mathbf{2}\rbrack\langle\mathbf{14}\rangle.
\end{align}
Plugging these in gives, finally,
\begin{align} \nonumber
\mathcal{M} &= 
   \frac{g_{Wff}^2}
        {2M_W^2 ((p_1+p_2)^2 - M_W^2)} \\ \nonumber
   & \qquad \times \biggl[
      2M_W^2 \lbrack \mathbf{23} \rbrack
            \langle \mathbf{14} \rangle
            +m_b m_u \langle \mathbf{12}\rangle \langle\mathbf{34}\rangle
            -m_t m_u \lbrack \mathbf{12}\rbrack \langle\mathbf{34}\rangle \\
   & \qquad \qquad 
     - m_b m_d \langle \mathbf{12} \rangle\lbrack\mathbf{34}\rbrack
     + m_t m_d \lbrack \mathbf{12} \rbrack\lbrack\mathbf{34}\rbrack
   \biggr].
   \label{eq:M(t->bqq)}
\end{align}
As expected, we only get spinor brackets with no momenta sandwiched between and the coefficients are only functions of the masses.  No momenta are included in the numerator at all.  Furthermore, this result is very interesting.  As we have begun to see in the previous two cases, the last four terms that contain the external masses only contain brackets that connect lines on the same side of the propagator.  We get an $m_t$ if we have a $\lbrack\mathbf{12}\rbrack$ and an $m_b$ if we have $\langle\mathbf{12}\rangle$.  Similarly, we have an $m_u$ if we have $\langle\mathbf{34}\rangle$ and an $m_d$ if we have $\lbrack\mathbf{34}\rbrack$.  We, furthermore, find that the if spinor products that connect the same side of the propagator are both square or both angle, we get a plus sign, whereas, if one is a square product and the other is an angle product, we get a minus sign.  We do not know how general these rules are.  At the very least, they are specific to a spin-$1$ boson that only couples to left-chiral fermions.  Presumably, a $4$-point amplitude mediated by the Z~boson will have a structure like this with only its left-chiral coupling, a related structure with only its right-chiral coupling and another structure with mixed couplings.  We will not do this calculation here since, in addition to not being a decay diagram, it is signifcantly more involved because the amplitude also includes a photon exchange diagram and the x-factor which we leave for a future publication.  

Once more, we comment on the great simplicity of our results when compared with Feynman diagrams.  The Feynman diagram result is much like Eq.~(\ref{eq:M(m->evv) FD}) where the leptons are replaced with quarks.  The number of calculations has not significantly changed for Feynman diagrams between these cases.  On the other hand, the constructive amplitude has increased in complexity as we have removed massless particles.  This may seem like a problem until one realizes that even with this greater number of terms, it is still significantly simpler than the Feynman diagram calculation.  At this point, for the quark decay of the top quark, we have $5$ terms in our propagator numerator.  Each of these is a simple calculation as seen in App.~\ref{app:explicit spinor products}.  On the other hand, we will have a similar level of complexity just to create the $4$-component spinors $\psi$ for each of the fermions using Feynman diagrams.  We will then have to multiply them by all the $4 \times 4$ gamma matrices and then finally, by the $4 \times 4$ propagator numerator of the $W$ boson.  In fact, even if our coupling were not chiral and we had the maximal number of terms in our constructive amplitude, we would have $10$ terms as described in Eqs.~(\ref{eq:t->b,q,q:prods1}) through (\ref{eq:t->b,q,q:prods4}) and the discussion surrounding them.  Let us perform a simple naive estimate of the number of calculations required, and compare between the two techniques.  For the constructive method, each element of the $2\times2$ matrices corresponds with one spin combination and each of these elements requires only 1 multiplication.  We need to do this for $\lbrack\mathbf{23}\rbrack, \langle\mathbf{14}\rangle, \langle\mathbf{12}\rangle, \lbrack\mathbf{12}\rbrack, \langle\mathbf{34}\rangle$ and $
\lbrack\mathbf{34}\rbrack$ for a total of 6 multiplications to obtain these brackets.  For each term in Eq.~(\ref{eq:M(t->bqq)}), we need to multiply two of these together and then by two masses for a total of 3 more multiplications per term.  Since there are five terms, which must be added at the end, we have a total of $6+3\times5+4=21$ calculations per spin combination per phase-space point.  
On the other hand, for the Feynman diagram, we have 4 multiplications plus 3 additions for each of the 4 rows of $\gamma^\alpha P_L\psi_{u}$ for a total of 28 calculations.  We then multiply this by $\bar{\psi}_{d}$ giving 4 more multiplications and 3 more additions for a total of 35 calculations to construct $\bar{\psi}_{d}\gamma^\alpha P_L\psi_{u}$ and 35 more calculations for $\bar{\psi}_{t}\gamma^{\beta}P_L\psi_{b}$.  Additionally, we have 2 multiplications and 1 addition for each of the 16 elements of the propagator numerator giving 48 calculations to construct $-g_{\alpha\beta}+p_{W\alpha}p_{W\beta}/M_W^2$.  We still need to multiply these pieces together and this requires another 35 calculations giving a naive total of $35+35+48+35=153$ calculations.  It appears to us that there is nearly an order of magnitude fewer calculations required to compute the amplitude using constructive techniques relative to Feynman diagrams.  Of course, this many calculations is no problem for computers, but when we consider scaling this up to more complex scattering processes, we begin to realize the potential power of this constructive approach.

After multiplying by the complex conjugate and summing over spins as described in App.~\ref{sec:spinor conjugation}, we compare with Feynman diagrams as output by CalcHEP and find agreement.  The expression is even longer and less illuminating than for tau decay, therefore, we do not include it. However, we note that the constructive method gives $14$ terms whereas Feynman diagrams, as given by CalcHEP, produces 64 terms.  In every example, we find constructive techniques produce simpler formulas that are equivalent to their more complicated Feynman-diagram counterparts.

\section{\label{sec:4-body decays}4-Body Decay}
We now turn to $4$-body decays.  We, of course, cannot consider all possible 4-body decays in the SM in this paper since there are too many.  Furthermore, $5$-point amplitudes are significantly more complicated, both because of the extra propagator and vertex, but also because there are typically more than one diagram and some of those may involve the x factor (photons and gluons), which itself adds significant complexity.  We will turn to these important matters in a later paper.  Instead, we consider a single 4-body decay that is still illustrative of the massive constructive method but still as simple as possible.  Therefore, we consider the decay of a Higgs boson to 4 neutrinos which only has one diagram mediated by Z~bosons.  The external states are either spinless (the Higgs) or massless (the neutrinos), but the internal lines of the single diagram are still massive (Z~bosons) with complicated spin-spinor structure.  This will be sufficient to see how the higher-point amplitudes work in principle while having a final result which is extremely simple and demonstrates the potential power of this method.  To the best of our knowledge, this is the first $5$-point amplitude calculated using the massive constructive method initiated by \cite{Arkani-Hamed:2017jhn}.

\subsection{\label{sec:h->vvvv}$\mathbf{h\to \nu\bar{\nu}\nu\bar{\nu}}$}

We begin by guessing the structure of the final amplitude. Since the Higgs is spinless and the neutrino and anti-neutrino are $-1/2$- and $+1/2$-helicity, respectively, the final amplitude must be a product of the helicity-spinors $|2\rangle$, $|3\rbrack$, $|4\rangle$ and $|5\rbrack$.  Therefore, we expect the amplitude to be $\langle24\rangle\lbrack35\rbrack$ up to a prefactor.  The mass dimension of a $5$-point amplitude is $4-5=-1$, therefore, since each bracket product contributes a mass dimension of $+1$ and each propagator contributes a mass dimension of $-2$, we find that the mass dimension of the coefficient is $+1$.  The only diagram that produces this amplitude has two intermediate Z~bosons, therefore, we expect the amplitude to be of the form, 
\begin{equation}
    \mathcal{M} \sim \frac{g_{hZZ}g_{Z\nu\nu}^2 \mathbf{M} \langle24\rangle\lbrack35\rbrack}
    {((p_2+p_3)^2-M_Z^2)((p_4+p_5)^2-M_Z^2)},
    \label{eq:hvvvv guess}
\end{equation}
where we have factored any masses out of the coupling constants and $\mathbf{M}$ is either $M_h$ or $M_Z$, the only two masses available.  It is also possible $\mathbf{M}$ could be an appropriately powered ratio of the two masses.  However, since we know the coupling to the Z~boson contains the weak mass in it, we expect $\mathbf{M} = M_Z$.  Before we turn to the actual calculation, we comment on the remarkable simplicity of the final result.  We expect such simplicity in purely massless theories such as gluodynamics, but this amplitude is in the full SM containing two massive spin-$1$ particles in the diagram.  In particular, the Z~boson whose spin-indices must be symmetrized produces several intermediate terms during the calculation, as we shall see.  Nevertheless, the spin structure of the external states already gives us a clue of the great simplicity of the final result.  We now turn to doing the calculation and checking our prediction. In order to only consider one diagram, we will do the case with two distinct neutrino flavors.  The case with only one neutrino species is closely related, but has two related diagrams.

The vertex between the Higgs and Z~bosons is given by $-g_{hZZ}\lbrack\mathbf{P}_{23}\mathbf{P}_{45}\rbrack\langle\mathbf{P}_{23}\mathbf{P}_{45}\rangle/M_Z$.  The vertices between the Z~bosons and the neutrinos are given by $g_{Z\nu\nu}\lbrack3\mathbf{P}_{23}\rbrack\langle\mathbf{P}_{23}2\rangle/M_Z$ and $g_{Z\nu\nu}\lbrack5\mathbf{P}_{45}\rbrack\langle\mathbf{P}_{45}4\rangle/M_Z$.  The implicit indices on the $\mathbf{P}_{23}$ and $\mathbf{P}_{45}$ must be symmetrized and then connected across the propagators giving us,
\begin{align} \nonumber
    \mathcal{M} &=
    - \frac{ g_{hZZ}g_{Z\nu\nu}^2 }
    {4M_Z^3 ((p_2+p_3)^2-M_Z^2 ) ((p_4+p_5)^2-M_Z^2 )} \\ \nonumber
    &\times 
     (\lbrack \mathbf{P}_{23}^{I_1}\mathbf{P}_{45}^{J_1} \rbrack
      \langle \mathbf{P}_{23}^{I_2}\mathbf{P}_{45}^{J_2} \rangle ) \\ \nonumber
    &\times
     (\lbrack 3\mathbf{P}_{23I_1} \rbrack
      \langle \mathbf{P}_{23I_2}2 \rangle
     +\lbrack 3\mathbf{P}_{23I_2} \rbrack
      \langle \mathbf{P}_{23I_1}2 \rangle ) \\
    &\times
     (\lbrack 5\mathbf{P}_{45J_1} \rbrack
      \langle \mathbf{P}_{45J_2}4 \rangle
     +\lbrack 5\mathbf{P}_{45J_2} \rbrack
      \langle\mathbf{P}_{45J_1}4 \rangle ).
\end{align}
Expanding and using the rules for contracted indices given in Eqs.~(\ref{eq:app:p=|>[|}) through (\ref{eq:App:|][|=-m}), we obtain,
\begin{equation}
    \mathcal{M} \propto
    \frac{M_Z\langle42\rangle\lbrack35\rbrack}{4} 
-\frac{\lbrack3| p_2 | 4\rangle\lbrack5| 
p_3 | 2\rangle}{4M_Z} -\frac{\lbrack3| p_5 
| 4\rangle\lbrack5| p_4 | 2\rangle}{4M_Z} 
+\frac{\langle4| p_5 p_3 | 2\rangle\lbrack3
       | p_2 p_4 | 5\rbrack}{4M_Z^3}, 
\end{equation}
where we have also used the masslessness of the neutrinos and we have left out the coupling constants and propagator denominators in order to fit the expression on one line.  We next use Schouten identities to reduce these products further.  We have already seen examples similar to the middle two terms.  Here, we show how we reduce the last term.  We use the rules of App.~\ref{sec:generalized Schouten} to make the replacement,
\begin{equation}
    \langle4| p_5 p_3 | 2\rangle\lbrack3| p_2 p_4 | 5\rbrack =
    -\lbrack5| p_4| 2\rangle\lbrack3| p_2 p_3 p_5|4\rangle + \lbrack5| p_4 p_3 p_5|4\rangle\lbrack3| p_2 | 2\rangle.
\end{equation}
We did not include this particular Schouten identity as one of our explicit examples, but it still follows Eq.~(\ref{eq:general Schouten 2}) and the mnemonic given in that subsection makes it easy to apply. The second term on the right is zero since the neutrino is massless.  The order of the momenta in the first term can be rearranged using $p_2 p_3 = 2 p_2 \cdot p_3 - p_3 p_2$ from Eq.~(\ref{eq:p1p2=p2p1-2p1.p2}) giving us, 
\begin{equation}
    \langle4| p_5 p_3 | 2\rangle\lbrack3| p_2 p_4 | 5\rbrack =
    -2p_2\cdot p_3\lbrack5| p_4| 2\rangle\lbrack3| p_5|4\rangle .
\end{equation}
We then apply the Schouten identity again to obtain,
\begin{equation}
    \langle4| p_5 p_3 | 2\rangle\lbrack3| p_2 p_4 | 5\rbrack =
    2p_2\cdot p_3
    \langle42\rangle\lbrack3| p_5p_4|5\rbrack,
\end{equation}
where we have implicitly used the masslessness of the neutrinos this time.  Rearranging the momenta again using $p_5 p_4 = 2 p_4 \cdot p_5 - p_4 p_5$, we finally obtain,
\begin{equation}
    \langle4| p_5 p_3 | 2\rangle\lbrack3| p_2 p_4 | 5\rbrack =
    4p_2\cdot p_3p_4\cdot p_5
    \langle42\rangle\lbrack35\rbrack.
\end{equation}
At this point, we use the on-shell condition setting $2 p_2 \cdot p_3 = (p_2+p_3)^2 = M_Z^2$ and, similarly, $2 p_4 \cdot p_5 = M_Z^2$.  We pause to note that we would have missed this important application of the on-shell condition if we left the momenta sandwiched between the spinors and we would have got the wrong result.  Therefore, we see the importance of reducing the spinor products.  It is more than being merely helpful, convenient or even simply useful for a creating nicer end results. It appears to be required to obtain the correct result, at least in this case and others we have studied.  After the amplitude is completed and actual momenta are entered for the Higgs and neutrinos, they will \textit{not} satisfy this on-shell condition.  These products will not equal the mass of the Z~boson.  Therefore, we cannot overstate the importance of this simplification taking place.  After applying the on-shell condition, we obtain,
\begin{equation}
    \langle4| p_5 p_3 | 2\rangle\lbrack3| p_2 p_4 | 5\rbrack =
    M_Z^4
    \langle42\rangle\lbrack35\rbrack.
\end{equation}
Plugging this in, as well as using the Schouten identities and on-shell condition on the middle two terms, gives us finally,
\begin{equation}
\mathcal{M}_{\nu_e\bar{\nu}_e\nu_\mu\bar{\nu}_\mu} =- 
   \frac{g_{hZZ}g_{Z\nu\nu}^2 M_Z \langle 24 \rangle \lbrack 35 \rbrack}
        {((p_2+p_3)^2-M_Z^2)((p_4+p_5)^2-M_Z^2)} ,
\label{eq:M_h->vvvv}
\end{equation}
where we have used the antisymmetry of the spinor products to put them in a standard order.  Amazingly, we guessed the correct answer at the beginning in Eq.~(\ref{eq:hvvvv guess}), even up to factors of $2$.  

We pause, once again, to compare this result with the same calculation done using Feynman diagrams.  There are two Z-boson propagators connecting two neutrino lines giving us something like,
\begin{align} \nonumber
   &\bar{\psi}_{\nu_1}\gamma^\alpha\left(g_LP_L+g_RP_R\right)\psi_{\nu_2}
    \left(-g_{\alpha\beta}+\frac{p_{Z_1\alpha}p_{Z_1\beta}}{M_Z^2}\right) \\ 
   &\qquad \times \left(-g^{\beta\epsilon}+\frac{p_{Z_2}^{\beta}p_{Z_2}^{\epsilon}}{M_Z^2}\right)
    \bar{\psi}_{\nu_3}\gamma_{\epsilon}\left(g_LP_L+g_RP_R\right)\psi_{\nu_4}.
\end{align}
This Feynman diagram is even worse than in Eq.~(\ref{eq:M(m->evv) FD}), requiring 48 more calculations to construct the other propagator numerator and 28 more calculations for the final matrix multiplication giving a naive total of $35+35+48+48+(35+28)=229$.  On the other hand, the constructive amplitude given in Eq.~(\ref{eq:M_h->vvvv}) has only $M_Z \langle 24 \rangle \lbrack 35 \rbrack$ to calculate, which requires approximately 3 products all together .
Therefore, we expect a naive improvement in efficiency of a factor of 76, nearly two orders of magnitude, in this 5-point-amplitude case if constructive amplitudes are used rather than Feynman diagrams.

We square this amplitude following the rules outlined in previous sections to obtain,
\begin{equation}
|\mathcal{M}|^2_{\nu_e\bar{\nu}_e\nu_\mu\bar{\nu}_\mu} =  
 \frac{g_{hZZ}^2 g_{Z\nu\nu}^4 4 M_Z^2 \, p_2 \cdot p_4 \, p_3\cdot p_5}
      {((p_2+p_3)^2-M_Z^2)^2((p_4+p_5)^2-M_Z^2)^2} .
\label{eq:M^2_vevevmvm}
\end{equation}
We find agreement with Feynman diagrams, as produced by CalcHEP.  To demonstrate the simplification in this case, we show the result of Feynman diagrams as given by CalcHEP,
\begin{equation}
  | \mathcal{M} |^2_{\nu_e\bar{\nu}_e\nu_\mu\bar{\nu}_\mu} =
    \frac{e^6 M_W^2 
          \left(M_h^2 \, p_2 \cdot p_4 + 2(p_2 \cdot p_4)^2 
         -2 p_1 \cdot p_2 \, p_2 \cdot p_4
         -2 p_1 \cdot p_4 \, p_2 \cdot p_4 \right)}
         {2 c_W^8 s_W^6 ( (p_2+p_3)^2-M_Z^2 )^2 ( (p_4+p_5 )^2-M_Z^2 )^2}.
   \label{eq:M^2_vevevmvm Feynman}
\end{equation}
Removing the $p_1$ by use of momentum conservation $p_1 = p_2 + p_3 + p_4 + p_5$ is a standard technique one might try, but it unfortunately results in a more complicated expression with six terms rather than four.  On the other hand, who would think, a priori, that additionally replacing the Higgs mass with momenta as in $M_h^2 = p_1^2 = (p_2 + p_3 + p_4 + p_5)^2$ would simplifiy this expression.  But, it is exactly the combination of these two identities that leads to Eq.~(\ref{eq:M^2_vevevmvm}).  However, with the constructive technique, we did not have to guess at these simplifications or try many different identities looking for the simplest result, our final result followed directly from the formalism. Although the level of improvement of the squared diagram is not as great in this case (the Feynman-diagram result simplified greatly due to the masslessness of the neutrinos), it is still substantially simpler.  Based on this and the previous $3$-body decay examples, we expect that the constructive method will reduce the complexity of the final expressions in most, if not all, cases in the future.

Although we were successful in guessing the structure of this final $5$-body result based purely on the helicities of the neutrinos, as we increase the number of external states, it will become more complicated.  As a simple yet important example of this, we note that if we considered Higgs decay to six neutrinos, we would have three $-1/2$-helicity neutrinos and three $+1/2$-helicity anti-neutrinos.  This would lead to the realization that the final amplitude must be composed of the following helicity-spinors: $|2\rangle$, $|3\rbrack$, $|4\rangle$, $|5\rbrack$, $|6\rangle$ and $|7\rbrack$.  This is correct.  However, when we consider what Lorentz invariant products we can form from these, we realize that there is a mismatch in the number of helicity-spinors of each type.  At first, we might erroneously conclude that this amplitude is zero since we cannot form a Lorentz invariant product from only these helicity-spinors.  However, there are Feynman diagrams for this process.  In fact, if we consider all three generations of fermions in the decay ($h \to \nu_e \bar{\nu}_e \nu_\mu \bar{\nu}_\mu \nu_\tau \bar{\nu}_\tau$), we see that there are twelve diagrams, all of which are some form of $h \to ZZ$ with each Z~boson producing a pair of neutrinos and finishing with one of those neutrinos emitting a Z~boson which produces the last pair of neutrinos.  So, it can be produced by a diagram of the form $h \to ZZ \to  \nu_e \bar{\nu}_e \nu_\mu \bar{\nu}_\mu Z \to \nu_e \bar{\nu}_e \nu_\mu \bar{\nu}_\mu \nu_\tau \bar{\nu}_\tau$ and related diagrams.  Therefore, if we know this amplitude exists but we cannot produce it with products of spinors alone, how can it be achieved in this constructive theory?  The answer is that it must have a momentum sandwiched between two of the spinors.  For example, we can form a Lorentz invariant product as $\langle23\rangle\lbrack45\rbrack\langle6|p|7\rbrack$, where $p$ is neither $p_6$ nor $p_7$, but must be one of the other momenta.  This is just one example; there are others.  This is very interesting since all the amplitudes shown so far have been reducible to a form that does not contain any momenta left in the numerator.  We have even gone so far as to claim that this is very important to producing a correct final result.  We have suggested that this is because the internal lines must be taken on-shell during intermediate steps and that this requires some of the external momenta to be taken complex.  Therefore, they must be removed before the end, where they are taken real again.  In deed, we have only found agreement with Feynman diagrams thus far, when we have removed the momenta completely from the numerator.  It appears to us that a more careful treatment of the complexification of the momenta is required to understand Higgs decay to six neutrinos, and likely many other processes as well.  We do not yet have a satisfactory solution for this problem but hope to provide one in future publications.

\section{\label{sec:conclusions}Summary and Conclusions}
In this paper, we have calculated many $2$- and $3$-body decays without the emission of a gluon or photon and one $4$-body decay of the SM using the massive constructive techniques described in \cite{Arkani-Hamed:2017jhn} and the $3$-point vertices of \cite{Christensen:2018zcq}.  We have squared these amplitudes and compared with the expressions coming from Feynman diagrams and found complete agreement.  As we have done this, we have developed many further techniques for massive constructive amplitudes.

In Sec.~\ref{sec:general techniques}, we developed the techniques required to square the amplitude and reduce the amplitude to a minimal form.  This reduction is necessary, as we discuss in that section, because the constructive amplitude method is a purely on-shell formalism for constructing an amplitude.  Since real physical momenta do not allow the internal particles to go on-shell (in most scenarios), the momenta are usually taken complex and the helicity-spinors (in the massless case) are adjusted accordingly.  However, it is not yet known how to do this for the massive spin-spinors.  Therefore, we note that until the complexification process is understood in detail, it is imperative that all amplitudes get reduced to a form that does not contain any momenta or internal spinors in the amplitude numerators.  In order to accomplish this reduction, we use standard momentum conservation, the on-shell conditions for both the spinors [see Eq.~(\ref{eq:pi|i>=-mi|i]})] and squared momenta ($p_i^2 = m_i^2$) and the anticommutation properties for the momenta [see Eq.~(\ref{eq:p1p2=p2p1-2p1.p2})].  We also must generalize the Schouten identities to forms with any number of momenta sandwiched between the spinors.  We find the general form in Eqs.~(\ref{eq:general Schouten 1}) and (\ref{eq:general Schouten 2}) and create a mnemonic to remember it.  All the Schouten identities used here satisfy ``four two three one minus four one three two with an extra minus sign for every momentum that is reversed'' with the numbers referring to the position in the product.  We further give several useful examples of the generalized Schouten identity that are used throughout this paper.  If the original product of spinor-chains contains one momentum, we give the result in Eqs.~(\ref{eq:Schouten:[ij]<k|p|l]}) through (\ref{eq:Schouten:<ij><k|p|l]}).  If there are two momenta both in one of the spinor-chains, we give the rules in Eqs.~(\ref{eq:Schouten:<i|pp|j><kl> 1}) through  (\ref{eq:Schouten:<i|pp|j>[kl]}).  If the two momenta are split between the two spinor-chains, we give the identity in Eqs.~(\ref{eq:Schouten:<i|p|j]<k|p|l] 1}) through (\ref{eq:Schouten:<i|p|j]<k|p|l] 2}). 

We also develop rules for squaring the amplitude in App.~\ref{sec:spinor conjugation}.  We begin by working out the explicit spinors with their spinor-indices both up and down in Eqs.~(\ref{eq:Spinor Def 1}) through (\ref{eq:Spinor Def 4}) and (\ref{eq:Spinor Def 5}) and comparing them.  By doing this, we learn that, although they are all related by complex conjugation as expected, half of them are related by complex conjugation with a relative minus sign.  We give a summary of the conjugation rules for spinors in Eqs.~(\ref{eq:app:|>^I*}) and (\ref{eq:app:|>_I*}).  In order to conjugate any amplitude, we also describe the conjugation of complete spinor-chains with momenta in Eqs.~(\ref{eq:app:(<>)*}) through (\ref{eq:app:([])*}) and (\ref{eq:<i|pppp|j>*}) and (\ref{eq:[i|pppp|j]*}).  

After conjugating and multiplying the amplitude by its conjugate, we need to simplify the result to a form with only momenta and momentum dot products ($p_i \cdot p_j$).  In order to do this, we need to know the outer products of spinors.  If the spinors are mixed, one angle spinor and one square spinor (for example, $|\mathbf{i}^{\mathrm{I}}\rangle \lbrack\mathbf{i}_{\mathrm{I}}| = p_i$), a momentum results as this is one route to defining the spinors as described in \cite{Arkani-Hamed:2017jhn}.  We review this definition in Eq.~(\ref{eq:app:p=|>[|}) and note the sign change if the spin-indices are raised and lowered in Eq.~(\ref{eq:app:p=-|>[|}).  We also work out the outer products if both spinors are the same type, both angle or both square.  We find that we obtain a mass times a Kronecker delta function as we show in Eqs.~(\ref{eq:App:|][|=m}) and (\ref{eq:App:|][|=-m}).  We also show similar inner products that give the mass in Eqs.~(\ref{eq:<ij>=mdelta}) and (\ref{eq:<ij>=-mdelta}).  In order to apply these identities, we sometimes need to reverse spinor-chains.  The simplest case, with no momenta in the middle, is already known.  However, we work the reversal out for any number of momenta in between the spinors in Eq.~(\ref{eq:<i|pppp|j> reversed}) and give several useful examples in Eqs.~(\ref{eq:<ij> reversed}) through (\ref{eq:<i|ppp|j] reversed}).  Finally, once the amplitude has been multiplied by its conjugate, the spinor-chains reversed and the identities applied, we describe how we obtain traces over the two-by-two momenta.  We describe a recursive method (similar to the method in Ref.~\cite{Franken:2019wqr}) for calculating any trace over momenta.

With the identities required worked out, we turn our attention first to squaring the amplitudes for $2$-body decays in Sec.~\ref{sec:2-body decays}.  The amplitude is already known as it is just the relevant $3$-point constructive vertex of the SM given in \cite{Christensen:2018zcq}.  In this section, we first work out explicit expressions for the amplitude for each spin of the external particles.  For example, for the Z~boson, we work out the decay amplitude when it is spin $+1, 0$ and $-1$ and similarly for its daughter particles.  We do this for each decay.  We then explicitly square the amplitude for each spin combination and add them all together to get the final squared amplitude (appropriate when we do not measure the spins of either the mother or the daughter particles).  After we complete this, we square the amplitude implicitly.  We do this by multiplying the amplitude by its complex conjugate, reversing spinor-chains where necessary, applying the outer product identities and computing the resulting traces.  This second method is much like is traditionally done with Feynman diagrams.  Once the squared amplitude is calculated this second way, we compare with the first method with explicit expressions for each spin combination and find agreement.  We also compare with the result coming from Feynman diagrams, as given by a standard analytic squared-amplitude calculator, CalcHEP\cite{Belyaev:2012qa}, and find agreement for every case described in this paper.  In the rest of this section, we do each of these steps in great detail for all $2$-body decays of the SM.  We do it for $Z\to \nu\bar{\nu}$ in Subsec.~\ref{sec:Z->vv} where we show its general amplitude in Eq.~(\ref{eq:M(Z->vv)}).  The explicit amplitude for each spin of the Z~boson is given in Eq.~(\ref{eq:M(Z->vv) explicit}) and its resulting square in Eq.~(\ref{eq:M^2(Z->vv) explicit}).  In this subsection, we also note that the symmetrization of the spin-$0$ Z~boson must be done with a factor of $1/\sqrt{2}$ rather than $1/2$.  This is required to achieve the correct normalization for the spin-$0$ state and to achieve the correct final squared amplitude result.  On the other hand, when we square the amplitude implicitly in the same subsection, we note that we must use a factor of $1/2$ for all the spins of the Z~boson, which we show in Eq.~(\ref{eq:M(Z->vv) 1/2}).  We then complete the implicit squaring with this factor of $1/2$ and show the final squared amplitude in Eq.~(\ref{eq:M^2(Z->vv)}).  We find agreement between these two methods and with Feynman diagrams.  We continue to use a factor of $1/\sqrt{2}$ for the other processes whenever we calculate the amplitude for an explicit spin-$0$ amplitude and a $1/2$ whenever the spin is left unspecified and always achieve agreement between these two methods.  In Subsec.~\ref{sec:Z->ff}, we calculate the decay of the Z~boson to two massive fermions.  We show the amplitude in Eq.~(\ref{eq:M(Z->ff)}) and its explicit form for each spin combination in Eqs.~(\ref{eq:M(Z->ff) explicit 1}) and (\ref{eq:M(Z->ff) explicit 2}).  We square each of these results and add them all together to achieve the total squared amplitude in Eq.~(\ref{eq:M^2_Zff}).  After this, we calculate the squared amplitude implicitly and obtain the same squared amplitude in Eq.~(\ref{eq:M^2(Z->ff) implicit}).  In Subsec.~\ref{sec:W->lv}, we find the decay of the W~boson to leptons, giving its amplitude in Eq.~(\ref{eq:M(W->lv)}), its explicit-spin expressions in Eq.~(\ref{eq:M(W->lv) explicit}) and its resulting square in the two ways, respectively, in Eqs.~(\ref{eq:M^2(W->lv) explicit}) and (\ref{eq:M^2(W->lv)}).  Similarly for the quark channel of W-boson decay, we give its amplitude in Eq.~(\ref{eq:M(W->qq)}), its explicit form in Eqs.~(\ref{eq:M(W->qq) explicit 1}) and (\ref{eq:M(W->qq) explicit 2}) and its square in Eqs.~(\ref{eq:M^2(W->qq)}) and (\ref{eq:M^2(W->qq) implicit}).  We end this section with the decay of the Higgs to two massive fermions in Subsec.~\ref{sec:h->ff}, where we give its amplitude in Eq.~(\ref{eq:M(h->ff)}), its explicit-spin form in Eq.~(\ref{eq:M(h->ff) explicit}) and its square in Eqs.~(\ref{eq:M^2(h->ff)}) and (\ref{eq:M^2(h->ff) implicit}).

In Sec.~\ref{sec:3-body decays}, we turn our attention to a constructive calculation of the $3$-body decays of the SM.  In order to keep this section managable, we only calculated the 3-body decays that do not involve the radiation of a photon or gluon from a $2$-body decay. Since this involves four external particles, this section involves the construction of $4$-point amplitudes by gluing two $3$-point vertices together with a propagator.  As already mentioned, this involves the removal of all momenta and internal spinors from the numerator of the amplitude.  We thus use a variety of identities including momentum conservation, the on-shell conditions and the generalized Schouten identities to accomplish this.  Since the squaring of the amplitude has already been shown in great detail in the previous section and it is not significantly different in this section, we do not show the details of the squaring in this section.  We do, however, describe the agreement with Feynman diagrams and show the squared amplitude where it is convenient.  In Subsec.~\ref{sec:m->vev}, we begin with a calculation of muon decay to leptons.  Before doing the detailed calculation, we use the structure of the external states to determine the possible structure of the final amplitude.  Since there are two massless neutrinos in the final state, we are able to limit the possible forms of the amplitude to only two.  Together, with the propagator denominator, we give this expected form in Eq.~(\ref{eq:M(m->vlv) expected}).  We also note that the dimension of the amplitude requires the coefficients of the two terms to be dimensionless ratios of the masses in the problem.  After discussing this, we begin the detailed calculation by stating the two vertices in Eq.~(\ref{eq:vertices m->vlv}) and multiplying them together and including the propagator denominator in Eq.~(\ref{eq:M(m->vlv) unsimplified}).  We then apply a long sequence of identities that slowly remove all traces of the momenta and internal spinors until we achieve the final form of the amplitude in Eq.~(\ref{eq:M(m->vlv)}).  As expected, the amplitude has exactly the form we predicted in Eq.~(\ref{eq:M(m->vlv) expected}) and the coefficients are dimensionless ratios of the masses.  We also comment on the great simplicity of our final result compared to Feynman diagrams and note that, naively, there are something like $3$ orders of magnitude fewer calculations required to compute the amplitude coming from the constructive method compared with the Feynman diagram approach.  This fact is potentially very promising for matrix-element generators.  If they were to implement this method into their code, the integration of phase space could likely be made orders of magnitude more efficient.  We follow this with the square of the amplitude in Eq.~(\ref{eq:M^2(m->evv)}), which agrees with Feynman diagrams, shown in Eq.~(\ref{eq:M^2(m->evv) Feynman}).  We again note that the result coming from constructive techniques is much simpler, containing only 3 terms in comparison with the 15 terms naturally coming from Feynman diagrams.  Again, the phase-space integration step could be made more efficient in principle.

In Subsec.~\ref{sec:m->vqq}, we turn our attention to the decay of a tau to quarks.  We again begin by determining the allowed structures for the amplitude given in Eq.~(\ref{eq:mu->nu,q,q:prods2}).  Unfortunately, there are many more.  We find that, due to the Schouten identity, there are at most five.  The symmetries of the external states is much less helpful this time.  The reason is that there is only one massless external particle this time, rather than the two with the leptonic decay of the previous subsection. Nevertheless, we can still calculate the amplitude constructively.  We give the vertices in Eq.~(\ref{eq:vertices m->vqq}) and we multiply them, including the propagator denominator, in Eq.~(\ref{eq:M(m->vqq) unsimplified}).  We then go through a similar, but unique, set of identities to finally remove the momenta and internal spinors from the numerator, achieving the reduced amplitude in Eq.~(\ref{eq:M(m->vqq)}).  It only has three terms, with the coefficients being dimensionless ratios of the masses.  Once again, the result is far simpler than that coming from Feynman diagrams and could result in orders of magnitude greater efficiency at the phase-space integration step if implemented into matrix-element generators.  We give its square in Eq.~(\ref{eq:M^2(m->vud)}) and find agreement with Feynman diagrams.  As in the previous subsection, we note that the result coming from constructive techniques is much simpler, this time containing only 6 terms in comparison with the 19 terms naturally coming from Feynman diagrams.  The last $3$-body decay in the SM is top-quark decay to quarks and we calculate it in Subsec.~\ref{sec:t->bqq}.  Since all the external particles are massive, the possible structures allowed for the amplitude is quite large. There are ten possible structures, so this does not help us much.  Following a similar, but unique, set of steps, we multiply the vertices, divide by the propagator denominator, and apply a series of identities to remove the momenta and internal spinors.  We finally achieve, in Eq.~(\ref{eq:M(t->bqq)}), the reduced amplitude for this process.  It has five unique terms with each coefficient a dimensionless ratio of masses.  We comment on the structure of this amplitude, noting how the coefficients are related to the structures they connect to.  We also note how setting some of the masses to zero relates this amplitude to the previous $3$-body decay amplitudes.  Finally, we note that the reason the amplitude was reducible to only five unique terms rather than the ten allowed was due to the left-chiral structure of the vertex. If both chiralities were allowed, it would likely contain all ten terms.  Nevertheless, even if it contained all ten terms, the calculation of each phase-space point would still likely be orders of magnitude more efficient than the same calculation using Feynman diagrams.  Although we do not include the full squared amplitude, we note that we obtain $14$ terms using the constructive technique and that we find it to be completely equivalent to the $64$ terms coming from Feynman diagrams given to us by CalcHEP.

In our final section (Sec.~\ref{sec:4-body decays}), we turn ourselves to calculating a $4$-body decay.  As far as we know, this is the first time a $5$-point amplitude has been calculated using the full spin-spinor structure of a massive constructive theory.  We only do the simplest possible case of this, which is Higgs decay to $4$ neutrinos.  Although the Higgs is spinless and the neutrinos are massless, this amplitude involves the spin-spinors of the two massive Z~bosons which mediate this decay.  Each internal line involves two spin-spinors on each side which are symmetrized in their spin-indices.  This results in many terms, which we will then simplify following the same rules used in previous sections.  Before we do this, we use the symmetry propeties of the external states to determine the most general structure allowed for the final amplitude.  Because all the final states are massless, in this case, there is only one possibility.  Together with the propagator denominators, we find that the amplitude is expected to be of the form given in Eq.~(\ref{eq:hvvvv guess}).  Amazingly, by considering the symmetry properties, we are able to completely determine the amplitude up to a single mass without doing any calculation at all.  The single mass must be either $M_Z$, $M_h$ or a dimension $1$ ratio of the two.  With this final result in mind, we begin the detailed calculation.  We multiply the four vertices together and begin applying the identities.  We skip a detailed explanation of steps that are similar to those taken in Sec.~\ref{sec:3-body decays} and focus on applications of the identities that are new for this process.  Finally, we achieve the reduced amplitude in Eq.~(\ref{eq:M_h->vvvv}) and see that it has exactly the expected form given in Eq.~(\ref{eq:hvvvv guess}) and the mass is that of the Z~boson.  We note how spectacularly simple this final result is for a theory with a massive intermediate boson.  In fact, if we compare to Feynman diagrams, which has grown in computational complexity by a factor of $4$ (due to summing over one new Lorentz index) and is naively in the hundreds of thousands, the constructive amplitude result still has on the order of ten calculations.  The reduction in the phase-space integration time cannot be overstated for this example. Additionally, we have compared the square of this constructive amplitude, given in Eq.~(\ref{eq:M^2_vevevmvm}), which only has one term, with the result of Feynman diagrams, given in Eq.~(\ref{eq:M^2_vevevmvm Feynman}), which has four terms.  The series of steps to simplify the Feynman result to the form given by the constructive method is not trivial.  It requires two steps.  The first is to replace the Higgs mass by momenta, $M_h^2=(p_2+p_3+p_4+p_5)^2$, and the second is to use momentum conservation $p_1=p_2+p_3+p_4+p_5$.  It is surprising that replacing the Higgs' mass with momenta would help simplify the expression while the second is a bit more natural.  Nevertheless, the combination of these two steps reduces it to the form given by the constructive theory.    Before ending this section, we comment on Higgs decay to $6$ neutrinos.  Although we do not attempt to calculate it here, we consider its symmetry structure due to the massless final neutrinos.  We note that if we do not allow momenta in the numerator, it is impossible to write a Lorentz invariant structure.  On the other hand, we note that there are Feynman diagrams that give this decay.  With this observation, we note that the final constructive form must have a momentum in the numerator.  This shows that reducing the amplitude to a form without momenta in the numerator, although extremely powerful, must not always be possible.  There must be amplitudes for which this is impossible.  We were, in fact, very fortunate that every amplitude calculated in this paper were reducible to this form.  This underlines the fact that to proceed systematically in the future, we will have to consider more general forms.

%
%
%
In multiple places throughout this paper, we have emphasized the importance of reducing the amplitude to a form that does not have any momenta or internal-line spinors in the numerator.  We have noted that, in every case that this has been achieved, we have found agreement with Feynman diagrams.  On the other hand, we have not found agreement yet for any amplitude for which we were unable to reduce the amplitude to such a simplified form.  We have further pointed out that there exist some amplitudes (Higgs decay to six neutrinos for instance) where it is impossible to reduce the amplitude to this form.  This leads us to the conclusion that a crucial next step, if this massive constructive theory is to succeed on a larger scale, is to develop an understanding of the complexification of the momenta and their resulting associated massive spin-spinors.  It is natural to suspect that the spin-spinors are treated in an analogous way to the helicity spinors, most likely a generalization of the formalism used in the massless theory.  To very briefly review\cite{Elvang:2015rqa}, for each massless amplitude, two adjacent particles can be chosen, call them particles $i$ and $j$ and their helicity spinors can be shifted as $|\hat{i}\rbrack = |i\rbrack + z |j\rbrack$ and $|\hat{j}\rangle = |j\rangle - z |i\rangle$ (for an $\lbrack i,j\rangle$ shift).  This shift automatically satisfies momentum conservation and keeps the external particles on-shell.  Moreover, this shift can also put the internal line on-shell if we take the pole value of the complex number $z$ to be $z_p = -\lbrack ik\rbrack/\lbrack jk\rbrack = \langle jl\rangle/\langle il\rangle$ where the momentum in the internal line is $(p_i+p_k) = -(p_j+p_l)$.  Using these rules, any massless tree-level amplitude can be calculated using the BCFW rules.   Based on this, one might assume that the same structure with spin-spinors would work, for example  $|\hat{i}^{\mathrm{I}}\rbrack = |i^{\mathrm{I}}\rbrack + z |j^{\mathrm{I}}\rbrack$ and $|\hat{j}^{\mathrm{J}}\rangle = |j^{\mathrm{J}}\rangle - z |i^{\mathrm{J}}\rangle$.  Although this does satsify momentum conservation, it does not keep the external particles on-shell.  Furthermore, it does not work when one particle is massless and the other is massive, since it would mix spin-spinors and helicity-spinors.  Nevertheless, we have found a way of generalizing this shift that satisfies momentum conservation as well as the on-shell condition for both the external particles and the internal line with the property that it reduces to the massless shift in the massless limit.  We have not yet resolved all the challenges with using it but we are encouraged that a formalism exists and plan to publish this work in the future.

Although we have worked out the $3$-point vertices of the SM \cite{Christensen:2018zcq}, the complete set of $4$-point vertices have not yet been determined.  Purely massless theories such as QCD do not require $4$-point vertices at all, but this is not true for massive theories.  Moreover, the $3$-point vertices given in \cite{Christensen:2018zcq} contain a few ambiguous couplings that still need to be removed.  We could accomplish both these tasks by comparing with Feynman diagrams, as we did in this paper.  However, we seek a method independent of fields and Feynman diagrams to accomplish these goals.  We propose that both can be done by demanding perturbative unitarity of all $2 \to 2$ scattering processes. It was also pointed out in Ref.~\cite{Durieux:2019eor} that these $3$-point vertices appear at the non-renormalizable level and can be identified via perturbative unitarity.  In a future work, we intend to analyze the high-energy growth of all spin channels of all $2 \to 2$ amplitudes of the SM including contributions from potential $4$-point vertices.  

Once the massive contructive rules are established for higher-point amplitudes, a greater number of amplitudes using this formalism needs to be worked out to determine whether the extreme simplifications found in constructive massless theories is also present to some degree in massive constructive theories.  We have shown several four-point amplitude examples and one five-point amplitude example where the massive constructive result is significantly simpler than the Feynman-diagram result. It would also be advantageous to determine if these novel, simplified expressions offer better numerical stability in phase-space calculations. However, we do not yet know how general this is and hope to investigate it further in the future. 

Finally, returning to the original point of the introduction, although the constructive method was established by developing far simpler formulas for the amplitudes, to a certain extent, the elegance of the complete theory has been lost.  Although the entire SM can be written in a single line with all the symmetries manifestly satisfied and the coupling constants related by those symmetries, the constructive SM, on the other hand, is a table of vertices.  The constructive vertices are simpler than the Feynman vertices.  Just consider the vertices of the spin-$1$ bosons or the gravitons to be convinced.  Nevertheless, the coupling constants of each vertex, although related by symmetries, can only be pinned down by calculating a variety of amplitudes.  We concede that this is currently a deficiency of massive constructive theories.  We do not know how to fix this at present but hope that a unifying principle for constructive vertices will be found in the future.

\appendix

\section{\label{sec:spinor conjugation}Spinor-chain Conjugation, Spinor Contractions and Trace Formulas}
The spin-spinors were originally defined in \cite{Arkani-Hamed:2017jhn} to be,
\begin{equation}
    |\mathbf{i}\rangle_\alpha^{\mathrm{I}} =
    \left(\begin{array}{cc}
    \sqrt{E+p}\ c & -\sqrt{E-p}\ s^* \\
    \sqrt{E+p}\ s & \sqrt{E-p}\ c
    \end{array}\right)
    \quad\mbox{and}\quad
    \lbrack\mathbf{i}|_{\dot{\alpha}\mathrm{I}} =
    \left(\begin{array}{cc}
    \sqrt{E+p}\ c & -\sqrt{E-p}\ s \\
    \sqrt{E+p}\ s^* & \sqrt{E-p}\ c
    \end{array}\right) ,
    \label{eq:Spinor Def 1}
\end{equation}
where $c \equiv \cos(\theta/2)$ and $s \equiv \sin(\theta/2) e^{i \phi}$. These two spin-spinors have been defined polarized along the direction of motion of each particle (in their helicity basis) and to be complex conjugates of each other. Other bases are allowed, but we find this basis convenient.
We also need the spin-spinors facing the other direction, which we obtain with the help of the epsilon tensor $\langle\mathbf{i}|^{\alpha I} = \epsilon^{\alpha\beta} |\mathbf{i}\rangle_\beta^I$ 
and $|\mathbf{i}\rbrack^{\dot{\alpha}}_{I} = \epsilon^{\dot{\alpha}\dot{\beta}} \lbrack\mathbf{i}|_{\dot{\beta}I}$ giving us,
\begin{equation}
    \langle\mathbf{i}|^{\alpha \mathrm{I}} =
    \left(\begin{array}{cc}
    \sqrt{E+p}\ s & \sqrt{E-p}\ c \\
    -\sqrt{E+p}\ c & \sqrt{E-p}\ s^*
    \end{array}\right)
    \quad\mbox{and}\quad
    |\mathbf{i}\rbrack^{\dot{\alpha}}_{\mathrm{I}} =
    \left(\begin{array}{cc}
    \sqrt{E+p}\ s^* & \sqrt{E-p}\ c\\
    -\sqrt{E+p}\ c & \sqrt{E-p}\ s
    \end{array}\right) ,
    \label{eq:Spinor Def 2}
\end{equation}
which are again the complex conjugates of each other, as expected.  However, we  also need to consider the situation when the spin-index is down for the angle spinor and up for the square spinor.  We achieve this by multiplying by the epsilon tensor on the right, such as, $|\mathbf{i}\rangle_{\alpha\mathrm{I}} = |\mathbf{i}\rangle_{\alpha}^{\ \mathrm{J}}\epsilon_{\mathrm{JI}}$ and $\lbrack\mathbf{i}|_{\dot{\alpha}}^{\ \mathrm{I}} = \lbrack\mathbf{i}|_{\dot{\alpha}\mathrm{J}}\epsilon^{JI}$.  We obtain,
\begin{equation}
    |\mathbf{i}\rangle_{\alpha \mathrm{I}} = 
    \left(\begin{array}{cc}
        -\sqrt{E-p}\ s^* & -\sqrt{E+p}\ c\\
        \sqrt{E-p}\ c & -\sqrt{E+p}\ s
    \end{array}\right)
    \quad\mbox{and}\quad
    \lbrack\mathbf{i}|_{\dot{\alpha}}^{\ \mathrm{I}} =
    \left(\begin{array}{cc}
        \sqrt{E-p}\ s & \sqrt{E+p}\ c\\
        -\sqrt{E-p}\ c & \sqrt{E+p}\ s^*
    \end{array}\right) ,
\end{equation}
which are minus complex conjugates of each other,
\begin{equation}
    \langle\mathbf{i}|^{\alpha}_{\ \mathrm{I}} = 
    \left(\begin{array}{cc}
        \sqrt{E-p}\ c & -\sqrt{E+p}\ s\\
        \sqrt{E-p}\ s^* & \sqrt{E+p}\ c
    \end{array}\right)
    \quad\mbox{and}\quad
    |\mathbf{i}\rbrack^{\dot{\alpha}\mathrm{I}} =
    \left(\begin{array}{cc}
        -\sqrt{E-p}\ c & \sqrt{E+p}\ s^*\\
        -\sqrt{E-p}\ s & -\sqrt{E+p}\ c
    \end{array}\right),
    \label{eq:Spinor Def 4}
\end{equation}
which are also minus the complex conjugate of one another.  Altogether, we find,
\begin{eqnarray}
    (|\mathbf{i}\rangle_{\alpha}^{\ \mathrm{I}})^* = \lbrack\mathbf{i}|_{\dot{\alpha}\mathrm{I}}
    &\quad,\quad&
    (\langle\mathbf{i}|^{\alpha\mathrm{I}})^* = |\mathbf{i}\rbrack^{\dot{\alpha}}_{\ \mathrm{I}}
    \label{eq:app:|>^I*}\\
    (|\mathbf{i}\rangle_{\alpha \mathrm{I}})^* = 
    -\lbrack\mathbf{i}|_{\dot{\alpha}}^{\ \mathrm{I}}
    &\quad,\quad&
    (\langle\mathbf{i}|^{\alpha}_{\ \mathrm{I}})^* = 
    -|\mathbf{i}\rbrack^{\dot{\alpha}\mathrm{I}}.
    \label{eq:app:|>_I*}
\end{eqnarray}
We see that when we conjugate either angle spinors with an upper spin-index or square spinors with a lower spin-index, we do not get a relative minus sign.  However, when we conjugate angle spinors with a lower spin-index or a square spinor with an upper spin-index, we do get a relative minus sign. 
Using these rules and the fact that the momenta are Hermitian, we obtain the rules,
\begin{eqnarray}
    (\langle\mathbf{i}^{\mathrm{I}}|p_1\cdots p_N
    |\mathbf{j}^{\mathrm{J}}\rangle)^* = 
    \lbrack\mathbf{j}_{\mathrm{J}}|p_N\cdots p_1|\mathbf{i}_{\mathrm{I}}\rbrack
    \label{eq:app:(<>)*}, \\
    (\langle\mathbf{i}^{\mathrm{I}}|p_1\cdots p_N
    |\mathbf{j}^{\mathrm{J}}\rbrack)^* = 
    -\langle\mathbf{j}_{\mathrm{J}}|p_N\cdots p_1|\mathbf{i}_{\mathrm{I}}\rbrack, \\
    (\lbrack\mathbf{i}^{\mathrm{I}}|p_1\cdots p_N
    |\mathbf{j}^{\mathrm{J}}\rangle)^* = 
    -\lbrack\mathbf{j}_{\mathrm{J}}|p_N\cdots p_1|\mathbf{i}_{\mathrm{I}}\rangle, \\
    (\lbrack\mathbf{i}^{\mathrm{I}}|p_1\cdots p_N
    |\mathbf{j}^{\mathrm{J}}\rbrack)^* = 
    \langle\mathbf{j}_{\mathrm{J}}|p_N\cdots p_1|\mathbf{i}_{\mathrm{I}}\rangle
    \label{eq:app:([])*}.
\end{eqnarray}
These rules are useful when we calculate a squared amplitude.

Furthermore, we include the helicity-spinors for the massless case for completeness,
\begin{align} \nonumber
    |i\rangle_\alpha &= 
      \sqrt{2E}\left(\begin{array}{c} c   \\ s \end{array}\right), \qquad
    \lbrack i|_{\dot{\alpha}} = 
      \sqrt{2E}\left(\begin{array}{c} c   \\ s^* \end{array}\right), \\ 
    \langle i|^\alpha &= 
      \sqrt{2E}\left(\begin{array}{c} s   \\ -c \end{array}\right), \qquad
    |i\rbrack^{\dot{\alpha}} = 
      \sqrt{2E}\left(\begin{array}{c} s^* \\ -c \end{array}\right).
    \label{eq:Spinor Def 5}
\end{align}
Since there is no spin-index, the rules are simpler.  The angle and square brackets are conjugates of each other giving us the rules,
\begin{eqnarray}
(\langle i|p_1\cdots p_N
    |j\rangle)^* &=& 
    \lbrack j|p_N\cdots p_1|i\rbrack,
    \label{eq:<i|pppp|j>*}\\
    (\langle i|p_1\cdots p_N
    |j\rbrack)^* &=& 
    \langle j|p_N\cdots p_1|i\rbrack,
    \label{eq:[i|pppp|j]*}
\end{eqnarray}
and vice versa.  Of course, mixed spinor-chains with one massive spinor and one massless spinor follow a similar pattern where the sign only depends on the position of the spin-index as given in Eqs.~(\ref{eq:app:|>^I*}) and (\ref{eq:app:|>_I*}).

We also note that,
\begin{align} \nonumber
    p_{i\alpha\dot{\beta}} &= |\mathbf{i}\rangle_\alpha^{\ \mathrm{I}}\lbrack\mathbf{i}|_{\dot{\beta}\mathrm{I}} =
    \left(\begin{array}{cc}
    \phantom{-} p_i^0+p_i^3   & \phantom{-} p_i^1-i p_i^2\\
    \phantom{-} p_i^1+i p_i^2 & \phantom{-} p_i^0-p_i^3
    \end{array}\right),
    \quad\mbox{and}, \\ 
    p_i^{\dot{\alpha}\beta} &= |\mathbf{i}\rbrack^{\dot{\alpha}}_{\ \mathrm{I}}\langle\mathbf{i}|^{\beta\mathrm{I}} =
    \left(\begin{array}{cc}
    \phantom{-} p_i^0-p_i^3   &          -  p_i^1+i p_i^2\\
             -  p_i^1-i p_i^2 & \phantom{-} p_i^0+p_i^3
    \end{array}\right),
    \label{eq:app:p=|>[|}
\end{align}
with,
\begin{eqnarray}
    p_i|\mathbf{i}\rangle = -m_i|\mathbf{i}\rbrack &\quad ,\quad & 
    p_i|\mathbf{i}\rbrack = -m_i|\mathbf{i}\rangle, \nonumber\\
    \langle\mathbf{i}|p_i = \lbrack\mathbf{i}|m_i &\quad ,\quad &
    \lbrack\mathbf{i}|p_i = \langle\mathbf{i}|m_i\ .
    \label{eq:pi|i>=-mi|i]}
\end{eqnarray}
Moreover, since the spin-indices are raised and lowered with an epsilon tensor, we also find,
\begin{equation}
    p_{i\alpha\dot{\beta}} = -|\mathbf{i}\rangle_{\alpha\mathrm{I}}\lbrack\mathbf{i}|_{\dot{\beta}}^{\ \mathrm{I}} 
    \quad\mbox{and}\quad
    p_i^{\dot{\alpha}\beta} = -|\mathbf{i}\rbrack^{\dot{\alpha}\mathrm{I}}\langle\mathbf{i}|^{\beta}_{\ \mathrm{I}}. 
    \label{eq:app:p=-|>[|}
\end{equation}

Additionally to the products in Eq.~(\ref{eq:app:p=|>[|}) and (\ref{eq:app:p=-|>[|}), we also find by direct computation,
\begin{eqnarray}
    |\mathbf{i}\rangle_{\alpha}^{\ \mathrm{I}}\langle\mathbf{i}|^{\beta}_{\ \mathrm{I}} = m_i\delta_\alpha^\beta 
    &\quad\mbox{,}\quad&
    |\mathbf{i}\rbrack^{\dot{\alpha}}_{\ \mathrm{I}}\lbrack\mathbf{i}|_{\dot{\beta}}^{\ \mathrm{I}}=m_i\delta_{\dot{\beta}}^{\dot{\alpha}},
    \label{eq:App:|][|=m}\\
    |\mathbf{i}\rangle_{\alpha\mathrm{I}}\langle\mathbf{i}|^{\beta\mathrm{I}} = -m_i\delta_\alpha^\beta 
    &\quad\mbox{and}\quad&
    |\mathbf{i}\rbrack^{\dot{\alpha}\mathrm{I}}\lbrack\mathbf{i}|_{\dot{\beta}\mathrm{I}}=-m_i\delta_{\dot{\beta}}^{\dot{\alpha}}.
    \label{eq:App:|][|=-m}
\end{eqnarray}
In both Eqs.~(\ref{eq:app:p=|>[|}) and (\ref{eq:app:p=-|>[|}) as well as Eqs.~(\ref{eq:App:|][|=m}) and (\ref{eq:App:|][|=-m}), we see that when dealing with the outer products of spinors, if we start with an angle bracket with an upper spin-index, we get a positive sign. Many of these fundamental outer products and building blocks have already explicitly appeared in the literature before and are collected here for easy reference. Also, if we start with a square bracket with a lower spin-index, we get a positive sign.  Starting with an angle bracket with a lower spin-index or a square bracket with an upper spin-index gives a minus sign.  Furthermore, if we have a mixture of angle and square brackets, we get a momentum, while if we have two angle brackets or two square brackets, we get a mass times a Kronecker delta function.  
When we consider inner products, we obtain
\begin{eqnarray}
    \langle \mathbf{i}_J\mathbf{i}^K\rangle = m_i\delta^K_J\quad &\mbox{,}&\quad
    \lbrack \mathbf{i}^J\mathbf{i}_K\rbrack = m_i \delta^J_K,
    \label{eq:<ij>=mdelta}\\
    \langle \mathbf{i}^K\mathbf{i}_J\rangle = -m_i\delta^K_J\quad &\mbox{and}&\quad
    \lbrack \mathbf{i}_K\mathbf{i}^J\rbrack = -m_i \delta^J_K.
    \label{eq:<ij>=-mdelta}
\end{eqnarray}

All of these identities will be very useful in reducing and generally simplifying spinor expressions.  However, before we can use them, we need the spinors to be in the correct place relative to one another.  For example, we may sometimes have contracted spinors which are not facing each other such as in $\langle \mathbf{i}^{I}|p_1\cdots p_k|\mathbf{j}^{J}\rangle \lbrack \mathbf{i}_{I}|p_{k+1}\cdots p_N|\mathbf{j}_{J}\rbrack$.  We cannot currently use any of our aforementioned identities on the spinors because they are not facing the right direction.  To use them, we have to reverse the order of one of these spinor-chains.  Any spinor-chain can be reversed following the rules,
\begin{eqnarray}
    \langle\mathbf{i}^{\mathrm{I}}|p_1\cdots p_N|\mathbf{j}^{\mathrm{J}}\rangle &=&
    (-1)^{N+1}
    \langle\mathbf{j}^{\mathrm{J}}|p_N\cdots p_1|\mathbf{i}^{\mathrm{I}}\rangle \label{eq:app:<...>=-1<...>}, \\
    \langle\mathbf{i}^{\mathrm{I}}|p_1\cdots p_N|\mathbf{j}^{\mathrm{J}}\rbrack &=&
    (-1)^{N+1}
    \lbrack\mathbf{j}^{\mathrm{J}}|p_N\cdots p_1|\mathbf{i}^{\mathrm{I}}\rangle, \\
    \lbrack\mathbf{i}^{\mathrm{I}}|p_1\cdots p_N|\mathbf{j}^{\mathrm{J}}\rbrack &=&
    (-1)^{N+1}
    \lbrack\mathbf{j}^{\mathrm{J}}|p_N\cdots p_1|\mathbf{i}^{\mathrm{I}}\rbrack.
\end{eqnarray}
This can be proved by induction.  We can easily show this is true for $N=0$.  The sign is due to the epsilon tensor.  For example,
\begin{equation}
    \langle\mathbf{i}^{I}\mathbf{j}^{J}\rangle =
    \langle\mathbf{i}^{I}|^{\alpha} \, |\mathbf{j}^{J}\rangle_{\alpha}
    = -\langle\mathbf{j}^{J}|^{\alpha} \, |\mathbf{i}^{I}\rangle_{\alpha}
    = -\langle\mathbf{j}^{J}\mathbf{i}^{I}\rangle,
\end{equation}
where we have used the epsilon tensor to raise the index on $|\mathbf{j}\rangle$ and lower the index on $\langle\mathbf{i}|$.  Now, we assume Eq.~(\ref{eq:app:<...>=-1<...>}) works for $N-1$.  We now show that it works for $N$.  Since both the left-end spinor and the right-end spinor are angle spinors, $N$ is even,
\begin{eqnarray}
    \langle\mathbf{i}^{\mathrm{I}}|p_1\cdots p_N|\mathbf{j}^{\mathrm{J}}\rangle = 
\langle\mathbf{i}^{\mathrm{I}}|p_1\cdots p_{N-1}|\mathbf{p_{N}}_{\mathrm{K}}\rbrack\langle \mathbf{p_N}^{\mathrm{K}}|\mathbf{j}^{\mathrm{J}}\rangle.
\end{eqnarray}
However, since the rules work up to $N-1$, we can reverse each of these to obtain,
\begin{eqnarray}
    \langle\mathbf{i}^{\mathrm{I}}|p_1\cdots p_N|\mathbf{j}^{\mathrm{J}}\rangle = 
(-1)^{N}
\lbrack\mathbf{p_{N}}_{\mathrm{K}}|p_{N-1}\cdots p_1|\mathbf{i}^{\mathrm{I}}\rangle
(-1)\langle \mathbf{j}^{\mathrm{J}}|\mathbf{p_N}^{\mathrm{K}}\rangle.
\end{eqnarray}
We can rearrange these to obtain,
\begin{eqnarray}
    \langle\mathbf{i}^{\mathrm{I}}|p_1\cdots p_N|\mathbf{j}^{\mathrm{J}}\rangle = 
(-1)^{N+1}
\langle \mathbf{j}^{\mathrm{J}}|\mathbf{p_N}^{\mathrm{K}}\rangle
\lbrack\mathbf{p_{N}}_{\mathrm{K}}|p_{N-1}\cdots p_1|\mathbf{i}^{\mathrm{I}}\rangle,
\end{eqnarray}
which finally gives,
\begin{eqnarray}
    \langle\mathbf{i}^{\mathrm{I}}|p_1\cdots p_N|\mathbf{j}^{\mathrm{J}}\rangle = 
(-1)^{N+1}
\langle \mathbf{j}^{\mathrm{J}}|p_N\cdots p_1|\mathbf{i}^{\mathrm{I}}\rangle.
\label{eq:<i|pppp|j> reversed}
\end{eqnarray}
The other rules are proved in a similar fashion.  Let us give a few simple examples,
\begin{eqnarray}
    \langle ij\rangle &=& -\langle ji\rangle,
    \label{eq:<ij> reversed}\\
    \langle i|p_m|j\rbrack &=& \lbrack j|p_m|i\rangle,
    \label{eq:<i|p|j]=[j|p|i>}\\
    \langle i|p_mp_n|j\rangle &=& -\langle j|p_np_m|i\rangle,
    \label{eq:<i|pp|j> reversed}\\
    \langle i|p_mp_np_l|j\rbrack &=& \lbrack j|p_lp_np_m|i\rangle,
    \label{eq:<i|ppp|j] reversed}\\ \nonumber
    &\vdots&
\end{eqnarray}

Interestingly, because the spin of each external particle is symmetrized if it is implicit, we note that we can also write identities such as,
\begin{eqnarray}
    \langle\mathbf{4}|p_1p_2|\mathbf{4}\rangle =
    -\langle\mathbf{4}|p_2p_1|\mathbf{4}\rangle,
\end{eqnarray}
if particle $4$ is an external particle.  Other similar identities follow.

Now that we have arranged all products of spinor-chains such that contracted spinors are next to each other, we can apply the identies to simplify them.  In particular, we will find products such as $\langle\mathbf{i}^{\mathrm{I}}|p_1\cdots p_k|\mathbf{j}^{\mathrm{J}}\rangle\lbrack\mathbf{j}_{\mathrm{J}}|p_{k+1}\cdots p_N|\mathbf{i}_{\mathrm{I}}\rbrack$.  For products like these, we will combine the middle two spinors into the momentum $p_j$ which is connected to the momenta $p_k$ and $p_{k+1}$.  We also combine the spinors on the two ends into the momentum $p_i$ which is then connected to the momenta $p_1$ and $p_N$.  Putting this altogether, we obtain a trace over all the momenta as,
\begin{equation}
    \langle\mathbf{i}^{\mathrm{I}}|p_1\cdots p_k|\mathbf{j}^{\mathrm{J}}\rangle\lbrack\mathbf{j}_{\mathrm{J}}|p_{k+1}\cdots p_N|\mathbf{i}_{\mathrm{I}}\rbrack = 
    \mbox{Tr}(p_1\cdots p_k p_j p_{k+1}\cdots p_N p_i).
\end{equation}
Similarly, we find,
\begin{equation}
    \langle\mathbf{i}^{\mathrm{I}}|p_1\cdots p_k|\mathbf{j}_{\mathrm{J}}\rbrack\langle\mathbf{j}^{\mathrm{J}}|p_{k+1}\cdots p_N|\mathbf{i}_{\mathrm{I}}\rbrack = 
    \mbox{Tr}(p_1\cdots p_k p_j p_{k+1}\cdots p_N p_i).
\end{equation}
Of course, if either outer product had the same type of spinor, we would get a mass times a trace of the remaining momenta.  For example,
\begin{equation}
    \langle\mathbf{i}^{\mathrm{I}}|p_1\cdots p_k|\mathbf{j}^{\mathrm{J}}\rangle\langle\mathbf{j}_{\mathrm{J}}|p_{k+1}\cdots p_N|\mathbf{i}_{\mathrm{I}}\rbrack = 
    m_j \mbox{Tr}(p_1\cdots p_k  p_{k+1}\cdots p_N p_i),
\end{equation}
and,
\begin{equation}
    \langle\mathbf{i}^{\mathrm{I}}|p_1\cdots p_k|\mathbf{j}_{\mathrm{J}}\rbrack\lbrack\mathbf{j}^{\mathrm{J}}|p_{k+1}\cdots p_N|\mathbf{i}_{\mathrm{I}}\rbrack = 
    m_j\mbox{Tr}(p_1\cdots p_k p_{k+1}\cdots p_N p_i).
\end{equation}
Of course, there are many cases.  We have only explicitly showed four.  We could have a mass coming from the ends while a momentum comes from the middle or we could have masses coming from both the middle and the ends.  Furthermore, we have illustrated the case where the spin-indices are arranged to give a positive sign, but they could be in the opposite arrangement giving minus signs.  All these cases are simple extensions of the cases we have shown here, following the rules given in Eqs.~(\ref{eq:app:p=|>[|}) through (\ref{eq:App:|][|=-m}).  Finally, it is also possible that we could have a product of three or more spinor-chains that are connected to each other in a nonseparable way.  But, this case simply follows the same rule giving a trace that includes the momentum from all the momenta in the product.  

Before calculating the traces, we emphasize that the order of the momenta matters.  There is, of course, the usual symmetry under cyclic permutation, but there is a subtelty that should be remembered when doing these traces.  Since the momenta can have either upper or lower indices as in Eq.~(\ref{eq:app:p=|>[|}), we must remember which form of the momentum we have in each position.  The convention we will use here is that we always begin traces with a momentum with lower indices.  So, in these two examples, we put $p_1$ as the first momentum of the trace because it has lower indices.  If the first momentum of the trace has upper indices, we will always use the cyclic permutation symmetry to move it to the end so that any time we write a trace, the first momentum is taken to have lower indices.   It is possible that this is only a problem in principle and may never matter in practice as we will see when we calculate the traces below.

Although we could calculate all the traces explicitly, there is a better way that is reminiscent of gamma matrices.  All the traces can be obtained inductively by removing pairs of momenta and relating the trace to one with two less momenta by use of the cyclic permutation symmetry and the anti-commutation property of the momenta\cite{Christensen:2018zcq, Chisholm1963, Kahane1968}, 
\begin{equation}
    p_{k\alpha\dot{\beta}}p_l^{\dot{\beta}\omega}+p_{l\alpha\dot{\beta}}p_k^{\dot{\beta}\omega} = 2p_k\cdot p_l \delta_{\alpha}^{\omega} \, .
    \label{eq:p1p2=p2p1-2p1.p2}
\end{equation}
This allows us to rearrange the order of any adjacent momenta in a trace as in,
\begin{equation}
    \mbox{Tr}(p_1\cdots p_i p_{i+1} \cdots p_N) = 
    2p_i\cdot p_{i+1} \mbox{Tr}(p_1\cdots p_{i-1}p_{i+2}\cdots p_N) - \mbox{Tr}(p_1\cdots p_{i+1}p_i\cdots p_N).
\end{equation}
Therefore, we can move any momenta to any position.  For example, if we have the same momentum separated by another momentum, such as,
\begin{equation}
    \mbox{Tr}(\cdots p_ip_jp_i\cdots),
\end{equation}
We can rewrite this as,
\begin{equation}
    \mbox{Tr}(\cdots p_ip_jp_i\cdots) = 
    2p_i\cdot p_j\mbox{Tr}(\cdots p_i\cdots) - M_i^2\mbox{Tr}(\cdots p_j\cdots),
\end{equation}
where we have used that $p_ip_i = M_i^2 \delta$, where $\delta$ is the Kronecker delta function on SL$(2,\mathbb{C})$ indices. We can do similar procedures when the momentum $p_i$ is separated by more than one momentum in between.  We just need to permute the momenta multiple times until the $p_i$ are next to each other.  Therefore, we only need consider traces over a product of unique momenta. 

Now, if all the momenta are unique, we can determine the trace with any number of momenta inside, though there are subtleties to consider.  We might think that we simply permute the first momentum until it is at the end, as in $\mbox{Tr}(p_2\cdots p_Np_1)$, and then move it to the front again using the cyclic symmetry property of the trace. However, we cannot do this because the momentum at the end of the trace has upper indices whereas the momentum at the beginning has lower indices, therefore, it isn't trivial to move it to the front like it is when tracing gamma matrices.  Instead, we need to raise and lower all the indices, resulting in a transpose of the entire product of momenta, as in $\mbox{Tr}(p_1p_N\cdots p_2)$.  We then need to move everything back to its original position in order to obtain our identity.  Unfortunately, at the end of this procedure, we get $\mbox{Tr}(p_1\cdots p_N)=\mbox{Tr}(p_1\cdots p_N)$, where everything else has exactly cancelled, and the identity we obtain is trivial and not helpful.

Another idea we might have is to move $p_1$ to the second to last position as in $\mbox{Tr}(p_2p_3\cdots p_1p_N)$ and then use the cyclic property to move $p_1p_N$ to the beginning.  We can do this since $p_1$ has lower indices here, giving us $\mbox{Tr}(p_1p_Np_2\cdots p_{N-1})$.  However, we now need to move $p_N$ back to the end.  Once again, at the end of this procedure, we obtain $\mbox{Tr}(p_1\cdots p_N)=\mbox{Tr}(p_1\cdots p_N)$, and this doesn't work either.

Nevertheless, we are able to find the trace of any number of momenta recursively by putting the trace into a totally antisymmetric form.  Though we don't have a master formula for any trace, we do have a master form for a totally antisymmetric trace and we will show that the process of putting the trace in a totally antisymmetric form results in smaller traces that we already know.  In order to make this clear, we will work several explicit examples with up to six momenta.  We begin with zero and one momenta, the foundation of our recursive method.  The trace of zero momenta is simply a trace of the identity and gives 2 while the trace of one momentum gives twice its energy,
\begin{eqnarray}
    \mbox{Tr}( \mathbb{I} ) &=& 2,
    \label{eq:tr()}\\
    \mbox{Tr}(p) &=&
    2E \, .
\end{eqnarray}
The trace over a single momentum is not Lorentz invariant.  In fact, we have never seen it, nor has any trace over an odd number of momenta come up in any calculations and we believe it never will due to the SL$(2,\mathbb{C})$ symmetry, although  we have not attempted a formal proof.  Therefore, we will focus solely on traces over an even number of momenta.  

We begin with two momenta.  Our first step is to put it in a form that is totally antisymmetric.  To make this procedure as clear as possible, we begin by writing this as a sum of two identical traces,
\begin{equation}
    \mbox{Tr}(p_1p_2) = \frac{1}{2!}\left[\mbox{Tr}(p_1p_2) + \mbox{Tr}(p_1p_2)\right].
\end{equation}
We next reorder the momenta in the second term obtaining,
\begin{eqnarray}
    \mbox{Tr}(p_1p_2) = \frac{1}{2!}\left[\mbox{Tr}(p_1p_2) + 2p_1\cdot p_2\mbox{Tr}(\mathbb{I}) - \mbox{Tr}(p_2p_1)\right].
\end{eqnarray}
Since we know the trace of the identity, we can insert it and pull the $p_1\cdot p_2$ out of the square brackets.  We also write the antisymmetric trace in a generic way that we will see again.  We have,
\begin{equation}
    \mbox{Tr}(p_1p_2) = 2\MP{1}{2} + \frac{1}{2!}\epsilon^{ij}\trt{i}{j},
\end{equation}
where $i$ and $j$ are over $1$ and $2$.  At this point, we must consider what the totally antisymmetric trace could possibly be.  The result must be Lorentz invariant and, besides the momenta, there are only two other structures that can be used to form Lorentz invariant products, namely the metric and the Levi-Civita epsilon tensor.  However, if we used the metric, we would get zero since it is symmetric in its indices and the product is already antisymmetrized.  That is $\epsilon^{ij}\MP{i}{j}=0$.  On the other hand, the Levi-Civita epsilon tensor has four indices and also cannot be used to form a nonzero Lorentz invariant product.  Therefore, we learn that this totally antisymmetric trace of two momenta must be zero,
\begin{equation}
    \frac{1}{2!}\epsilon^{ij}\trt{i}{j} = 0.
\end{equation}
  This can also easily be seen by explicit calculation.  With this, we are left with,
\begin{eqnarray}
    \mbox{Tr}(p_1p_2) &=&
    2p_1\cdot p_2.
\end{eqnarray}
This identity is already well known for the helicity-spinor product $\langle ji\rangle\lbrack ij\rbrack$ and also applies to the same product with spin-spinors with spin indices suitably contracted as in,
\begin{eqnarray}
    \langle ji\rangle\lbrack ij\rbrack =
    \mbox{Tr}(p_ip_j) =  2p_i\cdot p_j,
    \label{eq:app:<ji>[ij]=2pipj massless}\\
    \langle\mathbf{j}^{\mathrm{J}}\mathbf{i}^{\mathrm{I}}\rangle
    \lbrack\mathbf{i}_{\mathrm{I}}\mathbf{j}_{\mathrm{J}}\rbrack =
    \mbox{Tr}(p_ip_j) =  2p_i\cdot p_j .
    \label{eq:app:<ji>[ij]=2pipj}
\end{eqnarray}
Furthermore, other spin-index positions can be taken into account by lowering and raising the indices using the epsilon tensor and accounting for the appropriate signs.

Our next explicit example contains 4 momenta.  We find
\begin{eqnarray}
    \mbox{Tr}(p_1p_2p_3p_4) &=& \frac{1}{4!}\left[
    \, 4! \, \mbox{Tr}(p_1p_2p_3p_4)
    \right]
    \nonumber\\
    &=& \frac{1}{4!}\Big[
    \mbox{Tr}(p_1p_2p_3p_4) 
    \nonumber\\
    && + \, 2p_3\cdot p_4\mbox{Tr}(p_1p_2) - \mbox{Tr}(p_1p_2p_4p_3)
    \nonumber\\
    &&
    + \, 2p_2\cdot p_3\mbox{Tr}(p_1p_4) - \mbox{Tr}(p_1p_3p_2p_4)
    \nonumber\\
    &&
    + \, 2p_2\cdot p_3\mbox{Tr}(p_1p_4) - 2p_2\cdot p_4\mbox{Tr}(p_1p_3) + \mbox{Tr}(p_1p_3p_4p_2)
    \nonumber 
     +\cdots \Big] \nonumber \\
    &=&
    2\MP{1}{2}\MP{3}{4} - 2\MP{1}{3}\MP{2}{4} \nonumber \\
    && \qquad + \, 2\MP{1}{4}\MP{2}{3} + \frac{1}{4!}\epsilon^{ijkl}\trf{i}{j}{k}{l},
\end{eqnarray}
where we have used the already known trace of two momenta.  At this point, we must once again determine the totally antisymmetric trace.  Once again, since the result must be Lorentz invariant and any term containing the metric will vanish, we are left with the Levi-Civita epsilon tensor.  This time, there is a match between the number of indices on the epsilon tensor and the number of momenta.  Therefore, up to an overall constant, we expect $\epsilon^{ijkl}\trf{i}{j}{k}{l}\propto \epsilon^{ijkl}\epsilon_{\mu\nu\alpha\beta}p_i^\mu p_j^\nu p_k^\alpha p_l^\beta$.  Indeed, this is what we find explicitly with an overall factor of $2i$, resulting in
\begin{equation}
    \trf{1}{2}{3}{4} = 
    2\MP{1}{2}\MP{3}{4} - 2\MP{1}{3}\MP{2}{4} + 2\MP{1}{4}\MP{2}{3} + \frac{2i}{4!}\epsilon^{ijkl}\epsilon_{\mu\nu\alpha\beta}p_i^\mu p_j^\nu p_k^\alpha p_l^\beta
\end{equation}
We see that the antisymmetric piece is imaginary and resembles the trace over 4 gamma matrices and a $\gamma_5$.

We will do one more explicit example, that with six momenta. We begin by totally antisymmetrizing the momenta and plugging in the non-antisymmetrized traces, giving,
\begin{eqnarray}
    \trs{1}{2}{3}{4}{5}{6} &=& 
    2\MP{1}{2}\MP{3}{4}\MP{5}{6} - 2\MP{1}{2}\MP{3}{5}\MP{4}{6} + 2\MP{1}{2}\MP{3}{6}\MP{4}{5} \nonumber\\
    &+& \cdots \nonumber \\
    &+& \frac{1}{4!}\MP{1}{2}\epsilon^{ijkl}\trf{i}{j}{k}{l}\rvert_{3456} - \frac{1}{4!}\MP{1}{3}\epsilon^{ijkl}\trf{i}{j}{k}{l}\rvert_{2456} \nonumber \\
    &+& \cdots \nonumber \\
    &+& \frac{1}{4!}\MP{5}{6}\epsilon^{ijkl}\trf{i}{j}{k}{l}\rvert_{1234}
     + \frac{1}{6!}\epsilon^{ijklmn}\trs{i}{j}{k}{l}{m}{n},
\end{eqnarray}
where the $\rvert_{3456}$ means that the sum is over the momenta $p_3,p_4,p_5$ and $p_6$ and similarly for the other terms.  We already know the totally antisymmetrized trace of four momenta.  For the totally antisymmetrized trace of six momenta, we see, once again, that it cannot have a metric in it since that would vanish.  Therefore, we must use only the Levi-Civita epsilon tensor to contract the Lorentz indices of the momenta.  However, there is a mismatch between the six Lorentz indices of our six-momentum trace and the four Lorentz indices of the Levi-Civita epsilon tensor.  Therefore, we find that $\epsilon^{ijklmn}\trs{i}{j}{k}{l}{m}{n}=0$.  Plugging this all in, we find,
\begin{eqnarray}
    \trs{1}{2}{3}{4}{5}{6} &=& 
    2\MP{1}{2}\MP{3}{4}\MP{5}{6} - 2\MP{1}{2}\MP{3}{5}\MP{4}{6} + 2\MP{1}{2}\MP{3}{6}\MP{4}{5} 
    \nonumber\\
    && - 2\MP{1}{3}\MP{2}{4}\MP{5}{6} + 2\MP{1}{3}\MP{2}{5}\MP{4}{6} - 2\MP{1}{3}\MP{2}{6}\MP{4}{5}
    \nonumber\\
    && + 2\MP{1}{4}\MP{2}{3}\MP{5}{6} - 2\MP{1}{4}\MP{2}{5}\MP{3}{6} + 2\MP{1}{4}\MP{2}{6}\MP{3}{5}
    \nonumber\\
    && - 2\MP{1}{5}\MP{2}{3}\MP{4}{6} + 2\MP{1}{5}\MP{2}{4}\MP{3}{6} - 2\MP{1}{5}\MP{2}{6}\MP{3}{4}
    \nonumber\\
    && + 2\MP{1}{6}\MP{2}{3}\MP{4}{5} - 2\MP{1}{6}\MP{2}{4}\MP{3}{5} + 2\MP{1}{6}\MP{2}{5}\MP{3}{4}
    \nonumber\\
    && + \frac{2i}{4!}\MP{1}{2}\epsilon^{ijkl}\epsilon_{\mu\nu\alpha\beta}p_i^\mu p_j^\nu p_k^\alpha p_l^\beta\rvert_{3456} - \frac{2i}{4!}\MP{1}{3}\epsilon^{ijkl}\epsilon_{\mu\nu\alpha\beta}p_i^\mu p_j^\nu p_k^\alpha p_l^\beta\rvert_{2456}
    \nonumber\\
    &&+ \frac{2i}{4!}\MP{1}{4}\epsilon^{ijkl}\epsilon_{\mu\nu\alpha\beta}p_i^\mu p_j^\nu p_k^\alpha p_l^\beta\rvert_{2356} - \frac{2i}{4!}\MP{1}{5}\epsilon^{ijkl}\epsilon_{\mu\nu\alpha\beta}p_i^\mu p_j^\nu p_k^\alpha p_l^\beta\rvert_{2346}
    \nonumber\\
    &&+ \frac{2i}{4!}\MP{1}{6}\epsilon^{ijkl}\epsilon_{\mu\nu\alpha\beta}p_i^\mu p_j^\nu p_k^\alpha p_l^\beta\rvert_{2345} + \frac{2i}{4!}\MP{2}{3}\epsilon^{ijkl}\epsilon_{\mu\nu\alpha\beta}p_i^\mu p_j^\nu p_k^\alpha p_l^\beta\rvert_{1456}
    \nonumber\\
    && - \frac{2i}{4!}\MP{2}{4}\epsilon^{ijkl}\epsilon_{\mu\nu\alpha\beta}p_i^\mu p_j^\nu p_k^\alpha p_l^\beta\rvert_{1356} + \frac{2i}{4!}\MP{2}{5}\epsilon^{ijkl}\epsilon_{\mu\nu\alpha\beta}p_i^\mu p_j^\nu p_k^\alpha p_l^\beta\rvert_{1346}
    \nonumber\\
    &&- \frac{2i}{4!}\MP{2}{6}\epsilon^{ijkl}\epsilon_{\mu\nu\alpha\beta}p_i^\mu p_j^\nu p_k^\alpha p_l^\beta\rvert_{1345} + \frac{2i}{4!}\MP{3}{4}\epsilon^{ijkl}\epsilon_{\mu\nu\alpha\beta}p_i^\mu p_j^\nu p_k^\alpha p_l^\beta\rvert_{1256}
    \nonumber\\
    && - \frac{2i}{4!}\MP{3}{5}\epsilon^{ijkl}\epsilon_{\mu\nu\alpha\beta}p_i^\mu p_j^\nu p_k^\alpha p_l^\beta\rvert_{1246} + \frac{2i}{4!}\MP{3}{6}\epsilon^{ijkl}\epsilon_{\mu\nu\alpha\beta}p_i^\mu p_j^\nu p_k^\alpha p_l^\beta\rvert_{1245}
    \nonumber\\
    && + \frac{2i}{4!}\MP{4}{5}\epsilon^{ijkl}\epsilon_{\mu\nu\alpha\beta}p_i^\mu p_j^\nu p_k^\alpha p_l^\beta\rvert_{1236} - \frac{2i}{4!}\MP{4}{6}\epsilon^{ijkl}\epsilon_{\mu\nu\alpha\beta}p_i^\mu p_j^\nu p_k^\alpha p_l^\beta\rvert_{1235}
    \nonumber\\
    && + \frac{2i}{4!}\MP{5}{6}\epsilon^{ijkl}\epsilon_{\mu\nu\alpha\beta}p_i^\mu p_j^\nu p_k^\alpha p_l^\beta\rvert_{1234}.
\end{eqnarray}

At this point, we will not work out any further explicit traces.  Rather, we will just note that, in general, for a trace over any (even) number of momenta, we can simply permute the momenta into all possible orders.  As we do this and use the anticommutation properties of the momenta, we will reduce the trace to traces over a smaller number of momenta, which we already know and a totally antisymmetric trace over all the momenta,
\begin{equation}
    \mbox{Tr}(p_1p_2\cdots p_N) = 
    \mbox{(traces over less than $N$ momenta)}
    + \frac{1}{N!}\epsilon^{i_1i_2\cdots i_N}\mbox{Tr}(p_{i_1}p_{i_2}\cdots p_{i_N}).
\end{equation}
The first term is given recursively by smaller traces.  The second term is only nonzero if $N$ is divisible by $4$ and is a Lorentz invariant product of momenta times Levi-Civita epsilon tensors, which is totally antisymmetrized over the momenta.  That is,
\begin{equation}
    \frac{1}{N!}\epsilon^{i_1i_2\cdots i_N}\mbox{Tr}(p_{i_1}p_{i_2}\cdots p_{i_N})
    = \frac{2i}{N!}\epsilon^{i_1i_2\cdots i_N}\epsilon_{\mu_1\mu_2\mu_3\mu_4}\cdots \epsilon_{\mu_{N-3}\mu_{N-2}\mu_{N-1}\mu_N}p_{i_1}^{\mu_1}p_{i_2}^{\mu_2}\cdots p_{i_N}^{\mu_N}
\end{equation}
if $N$ is divisible by $4$ and it is $0$ otherwise.  We have tested this formula for up to 8 momenta, but expect it to hold for an arbitrary number of momenta.

\section{\label{sec:generalized Schouten}Generalized Schouten Identities}
As we simplify the amplitudes, we will make significant use of the Schouten identity.  In this subsection, we will review and work out a generalized form of it when there are momenta in the spinor-chains.  The Schouten identity\cite{Dixon:1996wi} for spinor products is based on that for epsilon tensors,
\begin{equation}
    \epsilon_{\alpha\beta}\epsilon_{\gamma\delta} = 
    \epsilon_{\alpha\gamma}\epsilon_{\beta\delta} -
    \epsilon_{\alpha\delta}\epsilon_{\beta\gamma},
\end{equation}
and is identical with dotted indices. Since our spinors are contracted with these, we can write,
\begin{equation}
    \langle\mathbf{ji}\rangle\langle\mathbf{lk}\rangle = 
    \epsilon^{\alpha\beta}
    \epsilon^{\gamma\delta}
    |\mathbf{i}\rangle_{\alpha}
    |\mathbf{j}\rangle_{\beta}
    |\mathbf{k}\rangle_{\gamma}
    |\mathbf{l}\rangle_{\delta} \, .
\end{equation}
Then, applying the identity on the epsilon tensors, we find,
\begin{eqnarray}
    \langle\mathbf{ji}\rangle\langle\mathbf{lk}\rangle &=& 
    \langle\mathbf{ki}\rangle\langle\mathbf{lj}\rangle -
    \langle\mathbf{li}\rangle\langle\mathbf{kj}\rangle, \\
    \lbrack\mathbf{ji}\rbrack\lbrack\mathbf{lk}\rbrack &=& 
    \lbrack\mathbf{ki}\rbrack\lbrack\mathbf{lj}\rbrack -
    \lbrack\mathbf{li}\rbrack\lbrack\mathbf{kj}\rbrack.
\end{eqnarray}
The same identity applies independent of whether the spinors are massive or massless.  The spin-indices just go along for the ride.  This is the simplest case of the Schouten identity, however, we note that there is a generalization of these identities. We will also need an expression when there are momenta sandwiched in between the spinors.  The momenta stick to their spinors and reverse their ordering when flipped from a right-facing spinor to a left-facing spinor and vice-versa.  In order to introduce the more general identity, we first work out the effect of raising an SL$(2,\mathbb{C})$ index on a chain of momenta ending in a spinor.  For example, consider,
\begin{equation}
    \epsilon^{\alpha\beta}p_{1\beta\dot{\beta}}p_2^{\dot{\beta}\gamma}\cdots p_N^{\dot{\delta}\delta}|l\rangle_{\delta},
\end{equation}
where we are raising the index $\beta$.  Each of the contractions can be rewritten as a product of epsilon tensors.  For example, $\delta^{\dot{\beta}}_{\dot{\zeta}}=\epsilon^{\dot{\beta}\dot{\omega}}\epsilon_{\dot{\omega}\dot{\zeta}}$.  This gives us,
\begin{equation}
\epsilon^{\alpha\beta}p_{1\beta\dot{\beta}}p_2^{\dot{\beta}\gamma}\cdots p_N^{\dot{\delta}\delta}|l\rangle_{\delta} =
    \epsilon^{\alpha\beta}p_{1\beta\dot{\beta}}\epsilon^{\dot{\beta}\dot{\omega}}\epsilon_{\dot{\omega}\dot{\zeta}}p_2^{\dot{\zeta}\gamma}\epsilon_{\gamma\kappa}\cdots \epsilon_{\dot{\mu}\dot{\delta}}p_N^{\dot{\delta}\delta}\epsilon_{\delta\theta}\epsilon^{\theta\lambda}|l\rangle_{\lambda}.
\end{equation}
However, we remember  from the text above Eq.~(\ref{eq:Spinor Def 2}) that both SL$(2,\mathbb{C})$ epsilon matrices must hit the momentum from the left when using matrix notation.  This requires us to transpose the epsilon on the right of every momentum, which introduces a sign.  The final epsilon flips the direction of the spinor giving us
\begin{equation}
\epsilon^{\alpha\beta}p_{1\beta\dot{\beta}}p_2^{\dot{\beta}\gamma}\cdots p_N^{\dot{\delta}\delta}|l\rangle_{\delta}=
    (-1)^N
    (\epsilon^{\alpha\beta}\epsilon^{\dot{\omega}\dot{\beta}}p_{1\beta\dot{\beta}})
    (\epsilon_{\dot{\omega}\dot{\zeta}}\epsilon_{\kappa\gamma}p_2^{\dot{\zeta}\gamma})
    \cdots 
    (\epsilon_{\dot{\mu}\dot{\delta}}\epsilon_{\theta\delta}p_N^{\dot{\delta}\delta})
    \langle l|^{\theta}
\end{equation}
where $N$ is the number of momenta in the product.  The epsilon tensors acting on the momenta raise and lower the indices but put them in the wrong order for matrix notation.  We have to also transpose the resulting matrix.  However, this is precisely the relationship between the momentum with upper and lower indices [see Eq.~(\ref{eq:app:p=|>[|})].  We raise and lower with epsilon tensors followed by a transposition to switch between one and the other.  Therefore, we have, 
\begin{equation}
\epsilon^{\alpha\beta}(
p_1p_2\cdots p_N|l\rangle)_{\beta}=
    (-1)^N
    (\langle l|
    p_{N}
    \cdots 
    p_{2}
    p_{1})^{\dot{\omega}},
    \label{eq:eps pppp|l>}
\end{equation}
where $N$ is the number of momenta in this product.  Again, whether the spinor is a helicity-spinor or a spin-spinor is irrelevant.  The spin-index does not change or affect this identity.  We get a similar result if we raise or lower the SL$(2,\mathbb{C})$ index on a left facing spinor.  In effect, whenever we raise or lower the SL$(2,\mathbb{C})$ index on a spinor, we get the reverse spinor, with all the momenta reversed as well and a sign flip for each of those momenta.  

We are now in a position to use the Schouten identity on any product of spinor-chains.  We begin with the completely general case.  We will then describe a mnemonic to remember the rule and give a few useful examples.  Suppose we begin with,
\begin{eqnarray}
\mbox{prod}&=&
\langle\mathbf{1}|p_{j_1}\cdots p_{j_m}p_{j_{m+1}}\cdots p_{j_M}|\mathbf{2}\rbrack
\langle\mathbf{3}|p_{k_1}\cdots p_{k_n}p_{k_{n+1}}\cdots p_{k_N}|\mathbf{4}\rangle, 
\label{eq:general Schouten 1}
\end{eqnarray}
where $\mathbf{1}, \mathbf{2}, \mathbf{3}$ and $\mathbf{4}$ represent general spinors.  Now, suppose we wanted to rearrange the spinor-chains at $p_{j_m}$ and $p_{k_n}$.  We could insert a delta function between the $p_{j_m}p_{j_{m+1}}$ and $p_{k_n}p_{k_{n+1}}$ and replace the delta functions by a product of epsilon tensors to obtain
\begin{equation}
    \mbox{prod}
=
    (-1)^{m+n}
    \epsilon^{\alpha\beta}\epsilon^{\gamma\delta}
    (p_{j_{m+1}}\cdots p_{j_M}|\mathbf{2}\rbrack)_{\alpha}
    (p_{j_m}\cdots p_{j_1}|\mathbf{1}\rangle)_{\beta}
    (p_{k_{n+1}}\cdots p_{k_N}|\mathbf{4}\rangle)_{\gamma}
    (p_{k_n}\cdots p_{k_1}|\mathbf{3}\rangle)_{\delta}
\end{equation}
where we have used the epsilon tensor on the left facing spinors to reverse their direction [as in Eq.~(\ref{eq:eps pppp|l>})].  As shown in that equation, we get a minus sign for every momentum we reversed and there are $m+n$ of them.  We next apply the Schouten identity directly to the epsilon tensors.  This splits the right side into two terms, each with its product of epsilon tensors.  Now that we have the two terms, we simply use the epsilon tensors to flip the direction of spinor $3$ and spinor $4$ and their accompanying momenta.  As before, we get a minus sign for every momentum we flip.  This leads us to the final result,
\begin{eqnarray}
    \mbox{prod}
&=&
    (-1)^{m+N-n}\Big[
    \langle\mathbf{4}|p_{k_N}\cdots p_{k_{n+1}} 
    p_{j_{m+1}}\cdots p_{j_M}|\mathbf{2}\rbrack
    \langle\mathbf{3}|p_{k_1}\cdots p_{k_n}
    p_{j_m}\cdots p_{j_1}|\mathbf{1}\rangle 
    \nonumber\\
&& -
    \langle\mathbf{4}|p_{k_{N}}\cdots p_{k_{n+1}}
    p_{j_m}\cdots p_{j_1}|\mathbf{1}\rangle
    \langle\mathbf{3}|p_{k_1}\cdots p_{k_n}
    p_{j_{m+1}}\cdots p_{j_M}|\mathbf{2}\rbrack
    \Big].
    \label{eq:general Schouten 2}
\end{eqnarray}
There are a few things we note about this result.  We flip the same number of momenta in both terms so that the extra sign is the same for both terms and we can factor it out.  We reverse the first $m$ momenta of the first spinor-chain and the last $N-n$ momenta of the second chain.  Therefore, our overall sign is $(-1)^{m+N-n}$.  Second, we did this case of the Schouten identity by splitting on an undotted index, but we could have done it with a dotted index with exactly the same result.  It does not matter whether the index is dotted or undotted, but it has to be the same in both spinor-chains.  We cannot split at an undotted index in one chain and a dotted index on the other chain.  This may make it seem overly complicated but, actually, all one needs to remember is that spinor-chains with either two angle brackets at its ends or two square brackets at its ends must have an even number of momenta in the middle.  Spinor-chains with a mixture of angle and square brackets at its ends, on the other hand, must have an odd number of momenta in the middle.  This applies both before and after applying the Schouten identity.  Therefore, you only need ensure that you apply the Schouten identity in a way that satisfies this rule.  Third, once you have chosen where to split the spinor-chains, whichever spinors you reverse, you simply reverse all the momenta that are connected to that spinor (after the split).  

In this general form, it can appear quite daunting to remember, but actually, since the form is very general, it is not so bad.  We find the following mnemonic to be useful: ``four two three one minus four one three two with an extra sign for every momentum reversed''.  The first part simply refers to the two other orders available.  Spinor 4 began connected to spinor 3.  So, now we are connecting it to the next one over first, namely spinor 2.  That is the origin of the ``four two three one''.  The other order is spinor 4 with spinor 1 (the next one to the left) and this order comes with the opposite sign and this gives rise to the ``minus four one three two''.   Finally, if we reverse any momenta, we add an extra sign for each of them giving rise to the ``with an extra sign for every momentum reversed''.  With this mnemonic, we find we can apply the Schouten identity to any product of spinor-chains that come up without difficulty.

We will consider a few useful examples that have come up multiple times in our calculations.  We begin with the well-known classic case of no momenta in the middle,
\begin{eqnarray}
    \langle ij\rangle \langle kl \rangle &=& \langle lj\rangle\langle ki\rangle - \langle li\rangle \langle kj\rangle, \\
    \lbrack ij\rbrack \lbrack kl \rbrack &=& \lbrack lj\rbrack\lbrack ki\rbrack - \lbrack li\rbrack \lbrack kj\rbrack.
\end{eqnarray}
We show the various cases with helicity-spinors, but the same identity applies to spin-spinors or mixtures of helicity and spin-spinors.
Next, let's consider just one momentum sandwiched in just one of the spinor-chains, 
\begin{eqnarray}
    \langle ij\rangle \langle k|p_m|l\rbrack &=&
    -\lbrack l|p_m|j\rangle \langle ki\rangle + \lbrack l|p_m|i\rangle \langle kj\rangle,
    \label{eq:Schouten:<ij><k|p|l]}\\
    \lbrack ij\rbrack \langle k|p_m|l\rbrack &=&
    \lbrack lj\rbrack \langle k|p_m|i\rbrack - 
    \lbrack li\rbrack \langle k|p_m|j\rbrack.
    \label{eq:Schouten:[ij]<k|p|l]}
\end{eqnarray}
In the first case, the $p_m$ had to contract with spinor $l$ in order to keep the appropriate number of momenta sandwiched between the spinors, namely an odd number when mixing angle and square spinors.  Since spinor $l$ was flipped, so was $p_m$ and therefore, we find an overall minus sign.  In the second example, the momentum $p_m$ had to contract with spinor $k$ in order to ensure the appropriate number of momenta in the chains.  In this case, spinor $k$ was not flipped, and neither was its momentum $p_m$, therefore, there was no additional sign.  

As we increase to two momenta, we see that the number of cases begins to increase significantly.  We could have both momenta in the first or the second chain or one in each.  In the first case, we could have both chains have angle brackets, or we could have one with angle brackets and one with square brackets.  Furthermore, there are multiple ways of breaking the momenta up.  Let's begin with both momenta in the same chain.  We find,
\begin{eqnarray}
    \langle i|p_mp_n|j\rangle \langle kl\rangle &=& \langle lj\rangle \langle k|p_np_m|i\rangle - \langle l|p_np_m|i\rangle \langle kj\rangle,
    \label{eq:Schouten:<i|pp|j><kl> 1}\\
    \langle i|p_mp_n|j\rangle \langle kl\rangle &=& \langle l|p_mp_n|j\rangle \langle ki\rangle - \langle li\rangle \langle k|p_mp_n|j\rangle,
    \label{eq:Schouten:<i|pp|j><kl> 2}\\
    \langle i|p_mp_n|j\rangle \lbrack kl\rbrack &=& 
    -\lbrack l|p_n|j\rangle \lbrack k|p_m|i\rangle + \lbrack l|p_m|i\rangle \lbrack k|p_n|j\rangle.
    \label{eq:Schouten:<i|pp|j>[kl]}
\end{eqnarray}
In the first two cases, we must keep an even number of momenta in the chains, so we must split at one end of the momentum chain or the other, while in the third case, we must end with an odd number of momenta in each chain since they will have both angle and square spinors, so we must split in the middle of the momenta.  Concerning the signs, in the first case, we chose to keep both momenta with spinor $i$, therefore, we had to reverse both of them.  But, since there are an even number of momenta reversed, no extra sign is required.  In the second case, both momenta are kept with spinor $j$, so neither is reversed and there is, again, no additional sign.  In the third case, we reverse one momentum, namely $p_m$ since it is contracted with spinor $i$, therefore, there is an extra sign in the third case.  Lastly, we consider the case where the momenta are one each in the chains,
\begin{eqnarray}
    \langle i|p_m|j\rbrack \langle k|p_n|l\rbrack &=&
    -\lbrack lj\rbrack \langle k|p_np_m|i\rangle +
    \lbrack l|p_m|i\rangle \langle k|p_n|j\rbrack,
    \label{eq:Schouten:<i|p|j]<k|p|l] 1}\\
    \langle i|p_m|j\rbrack \langle k|p_n|l\rbrack &=&
    -\lbrack l|p_np_m|j\rbrack \langle ki\rangle +
    \lbrack l|p_n|i\rangle \langle k|p_m|j\rbrack.
    \label{eq:Schouten:<i|p|j]<k|p|l] 2}
\end{eqnarray}
As we saw in Eq.~(\ref{eq:<i|p|j]=[j|p|i>}), the direction that the spinor-chains on the left face does not matter.  We reverse one momentum in both cases so we find an extra minus sign.

\section{\label{app:explicit spinor products}Useful Explicit Spinor Products}
In this appendix, we use the definitions of the spinors in Eqs.~(\ref{eq:Spinor Def 1}) through (\ref{eq:Spinor Def 4}) and (\ref{eq:Spinor Def 5}) to form several explicit spinor products that are useful in Sec.~\ref{sec:2-body decays}.

We begin with both spinors being massive.  If they are angle spinors, we have,
\begin{equation}
    \langle \mathbf{i}^{\mathrm{I}}\mathbf{j}^{\mathrm{J}} \rangle = \left(\begin{array}{cc}
    \sqrt{(E_i+p_i)(E_j+p_j)}(s_i c_j - c_i s_j) &
     -\sqrt{(E_i+p_i)(E_j-p_j)}(c_i c_j  + s_i s^*_j)\\
    \sqrt{(E_i-p_i)(E_j+p_j)}(s^*_i s_j  + c_i c_j)&
     -\sqrt{(E_i-p_i)(E_j-p_j)}(c_i s^*_j -s^*_i c_j)
    \end{array}\right),
    \label{eq:<i^Ij^J> explicit}
\end{equation}
whereas, if they are square spinors, we obtain,
\begin{equation}
    \lbrack \mathbf{i}^{\mathrm{I}}\mathbf{j}^{\mathrm{J}} \rbrack =
    \left(
\begin{array}{cc}
 \sqrt{\left(E_i-p_i\right)\left(E_j-p_j\right)}
   (c_i s_j-c_j s_i) &
   \sqrt{\left(E_i-p_i\right)\left(E_j+p_j\right)}
   (c_i c_j+s_i s_j^*) \\
 -\sqrt{\left(E_i+p_i\right)\left(E_j-p_j\right)}
   (c_i c_j+s_i^* s_j) &
   \sqrt{\left(E_i+p_i\right)\left(E_j+p_j\right)}
   (c_i s_j^*-c_j s_i^*) \\
\end{array}
\right).
    \label{eq:[i^Ij^J] explicit}
\end{equation}

We will also sometimes need the spinor product when one particle is massive and the other massless.  Once again, we begin with angle brackets obtaining,
\begin{equation}
    \langle \mathbf{i}^{\mathrm{I}} \mathrm{j} \rangle =
    \sqrt{2E_j}
    \left(\begin{array}{cc}
    \sqrt{(E_i+p_i)}
    (s_i c_j - c_i s_j) \\
    \sqrt{(E_i-p_i)}
    (s^*_i s_j +
    c_i c_j)
    \end{array}\right),
    \label{eq:<i^Ij> explicit}
\end{equation}
and,
\begin{equation}
    \langle \mathbf{{\mathrm{i}}j}^{\mathrm{J}} \rangle = 
    \sqrt{2E_i}
    \left(\begin{array}{cc}
    \sqrt{(E_j+p_j)}(s_i c_j -c_i s_j) &\hspace{0.1in}
    -\sqrt{(E_j-p_j)}(c_i c_j +s_i s^*_j)
    \end{array}\right).
\end{equation}
If we have square brackets, we find,
\begin{equation}
\lbrack \mathbf{i}^{\mathrm{I}} \mathrm{j} \rbrack = 
    \sqrt{2E_j}
\left(\begin{array}{cc}
        \sqrt{(E_i-p_i)}(c_i c_j + s_i s^*_j ) \\
        \sqrt{(E_i+p_i)}(c_i s^*_j - s^*_i c_j) 
    \end{array}\right),
\end{equation}
and,
\begin{equation}
    \lbrack \mathbf{{\mathrm{i}}j}^{\mathrm{J}} \rbrack =
    \sqrt{2E_i}    
    \left(\begin{array}{cc}
        -\sqrt{(E_j-p_j)}(c_i c_j+s^*_i s_j) &
        \sqrt{(E_j+p_j)}(c_i s^*_j -s^*_i c_j)
    \end{array}\right).
    \label{eq:[ij^J] explicit}
\end{equation}

Finally, if both particles are massless we obtain,
\begin{equation}
    \langle \mathrm{i {\mathrm{j}}} \rangle = 
    \sqrt{4E_i E_j}(s_i c_j - c_i s_j ),
    \label{eq:<ij> massless}
\end{equation}
and,
\begin{equation}
    \lbrack \mathrm{i {\mathrm{j}}} \rbrack = 
    \sqrt{4E_i E_j}(
    c_i s^*_j - s^*_i c_j 
    ).
    \label{eq:[ij] massless}
\end{equation}


\begin{thebibliography}{2019}

\bibitem{Parke:1986gb} 
  S.~J.~Parke and T.~R.~Taylor,
  ``An Amplitude for $n$ Gluon Scattering,''
  Phys.\ Rev.\ Lett.\  {\bf 56}, 2459 (1986).
  doi:10.1103/PhysRevLett.56.2459

\bibitem{Britto:2005fq} 
  R.~Britto, F.~Cachazo, B.~Feng and E.~Witten,
  ``Direct proof of tree-level recursion relation in Yang-Mills theory,''
  Phys.\ Rev.\ Lett.\  {\bf 94}, 181602 (2005)
  doi:10.1103/PhysRevLett.94.181602
  [hep-th/0501052].
  
\bibitem{Eden:1966dnq} 
  R.~J.~Eden, P.~V.~Landshoff, D.~I.~Olive and J.~C.~Polkinghorne,
  ``The analytic S-matrix,''

\bibitem{Dixon:1996wi} 
  L.~J.~Dixon,
  ``Calculating scattering amplitudes efficiently,''
  hep-ph/9601359.

\bibitem{DelDuca:1999rs} 
  V.~Del Duca, L.~J.~Dixon and F.~Maltoni,
  ``New color decompositions for gauge amplitudes at tree and loop level,''
  Nucl.\ Phys.\ B {\bf 571}, 51 (2000)
  doi:10.1016/S0550-3213(99)00809-3
  [hep-ph/9910563].

\bibitem{Dixon:2013uaa} 
  L.~J.~Dixon,
  ``A brief introduction to modern amplitude methods,''
  doi:10.5170/CERN-2014-008.31
  arXiv:1310.5353 [hep-ph].

\bibitem{Feng:2011np} 
  B.~Feng and M.~Luo,
  ``An Introduction to On-shell Recursion Relations,''
  Front.\ Phys.\ (Beijing) {\bf 7}, 533 (2012)
  doi:10.1007/s11467-012-0270-z
  [arXiv:1111.5759 [hep-th]].

\bibitem{Belyaev:2012qa} 
  A.~Belyaev, N.~D.~Christensen and A.~Pukhov,
  ``CalcHEP 3.4 for collider physics within and beyond the Standard Model,''
  Comput.\ Phys.\ Commun.\  {\bf 184}, 1729 (2013)
  doi:10.1016/j.cpc.2013.01.014
  [arXiv:1207.6082 [hep-ph]].
  
\bibitem{Elvang:2015rqa} 
  H.~Elvang and Y.~t.~Huang,
  ``Scattering Amplitudes in Gauge Theory and Gravity,''

\bibitem{Arkani-Hamed:2017jhn} 
  N.~Arkani-Hamed, T.~C.~Huang and Y.~t.~Huang,
  ``Scattering Amplitudes For All Masses and Spins,''
  arXiv:1709.04891 [hep-th].
  
\bibitem{Weinberg:1995mt} 
  S.~Weinberg,
  ``The Quantum theory of fields. Vol. 1: Foundations,''

\bibitem{Christensen:2018zcq} 
  N.~Christensen and B.~Field,
  ``Constructive standard model,''
  Phys.\ Rev.\ D {\bf 98}, no. 1, 016014 (2018)
  doi:10.1103/PhysRevD.98.016014
  [arXiv:1802.00448 [hep-ph]].
  
\bibitem{Boels:2017gyc} 
  R.~H.~Boels and H.~Luo,
  ``A minimal approach to the scattering of physical massless bosons,''
  JHEP {\bf 1805}, 063 (2018)
  doi:10.1007/JHEP05(2018)063
  [arXiv:1710.10208 [hep-th]].

\bibitem{Ochirov:2018uyq} 
  A.~Ochirov,
  ``Helicity amplitudes for QCD with massive quarks,''
  JHEP {\bf 1804}, 089 (2018)
  doi:10.1007/JHEP04(2018)089
  [arXiv:1802.06730 [hep-ph]].

\bibitem{Afkhami-Jeddi:2018apj} 
  N.~Afkhami-Jeddi, S.~Kundu and A.~Tajdini,
  ``A Bound on Massive Higher Spin Particles,''
  JHEP {\bf 1904}, 056 (2019)
  doi:10.1007/JHEP04(2019)056
  [arXiv:1811.01952 [hep-th]].

\bibitem{Alwall:2011uj} 
  J.~Alwall, M.~Herquet, F.~Maltoni, O.~Mattelaer and T.~Stelzer,
  ``MadGraph 5 : Going Beyond,''
  JHEP {\bf 1106}, 128 (2011)
  doi:10.1007/JHEP06(2011)128
  [arXiv:1106.0522 [hep-ph]].

\bibitem{Maltoni:2002qb} 
  F.~Maltoni and T.~Stelzer,
  ``MadEvent: Automatic event generation with MadGraph,''
  JHEP {\bf 0302}, 027 (2003)
  doi:10.1088/1126-6708/2003/02/027
  [hep-ph/0208156].

\bibitem{Bellm:2015jjp} 
  J.~Bellm {\it et al.},
  ``Herwig 7.0/Herwig++ 3.0 release note,''
  Eur.\ Phys.\ J.\ C {\bf 76}, no. 4, 196 (2016)
  doi:10.1140/epjc/s10052-016-4018-8
  [arXiv:1512.01178 [hep-ph]].
  
\bibitem{Gleisberg:2008ta} 
  T.~Gleisberg, S.~Hoeche, F.~Krauss, M.~Schonherr, S.~Schumann, F.~Siegert and J.~Winter,
  ``Event generation with SHERPA 1.1,''
  JHEP {\bf 0902}, 007 (2009)
  doi:10.1088/1126-6708/2009/02/007
  [arXiv:0811.4622 [hep-ph]].

\bibitem{Kilian:2007gr} 
  W.~Kilian, T.~Ohl and J.~Reuter,
  ``WHIZARD: Simulating Multi-Particle Processes at LHC and ILC,''
  Eur.\ Phys.\ J.\ C {\bf 71}, 1742 (2011)
  doi:10.1140/epjc/s10052-011-1742-y
  [arXiv:0708.4233 [hep-ph]].
  
\bibitem{Mathematica}
  Wolfram Research, Inc., Mathematica, Version 12.0, Champaign, IL (2019).

\bibitem{Chisholm1963}
  J.S.R.~Chisholm,
  ``Relativistic scalar products of $\gamma$ matrices,''
  Il\ Nuovo\ Cimento\ {\bf 30}, 426 (1963)
  doi:10.1007/BF02750778
  
\bibitem{Kahane1968}
  J.~Kahane,
  ``Algorithm for Reducing Contracted Products of $\gamma$ Matrices,''
  J.~Math.~Phys. {\bf 9}, 1732 (1968)
  doi:10.1063/1.1664506


\bibitem{Shadmi:2018xan} 
  Y.~Shadmi and Y.~Weiss,
  ``Effective Field Theory Amplitudes the On-Shell Way: Scalar and Vector Couplings to Gluons,''
  JHEP {\bf 1902}, 165 (2019)
  doi:10.1007/JHEP02(2019)165
  [arXiv:1809.09644 [hep-ph]].

\bibitem{Ma:2019gtx} 
  T.~Ma, J.~Shu and M.~L.~Xiao,
  ``Standard Model Effective Field Theory from On-shell Amplitudes,''
  arXiv:1902.06752 [hep-ph].
  
\bibitem{Aoude:2019tzn} 
  R.~Aoude and C.~S.~Machado,
  ``The Rise of SMEFT On-shell Amplitudes,''
  JHEP {\bf 1912}, 058 (2019)
  doi:10.1007/JHEP12(2019)058
  [arXiv:1905.11433 [hep-ph]].
  
\bibitem{Durieux:2019eor} 
  G.~Durieux, T.~Kitahara, Y.~Shadmi and Y.~Weiss,
  ``The electroweak effective field theory from on-shell amplitudes,''
  arXiv:1909.10551 [hep-ph].
  
\bibitem{Franken:2019wqr} 
  R.~Franken and C.~Schwinn,
  ``On-shell constructibility of Born amplitudes in spontaneously broken gauge theories,''
  arXiv:1910.13407 [hep-th].

\end{thebibliography}
\end{document}